\documentclass[preprint,12pt]{elsarticle}

\usepackage{amssymb}

\usepackage{subcaption}
\journal{arXiv}

\begin{document}

\begin{frontmatter}



\title{Comparative study on Mechanical characteristics of Functionally graded and Coreshell nanospheres: An Atomistic approach}

\author[label1]{Prottay Malakar}
\author[label1]{Md Al Rifat Anan}
\author[label1]{Mahmudul Islam}
\author[label3]{Md Shajedul Hoque Thakur}
\author[label1,label2]{Satyajit Mojumder \corref{cor1}}
\cortext[cor1]{Corresponding Author}
\ead{satyajitmojumder2022@u.northwestern.edu}
\affiliation[label1]{
	organization={Department of Mechanical Engineering, Bangladesh University of Engineering and Technology},
	city={Dhaka},
	postcode={1000},
	country={Bangladesh}}
\affiliation[label3]{
	organization={Department of Materials Science and NanoEngineering,Rice University},
	city={Houston},
	postcode={77005},
	state={TX},
	country={USA}}
\affiliation[label2]{
	organization={ Theoretical and Applied Mechanics Program, Northwestern University},
	city={Evanston},
	postcode={60208},
	state={IL},
	country={USA}}

%\author{}

%\affiliation{organization={},%Department and Organization

\begin{abstract}
Functionally Graded Material (FGM) is a type of advanced material consisting of two (or more) distinct substances with a constantly changing composition profile. FGM technologies have moved from their traditional use to advanced micro and nanoscale electronics and energy conversion systems along with the advent of nanotechnology. MD simulations are used in this analysis to examine the effect of compressive load on Ag-Au FGM and Core-shell nanospheres. The plasticity process is often started by the nucleation of partial dislocations from the contact surfaces, and these dislocations spread towards the nanosphere's center. Also, we have found the formation of pyramidal-shaped partial dislocations on the pseudo-plastic regime. For a given wt\% range of Ag in Au, Coreshell nanospheres have stronger mechanical strength than FGM nanospheres, and we have also observed two distinct patterns in ultimate stress variation for FGM and Coreshell nanospheres. The dislocation analysis suggests a correlation between this stress variation and the Shockley \& Hirth partial dislocation density.
\end{abstract}


%%\item Research highlight 1
%%\item Research highlight 2


\end{frontmatter}


\section{Introduction}
Functionally graded materials (FGMs) are relatively new and promising advanced high-performance materials among the class of advanced composite materials, where the material composition is continuously varied along a dimension following a particular function to regulate the mechanical and physical properties\cite{shanmugavel2012overview}. FGMs are still in the early stages of development, but they are expected to have a significant impact on the design and development of novel components and structures that perform better. Many studies have been conducted to better understand the thermal and mechanical properties of FGMs. Continuum mechanics and the finite element technique (FEM) are primarily used in this process\cite{moita2018material,moleiro2019deformations,song2018bending,zaoui2019new,woo2003thermomechanical,belabed2014efficient,sachdeva2018functionally}. According to Udupa et al. \cite{udupa2014functionally} and Howard\cite{howard1996delamination}, FGMS can interact as an interface layer between two incompatible materials, improving bond strength, decreasing stress concentration, and providing multifunctionality to control deformation, dynamic response, corrosion, wear, and other properties.

Recently, there has been a surge of interest in learning more about the plastic deformation of FGM nanoparticles. Despite this, the corresponding literature is still inadequate due to the difficulties associated with experimental mechanical testing. The mechanical properties of nanoparticles, specifically the nanosphere, have been measured using the compression or indentation approach. Gerberich et al. made one of the earliest observations by compressing gold and silicon nanospheres and analyzing the plastic deformation with transmission electron microscopy\cite{gerberich2003superhard,corcoran1997anomalous,mook2007compressive,gerberich2005reverse,deneen2006situ}. He tested the silicon nanosphere's hardness, which is up to four times that of its bulk value \cite{gerberich2003superhard}. The modulus of elasticity of gold nanoparticles reduced with increasing indentation depth, according to Ramos et al\cite{ramos2013hardness}. Mordehai et al. found that the nucleation process of dislocation in gold nanoparticles begins at the surface and spreads to the adjacent surfaces of the faceted particles, leading through defect-free particles during indentation, which is consistent with Gerberich's mechanism for silicon nanospheres \cite{gerberich2003superhard,mordehai2011nanoindentation}. The structure of the nanoparticle is thought to have an important impact in dislocation propagation. The nucleation of first Shockley partial dislocations is observed from the contact edge, followed by a pseudotwinning process\cite{prakash2015atom}. All of these studies on the properties of nanoparticles are based on theoretical models and finite element methods (FEM), both of which are related to the classical mechanics approach. 

Molecular dynamics (MD) simulations can now be an effective way to examine the mechanical and thermal properties of nanoparticles and understand that too. Several MD studies have already been carried out to understand the mechanical and thermal properties of nanowires, nanoplates, nanospheres and nanobeams of the same or different homogenous and non-homogeneic composites. By compressing in the [001] crystallographic direction, Bian and Wang \cite{bian2013atomistic} conducted molecular dynamics simulations on a copper nanosphere and thoroughly characterized the nucleation of dislocations process. They demonstrated how Shockley partial dislocations formed at the nanosphere-indenter contact edges and propagated. However, the earlier research was limited to a single copper nanosphere size (20nm). Further, Salah, Gerard and Pizzagalli performed molecular dynamics simulations on aluminium nanosphere under compression to show the influence of size on plasticity mechanism and yield stress\cite{salah2017influence}. The elastic modulus and strength of gold nanoparticles are dimension-dependent \cite{armstrong2012size,mordehai2011size}. Experiments confirmed the possession of higher modulus and stress in gold nanoparticles compared with bulk gold \cite{kim2008robust}. Through in-situ high-resolution transmission electron microscopy, it was discovered that silver nanoparticles as small as 10 nm may deform like a liquid droplet at ambient temperature\cite{sun2014liquid}. In addition, some ultra-small silver nanoparticles showed reversible plasticity dislocation\cite{carlton2012situ}. The yield stress of gold nanoparticles fluctuates according to surface atomic-scale morphologies, according to Yang et al.\cite{yang2017impact} Under [0 0 1] compression, face-centered-cubic (fcc) nanoparticle plasticity normally starts with partial dislocation nucleation from the contact edge and the formation of distinctive pyramid hillocks \cite{bian2013atomistic,salah2017influence}.

However, there has not been much research on the influence of dislocation propagation on stress in FGM nanospheres.  The goal of this research is to look at the initiation and propagation of dislocations in Ag-Au FGMs and Coreshell nanospheres.  We will also look into how the stress-strain curve is affected by dislocation propagation. In this study, we have put Ag-Au FGM and Coreshell nanospheres through a compression test at a specific temperature and pressure using MD simulations. The following section examines the deformation mechanism of nanospheres and the development of dislocations in further detail. We have also talked about how dislocation density affects nanosphere mechanical characteristics. There has been relatively little investigation of the mechanical properties of FGM and Coreshell nanospheres at any scale, according to prior studies. Our findings will help scientists better understand the alloying effect and deformation mechanism of FGM and Coreshell nanospheres.

\section{Methodology}
\begin{figure}[htbp]
	\centerline{\includegraphics[width=0.8\textwidth]{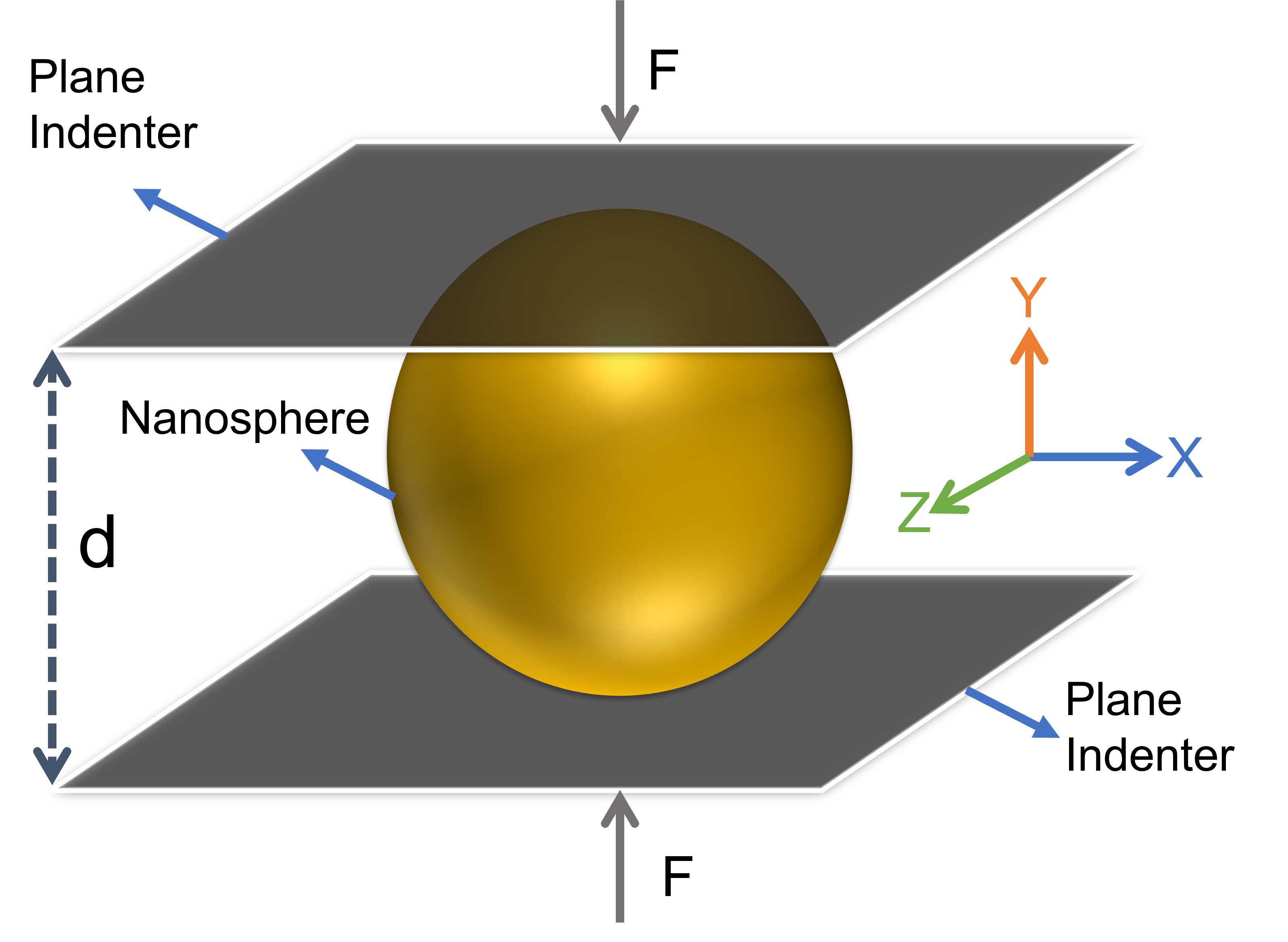}}
	\caption{Compression Setup of a nanosphere by Plane Indenter.}
	\label{fig1}
\end{figure}
In this study, the MD Simulation method has been considered to investigate the contact stress of the Ag-Au alloy nanospheres and the effect of dislocations \& alloying structure on them. These simulations are conducted by using Large-Scale Atomic/Molecular Massively Parallel Simulator, LAMMPS\cite{plimpton1995fast}. The selection of interatomic potential has a significant influence on the accuracy of simulation results. We has been used metal alloy embedded method potential (EAM/alloy) developed by Zhou et al\cite{zhou2004misfit}.This potential can successfully simulate the dislocations and their nucleation and propagation accurately\cite{zhou2004misfit}. Previously, Embedded method potential (eam) was used for different atomistic investigations of monatomic elements\cite{gu2012md}. In the EAM, the total energy of the crystal can be shown as below:

\begin{equation}
	E = \frac{1}{2}\sum_{i,j,j\neq i} \phi_{i,j}(r_{i,j}) + \sum_{i}F_{i}(\rho_{i})
\end{equation}

Where the pair energy $\phi_{i,j}$ is considered between two atoms $i \& j$ which are separated by distance $r_{i,j}$ . $F_{i}$ is the embedded energy and $\rho_{i}$ is the electron density associated with the atom $i$. The electron density $\rho_{i}$  can be calculated by using the equation, 

\begin{equation}
	\rho_{i} = \sum_{j,j\neq i}f_{j}(r_{i,j})
\end{equation}
This equation indicates the total sum of electron density of atom $i$ which is $r_{i,j}$ distance away from the different atoms $j$. For an alloy, the metal alloy EAM potential contains not only these three basic functions but also the pair energy between different alloying elements. This indicates that the alloy EAM potentials can be developed from elementary EAM alloys and such potentials can be used in MD simulations to keep consistency with experimental results\cite{zhou2001atomic}.

%%Mainly, for a monatomic element, the EAM potential is a collation of three basic functions of pair energy, embedding energy, and electron density. This potential is precious to analyze the stresses and dislocation in different monatomic elements. Many researchers developed different EAM potentials for different elements and additionally combined them to study alloys. But literature shows that without normalizing the potentials, the combination of monatomic elements affects the predicted alloy properties\cite{johnson1989alloy}.  %%

In this investigation, Core-shell nanospheres are generated by replacing core atoms with Ag atoms from pure Au nanospheres according to the required mass percentage. On the other hand, FGM nanospheres are modeled by using a slight modification of the source code of the nano-hub tool developed by Thakur el al\cite{thakur2019lammps}. This tool's validity has been demonstrated in previous research by Islam et al\cite{islam2020mechanical}. The nanospheres used for this study are shown in Fig.\ref{fig1}. For modeling the radially graded Ag-Au FGM nanosphere, we have strictly maintained three steps: (1) A pure Ag nanosphere is prepared with proper dimension (15 nm dia) with the lattice constant of 0.40853 nm. (2) The nanosphere is then subdivided into some annular spherical chunks of half lattice constant thickness. (3) In each chunk, the Ag atoms are replaced by Au atoms according to the grading function. Here, We have used the power-law function as our grading function. The details of the function are described by using the equation,

%\begin{table}[htbp]
%	\caption{Mass fraction distribution of Au in Ag nanosphere for functionally graded alloy}
%	\begin{center}
%		\resizebox{\textwidth}{!}{
%		\begin{tabular}{|p{2cm}|p{4cm}|p{7cm}|}
%			\hline
%			\textbf{Type of Graded Material} &	\textbf{Grading Function}& \textbf{Mass function Au in Ag, $g(r)$} \\
%			\hline
%			\centering FGM & \centering Power-law function \newline & $g(r) = (\frac{r}{R})^p$ \\
%			\hline
%		\end{tabular}}
%		\label{tab1}
%	\end{center}
%\end{table}
\begin{equation}
    g(r) = (\frac{r}{R})^p
\end{equation}

\begin{figure}[htbp]
\centering
	\begin{subfigure}{0.8\textwidth}
		\centerline{\includegraphics[width=1\textwidth]{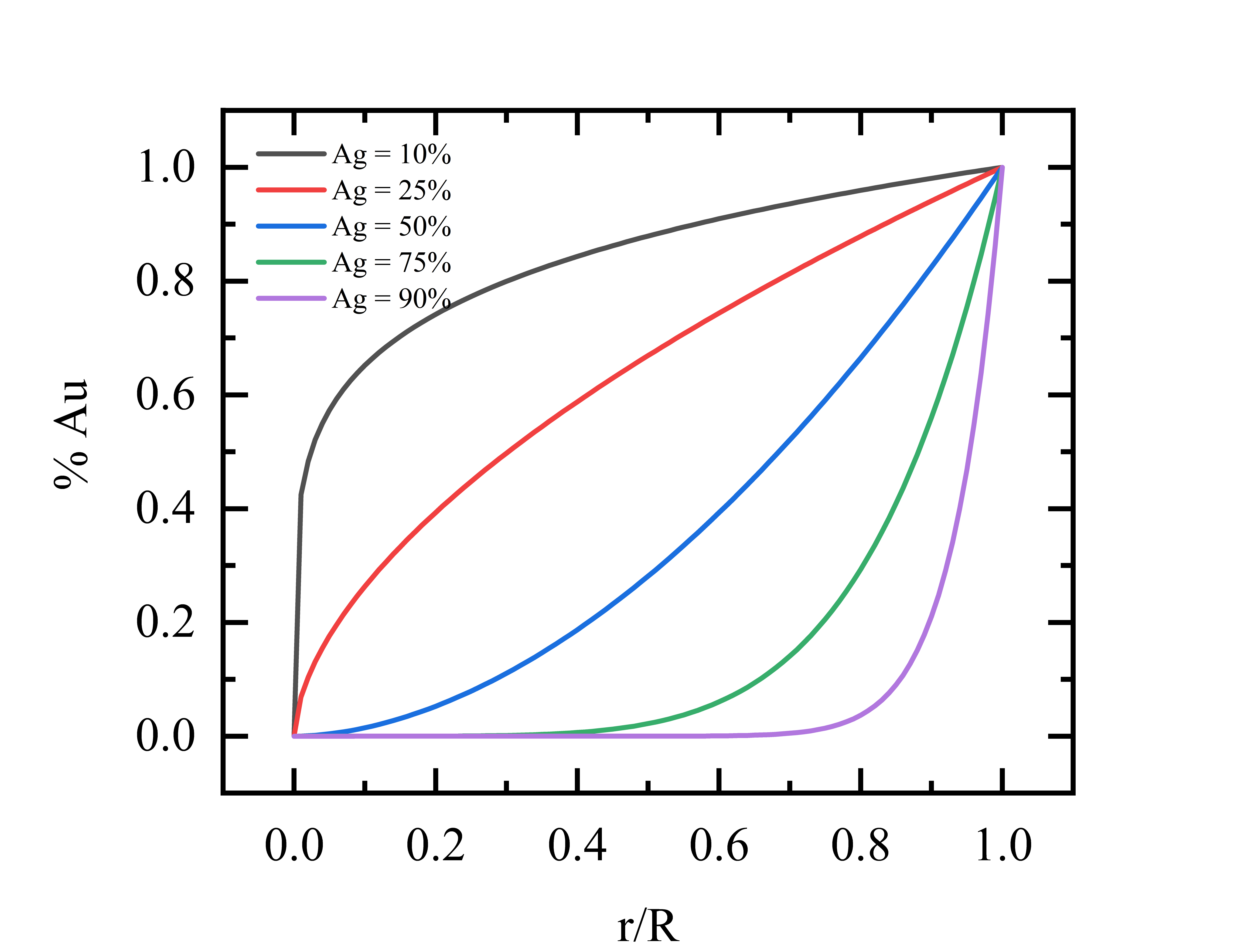}}
		\caption{ }
		\label{fig2a}
	\end{subfigure}
	\begin{subfigure}{0.8\textwidth}
		\centerline{\includegraphics[width=1\textwidth]{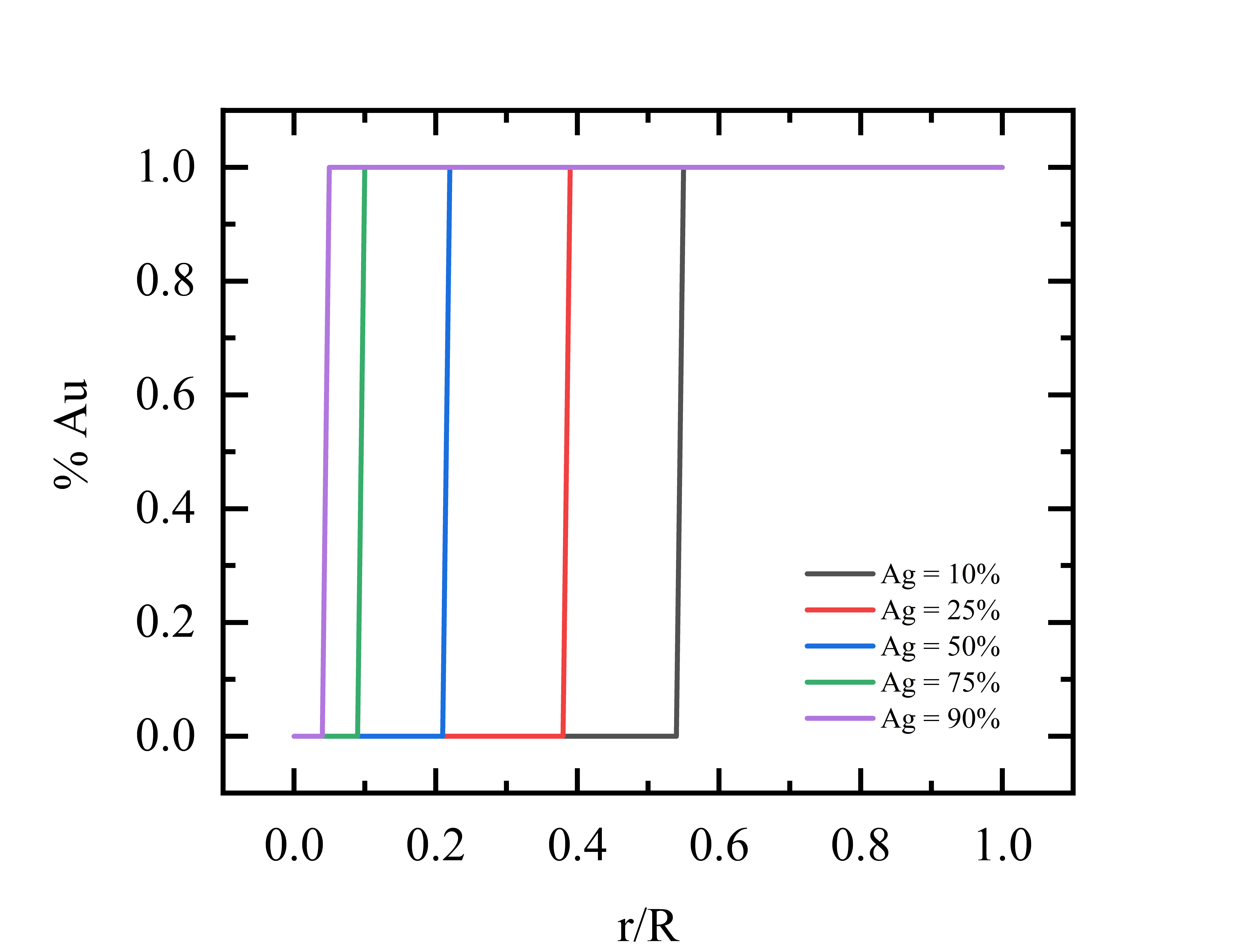}}
		\caption{ }
		\label{fig2b}
	\end{subfigure}
	\caption{Mass function distribution of Au in Ag of (a) FGM and (b) Coreshell nanospheres for different percentages of Ag.}
	\label{fig2}
\end{figure}

Here, the distribution of the mass function of Au in Ag, $g(r)$ is determined by the parameter p, according to the percentage of Ag and Au along the radial direction. Fig. \ref{fig2} shows the Mass function distribution of Au in Ag of FGM and Core-shell nanospheres for different percentages of Ag.

\begin{figure}[htbp]
\centering
	\begin{subfigure}{0.45\textwidth}
		\centerline{\includegraphics[width=0.6\textwidth]{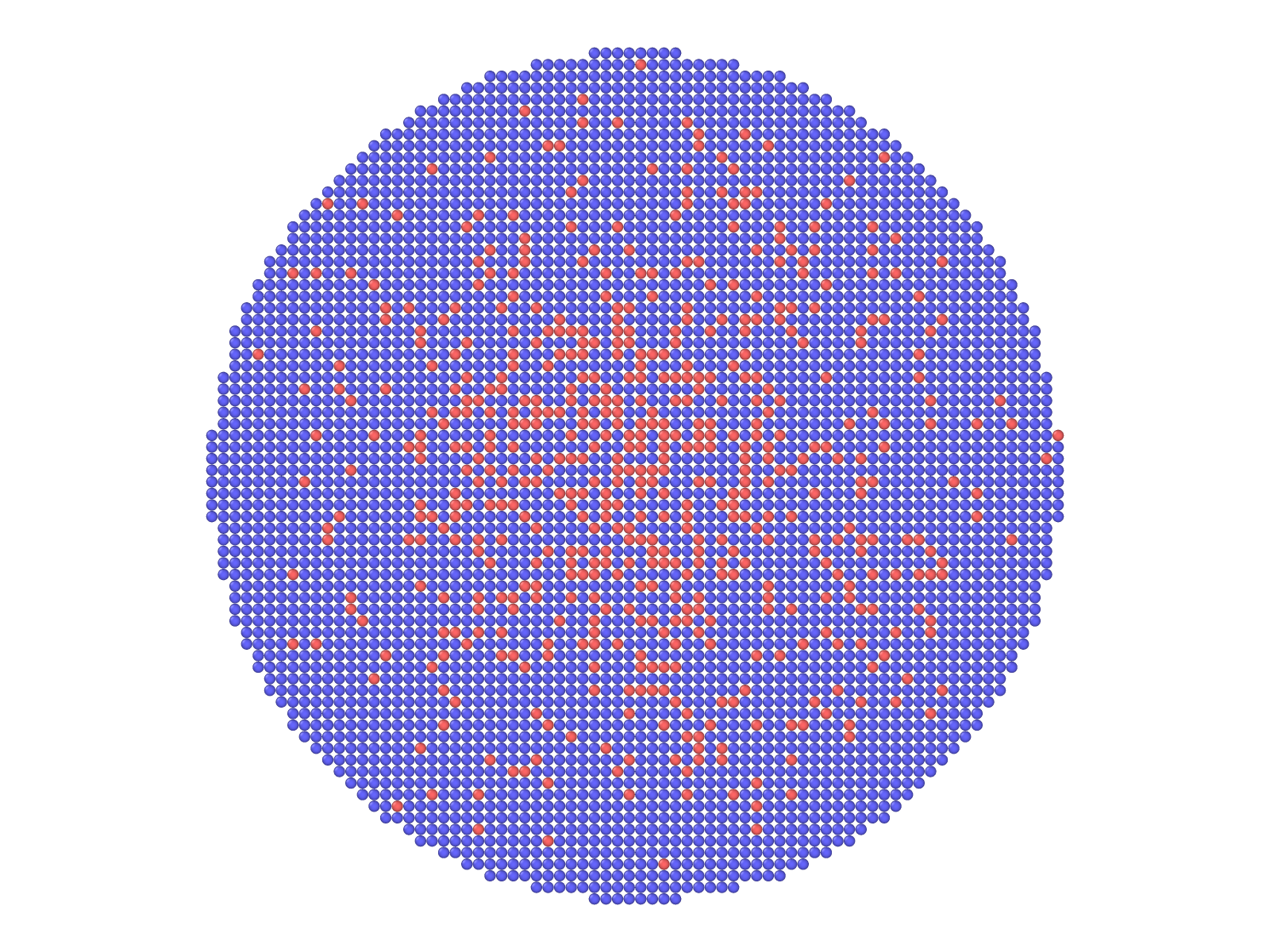}}
		\caption*{10\%}
		\centerline{\includegraphics[width=0.6\textwidth]{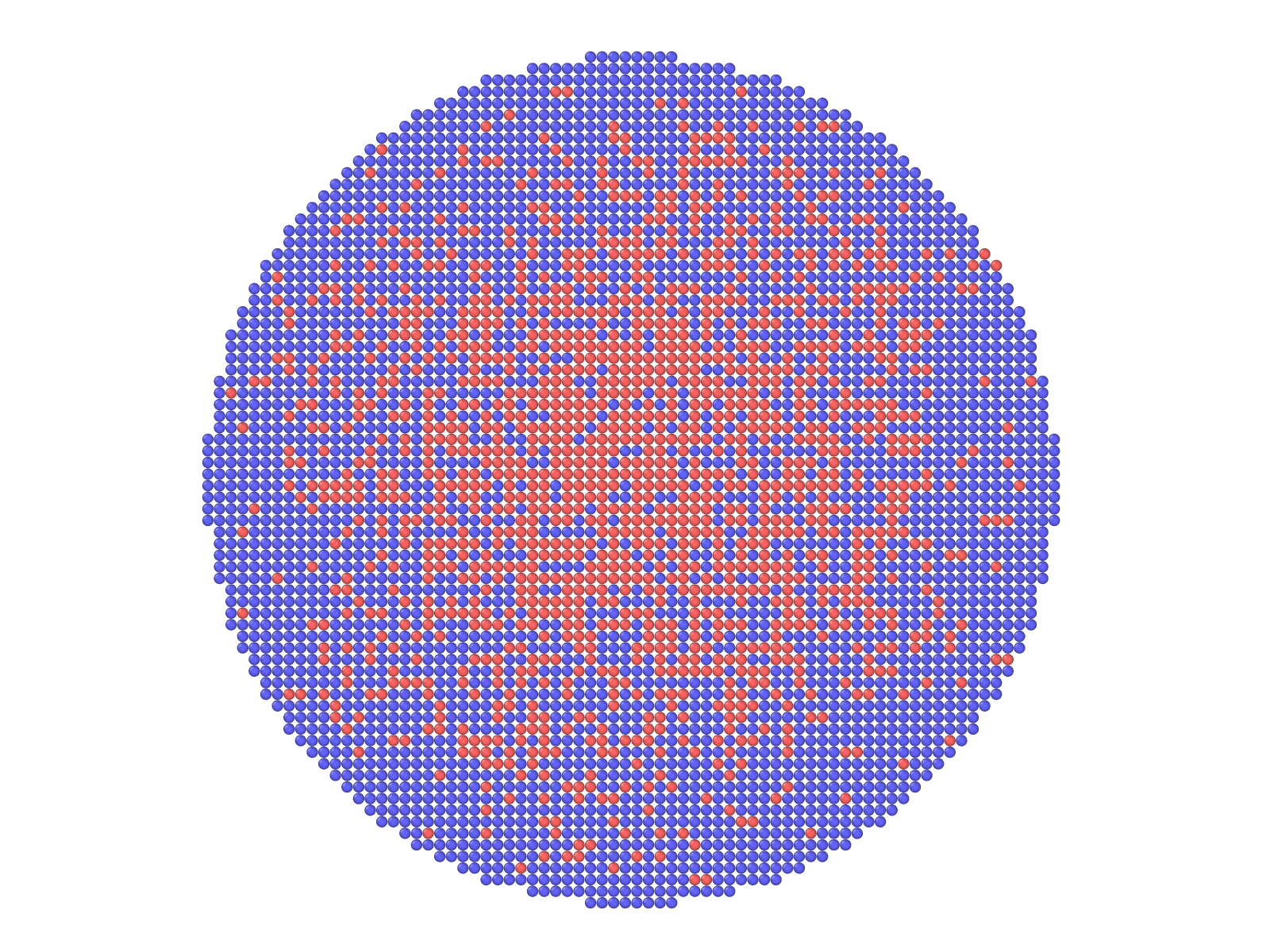}}
		\caption*{25\%}
		\centerline{\includegraphics[width=0.6\textwidth]{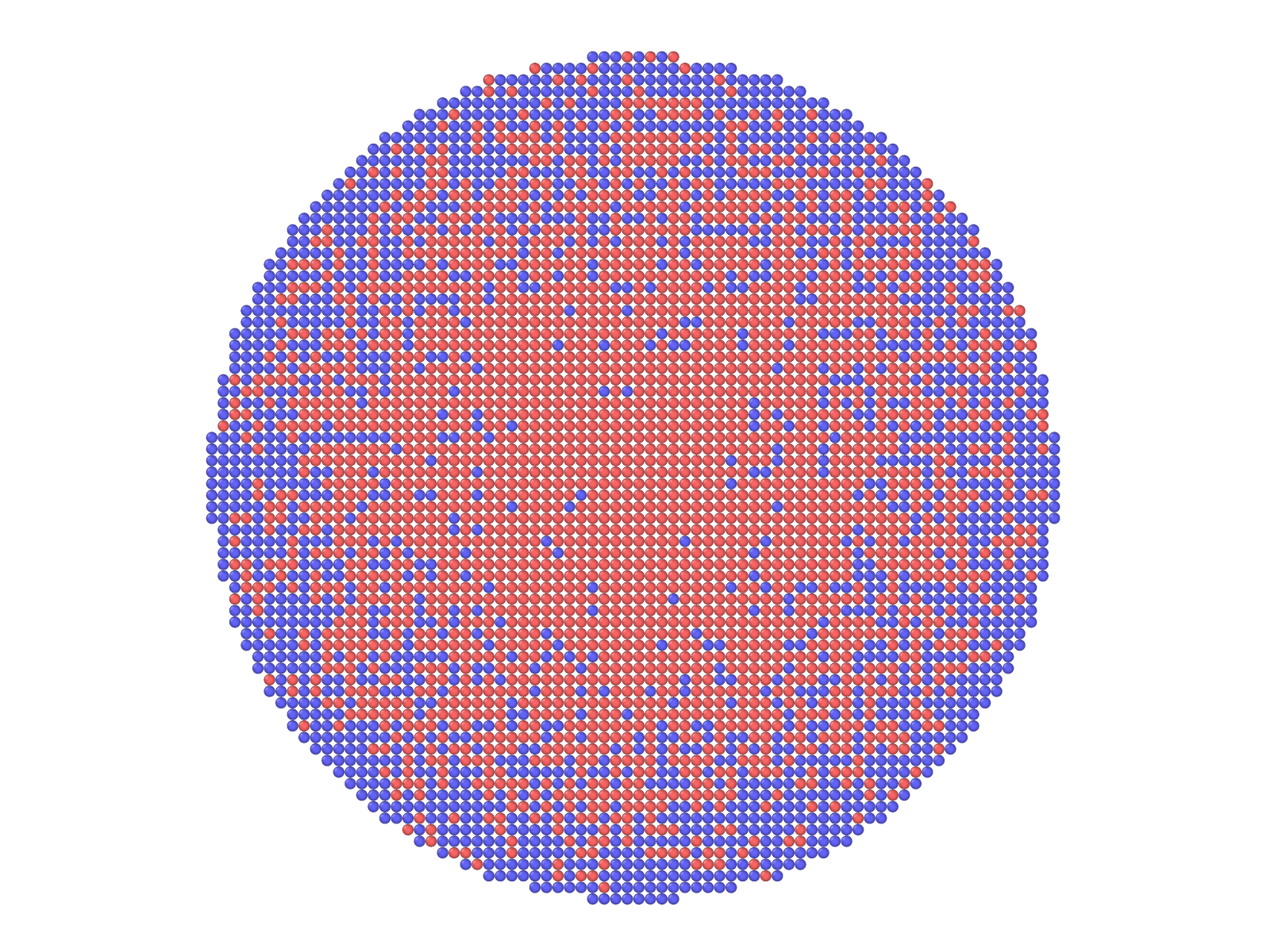}}
		\caption*{50\%}
		\centerline{\includegraphics[width=0.6\textwidth]{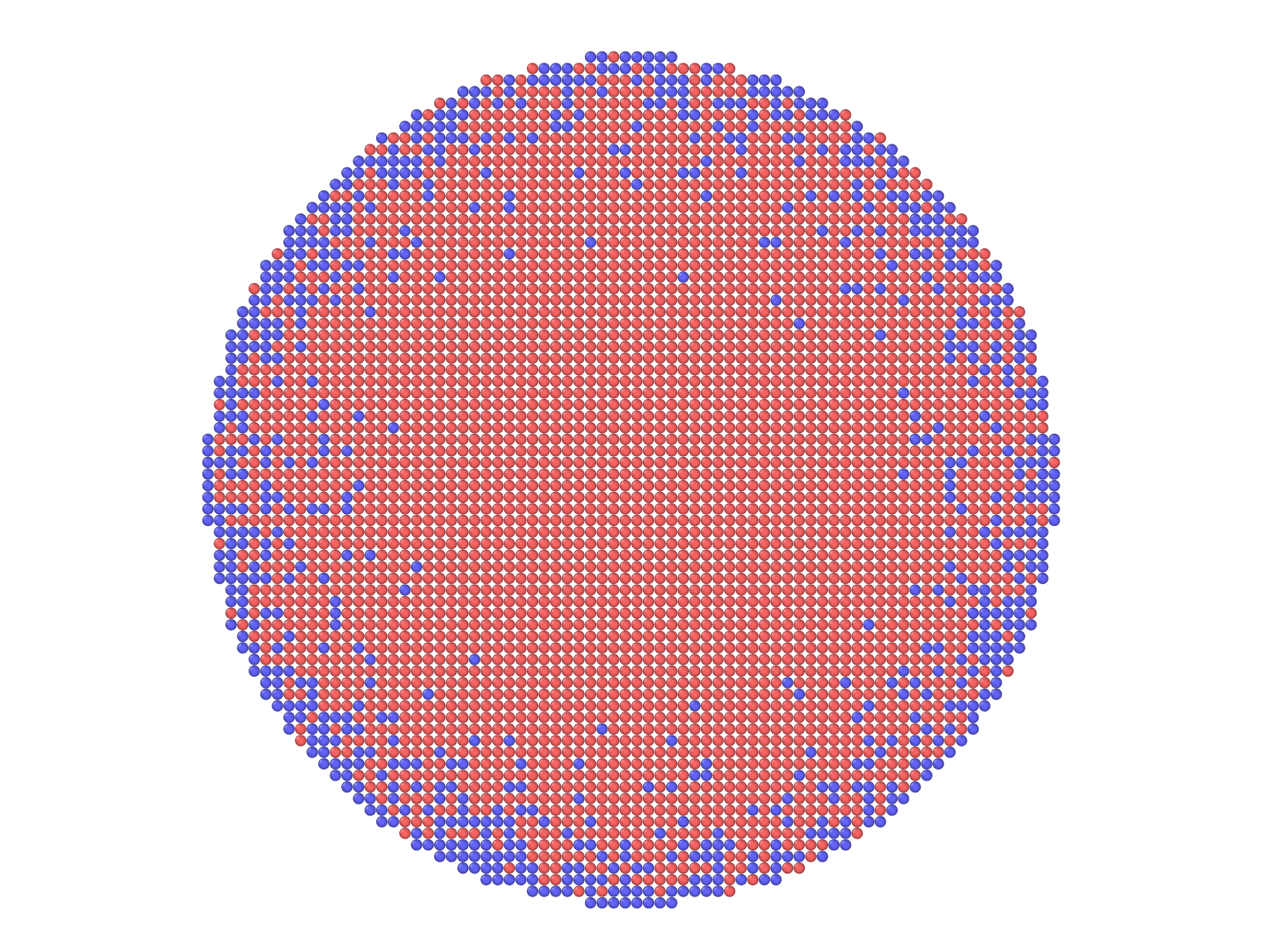}}
		\caption*{75\%}
		\centerline{\includegraphics[width=0.6\textwidth]{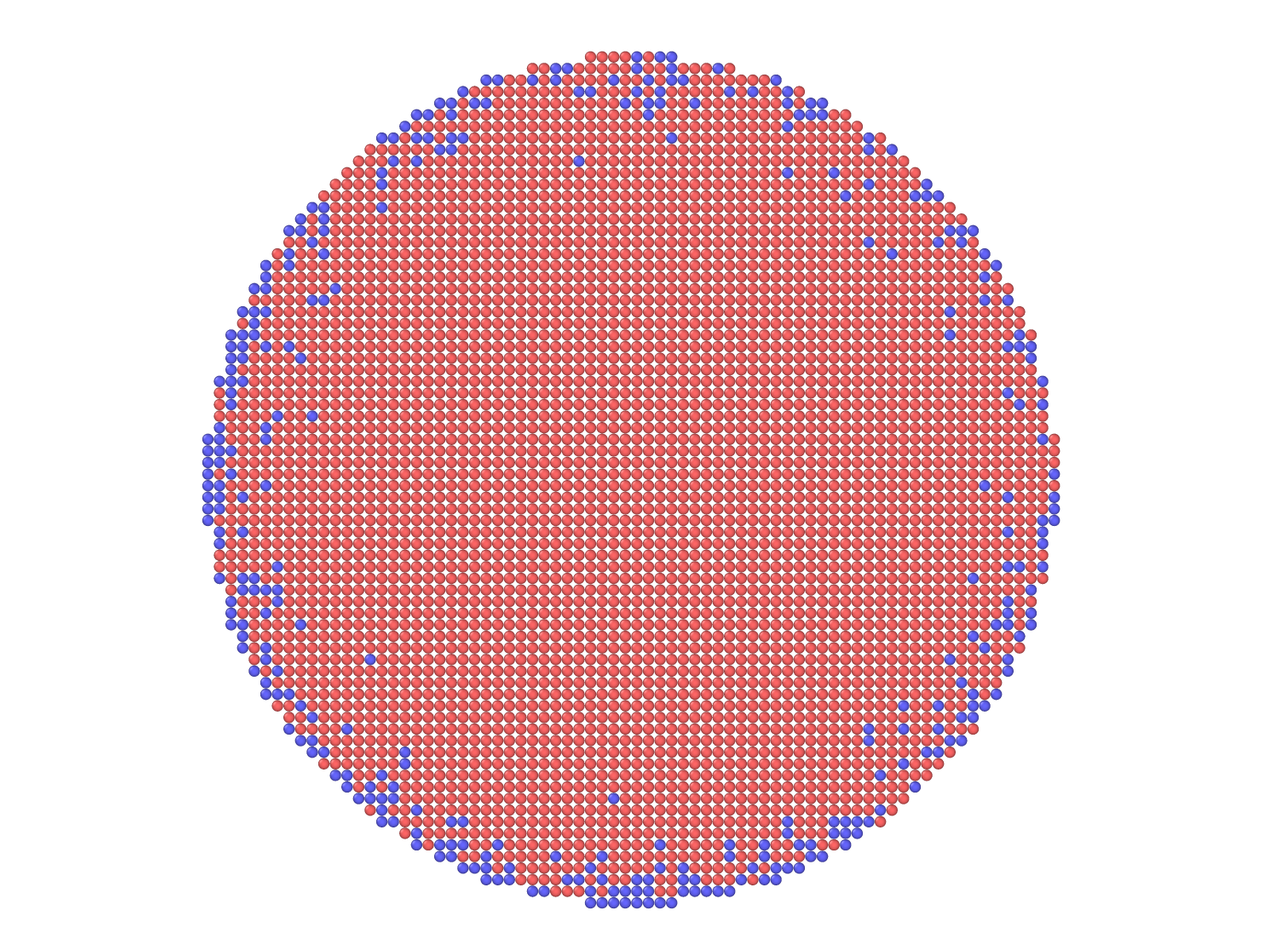}}
		\caption*{90\%}
		\caption{ }
		\label{fig3a}
	\end{subfigure}
	\begin{subfigure}{0.45\textwidth}
		\centerline{\includegraphics[width=0.6\textwidth]{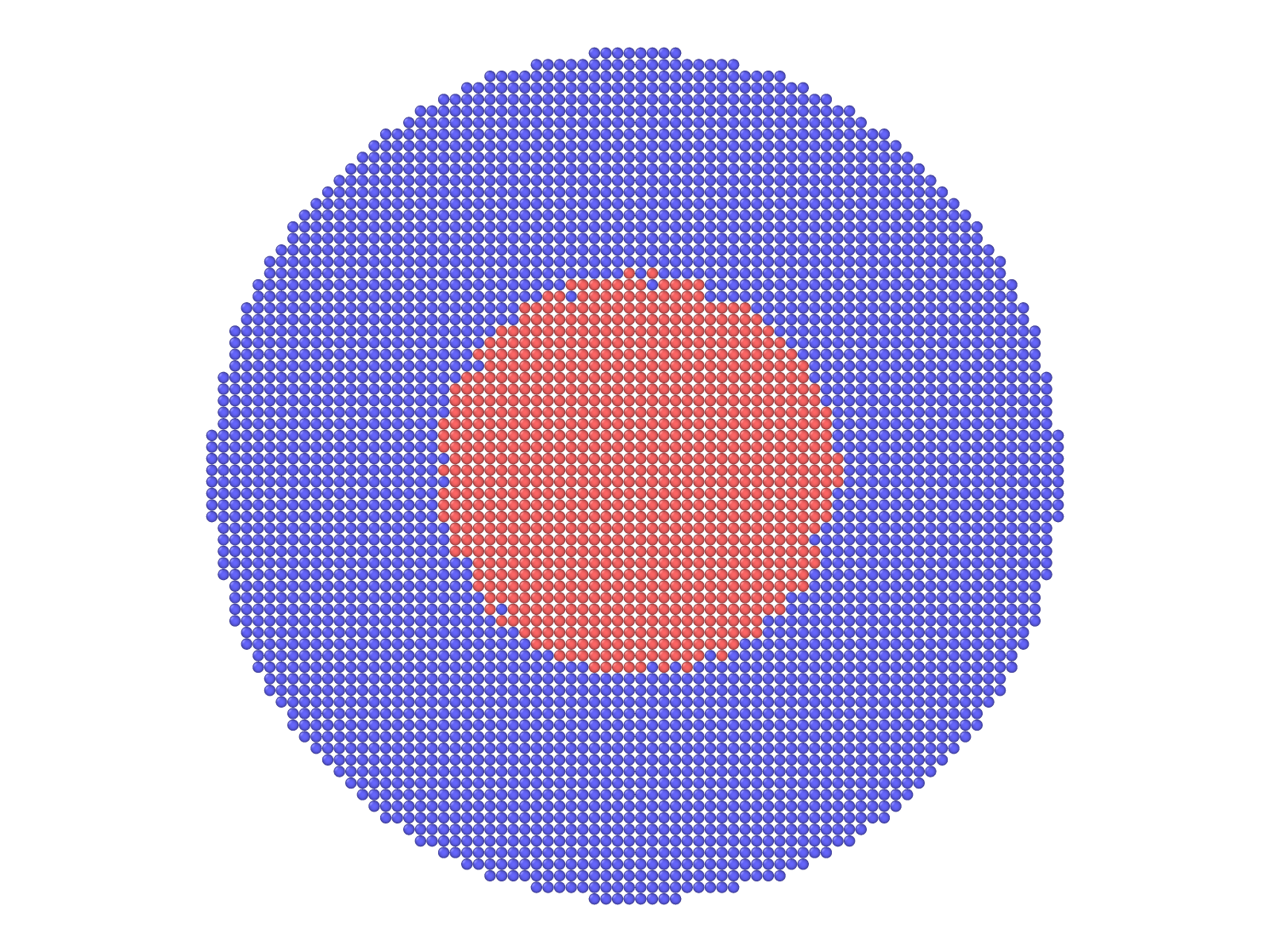}}
		\caption*{10\%}
		\centerline{\includegraphics[width=0.6\textwidth]{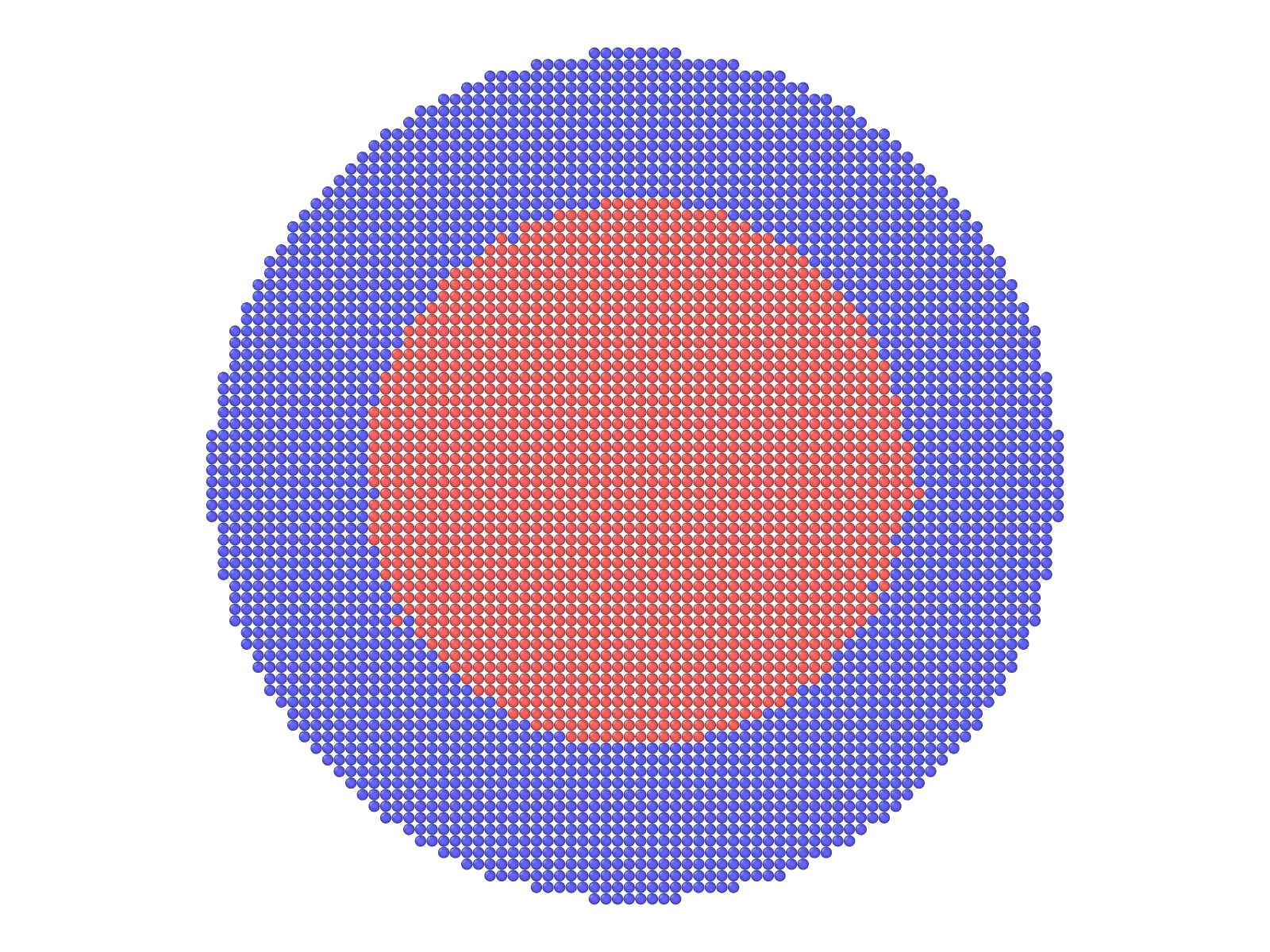}}
		\caption*{25\%}
		\centerline{\includegraphics[width=0.6\textwidth]{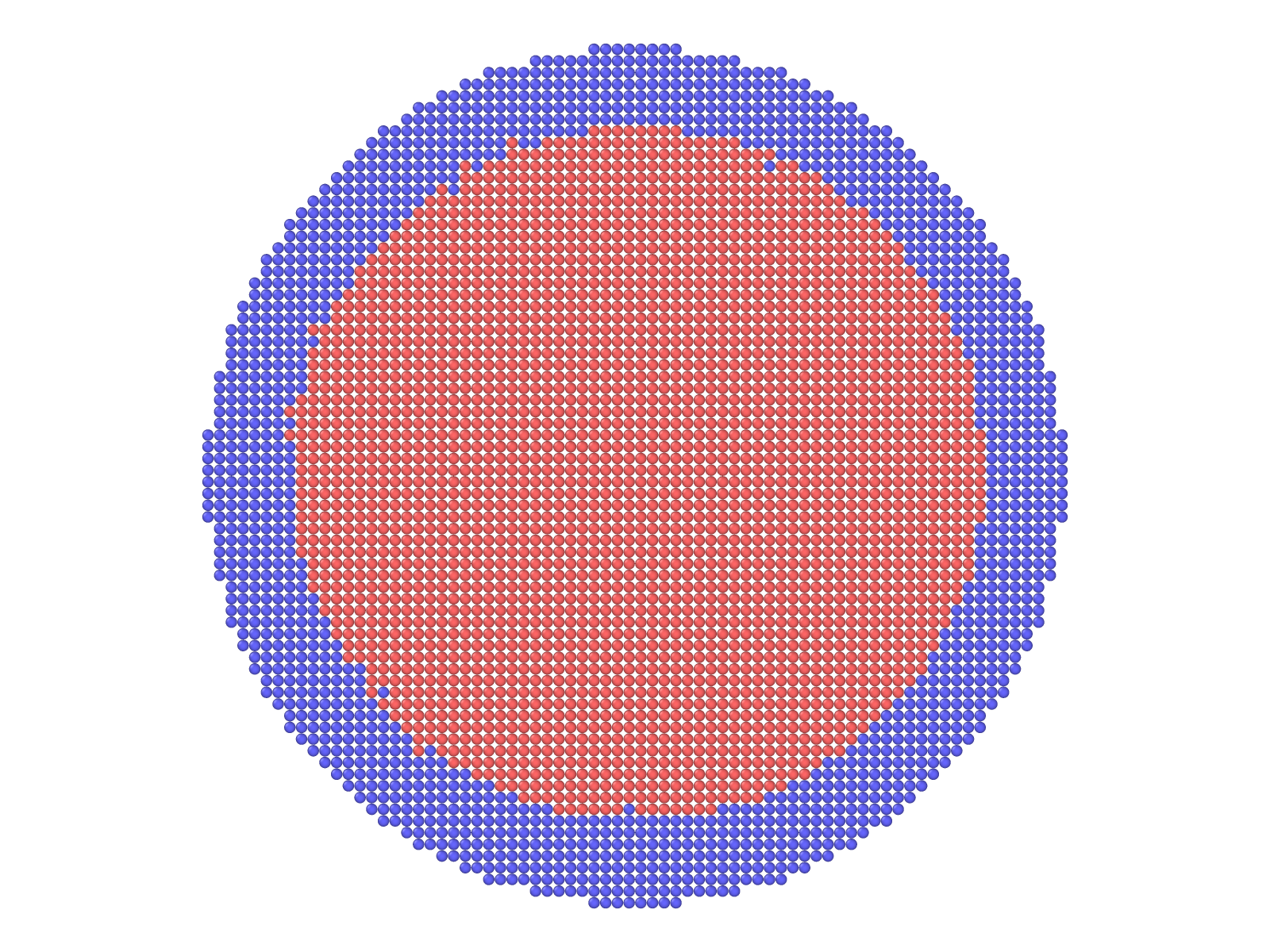}}
		\caption*{50\%}
		\centerline{\includegraphics[width=0.6\textwidth]{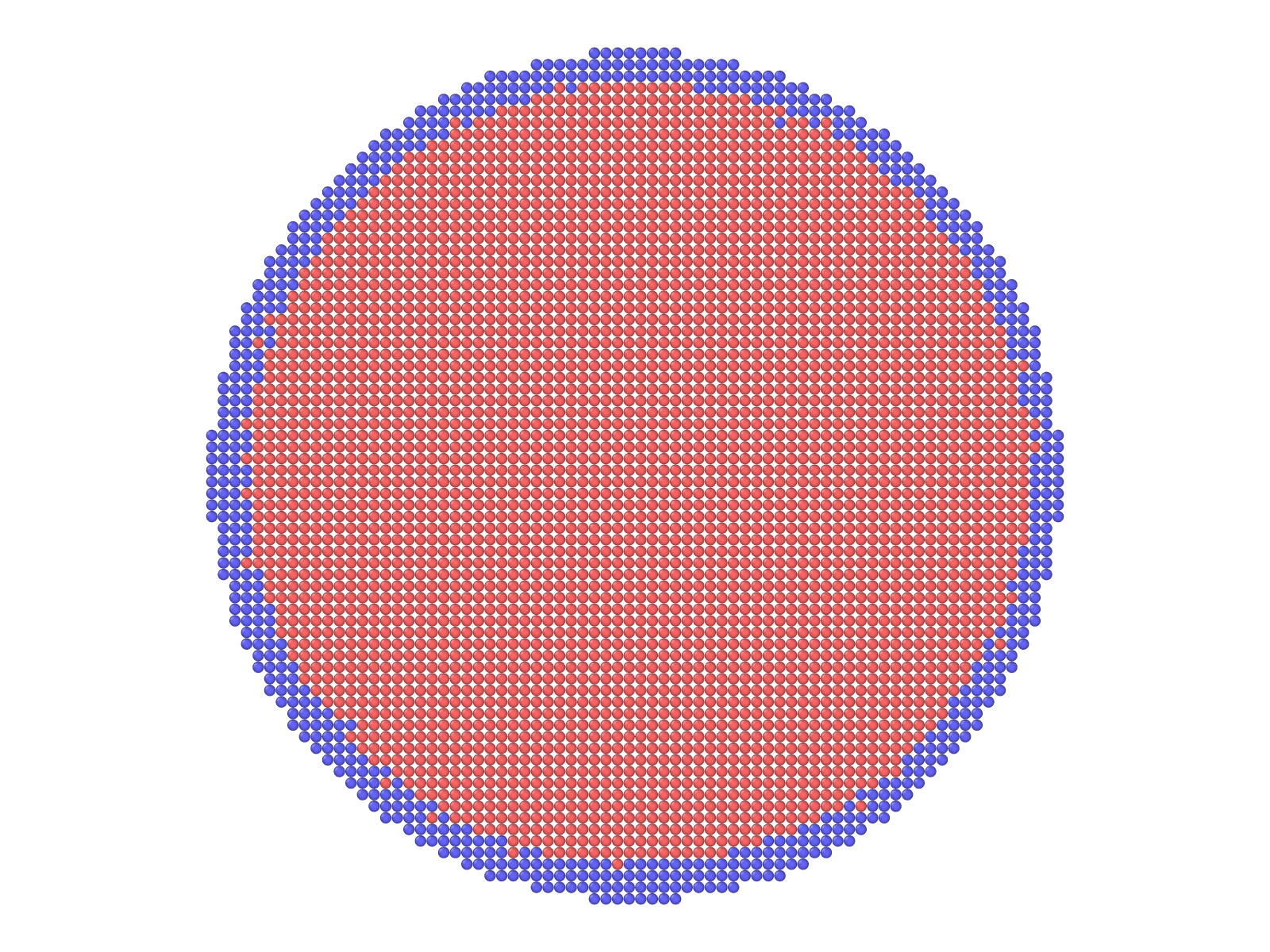}}
		\caption*{75\%}
		\centerline{\includegraphics[width=0.6\textwidth]{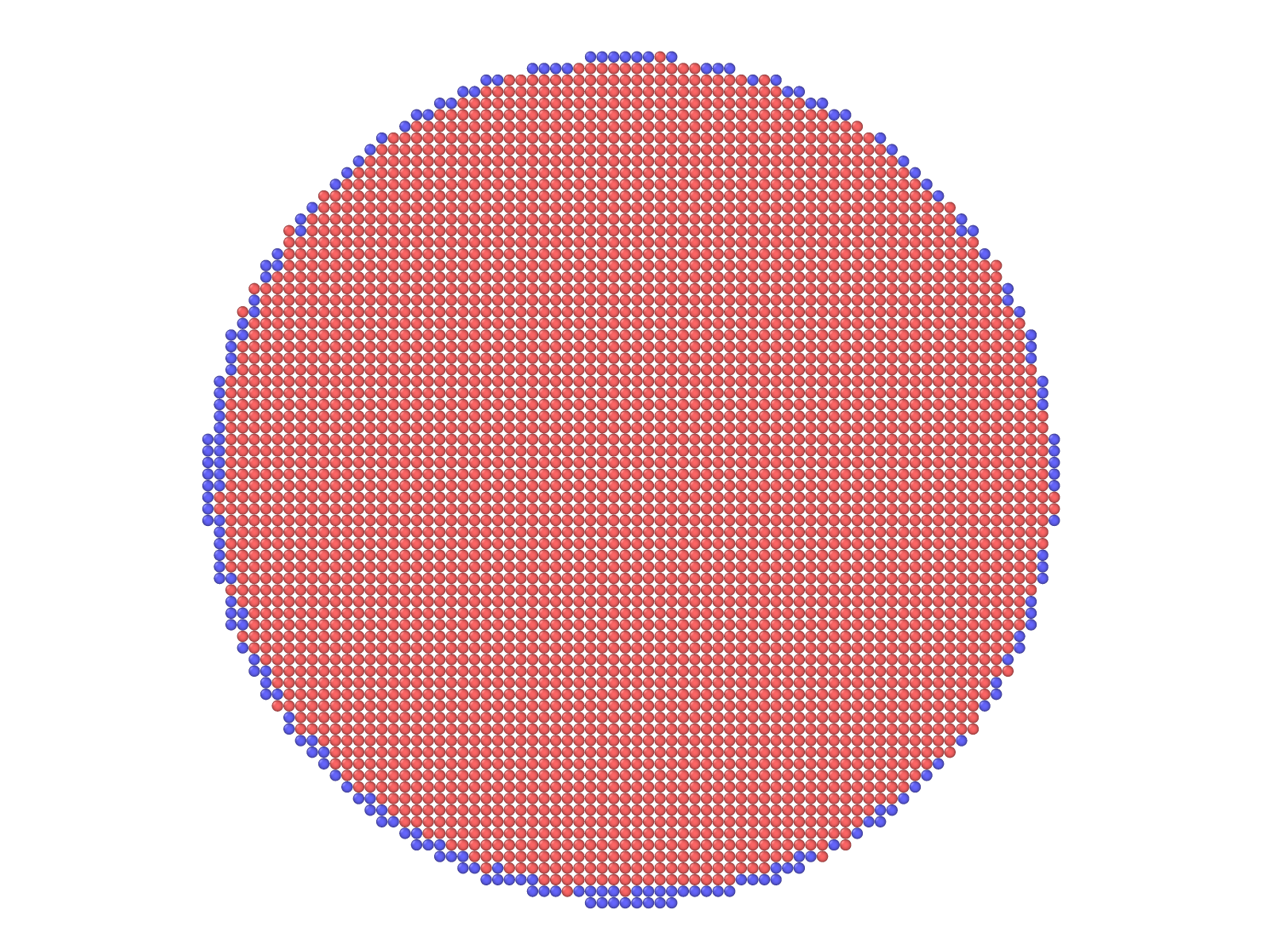}}
		\caption*{90\%}
		\caption{ }
		\label{fig3b}
	\end{subfigure}
	\label{fig3}
	\caption{Midplane view of (a) FGM and (b) Coreshell nanosphere for different percentages of Ag.}
\end{figure}

The simulation is conducted by maintaining some steps. At first, In each simulation, the system is minimized to a local minimum energy by using the Polak-Ribiere version of the conjugate gradient (CG) algorithm\cite{polak1969note}. After minimization, the system iss equilibrated in canonical ensemble at a temperature of 10 K for 150ps by using Nose-hoover thermostat\cite{hoover1985canonical}. After this step, the nanosphere is placed between two planner repulsive force field in order to perform uniaxial compression test. We have modeled this simulation according to the experimental process. Experimentally, the compression is performed by using indenters whose curvature is large enough compared to the nanosphere’s radius. In this simulation, those indenters are modelled by using this planner repulsive force field\cite{hale2011phase,bian2013atomistic}. The external forces acting on the nanosphere by the planner repulsive force fields are calculated by using the analytical function is described by:

\begin{equation}
	F(y) = -\alpha k(y-y_{i})^{2}
\end{equation}

Where, $y_{i}$ is the position of the indenter along [010] and the $y$ is the coordinate of the atom along the same axis. Here, $k$ is the effective stiffness constant of the indenters which value is considered $1000 ev/$ \AA in this simulation. The value of $\alpha$ is the term which determined the direction of force for the indenters. For upper indenter, $\alpha=1$ when, $y\geq y_{i}$ and $\alpha=0$ when,  $y < y_{i}$, while for the bottom indenter, $\alpha=-1$ when, $y \leq y_{i}$ and $\alpha=0$ when,  $y > y_i$. Then the indenters were moving towards each other with a strain rate of $10^9$ $s^{-1}$ along [010] direction at 10K. No periodic boundary conditions were used in all [100],[010], and [001] directions. We need to compute the contact surface area between the nanosphere and the indenters at each deformation step to calculate the contact stress.

The contact surface area is computed between the nanosphere and the plane indenter according to the following procedures: 

\begin{itemize}
	\item 	A nanosphere atom is considered to belong to the contact surface if $\mid y-y_{i} \mid <0.1$\AA .
	\item For calculating the radius of contact area, molecules in contact with plane indenter assumed between $y_{min}$ and $y_{min}+0.1$\AA for lower surface and $y_{max}$ and $y_{max}-0.1$ \AA for upper surface.
 	\item The considered molecules are projected in $xz$ plane and effective radius, $a$ was calculated by following expression:
	\begin{equation}
		\frac{a^{2}}{2} = \frac{1}{N} \sum_{i}[(x_i-x_c)^2+(z_i-z_c)^2]
	\end{equation}
	Here, $x_c$ and $z_c$ are the $x$ and $z$ coordinates of the center of mass of the projected molecules.
	\item So, the contact surface area is:
	\begin{equation}
		A = \pi a^2
	\end{equation}
\end{itemize}

All of the calculation steps has been done by using python Numpy and MDanalysis\cite{michaud2011mdanalysis} packages. Ovito\cite{stukowski2009visualization} is used for visualized the Molecular dynamics data. 
Dislocation density of the nanospheres has been calculated by using Dislocation Extraction Algorithm (DXA), develpoed by  Stukowski et al.\cite{stukowski2012automated} The DXA converts an atomistic model of a displaced crystal into a line-based dislocation network representation. It calculates each dislocation's Burgers vector and locates dislocation junctions. Partially dislocations, as well as some secondary grain boundary dislocations, are recognized by this algorithm.

\section{Results and discussion}
\subsection{Effect on alloying percentage}
\begin{figure}[htbp]
	\centering
	\begin{subfigure}{0.8\textwidth}
		\centerline{\includegraphics[width=1\textwidth]{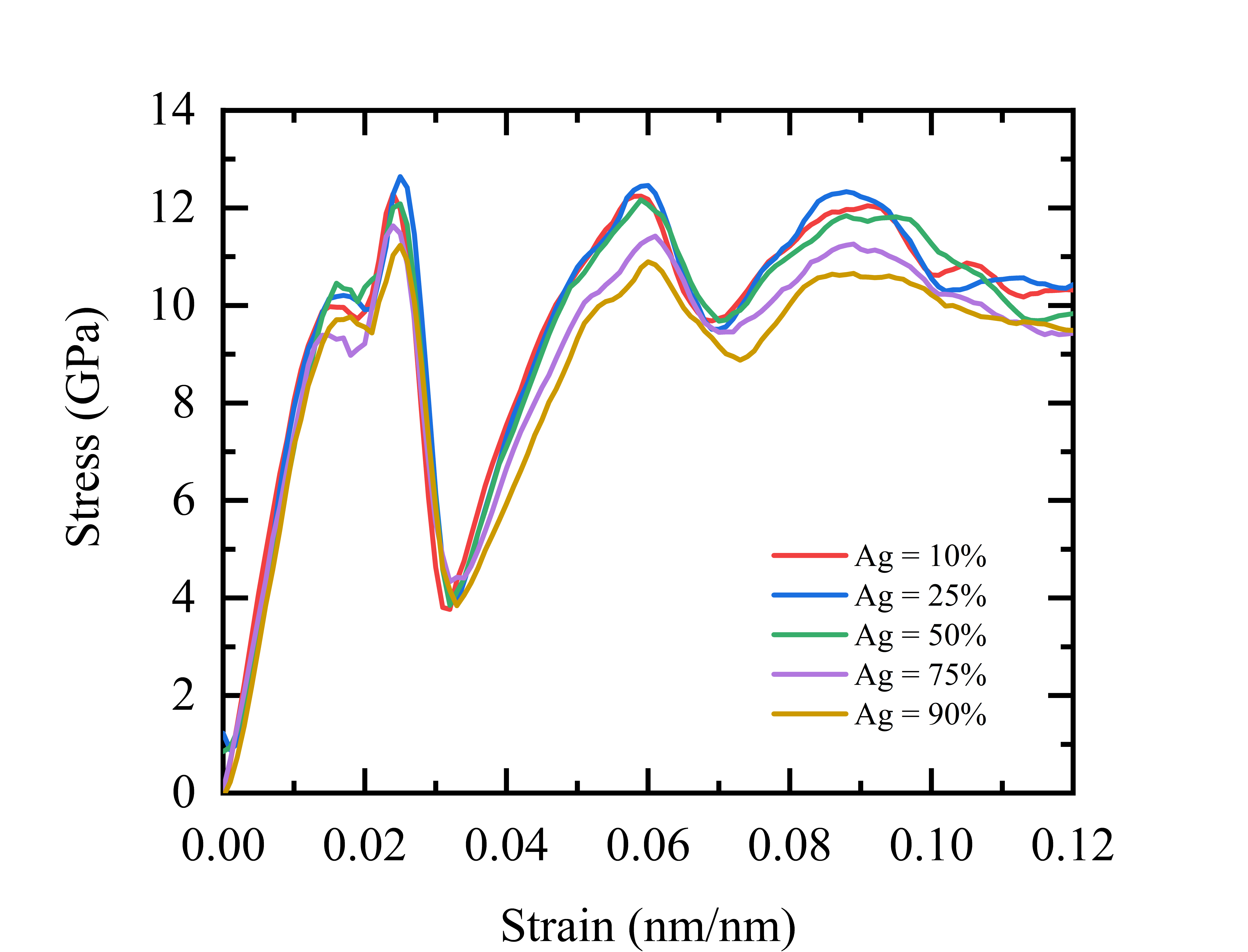}}
		\caption{ }
		\label{fig4a}
	\end{subfigure}
	\begin{subfigure}{0.8\textwidth}
		\centerline{\includegraphics[width=1\textwidth]{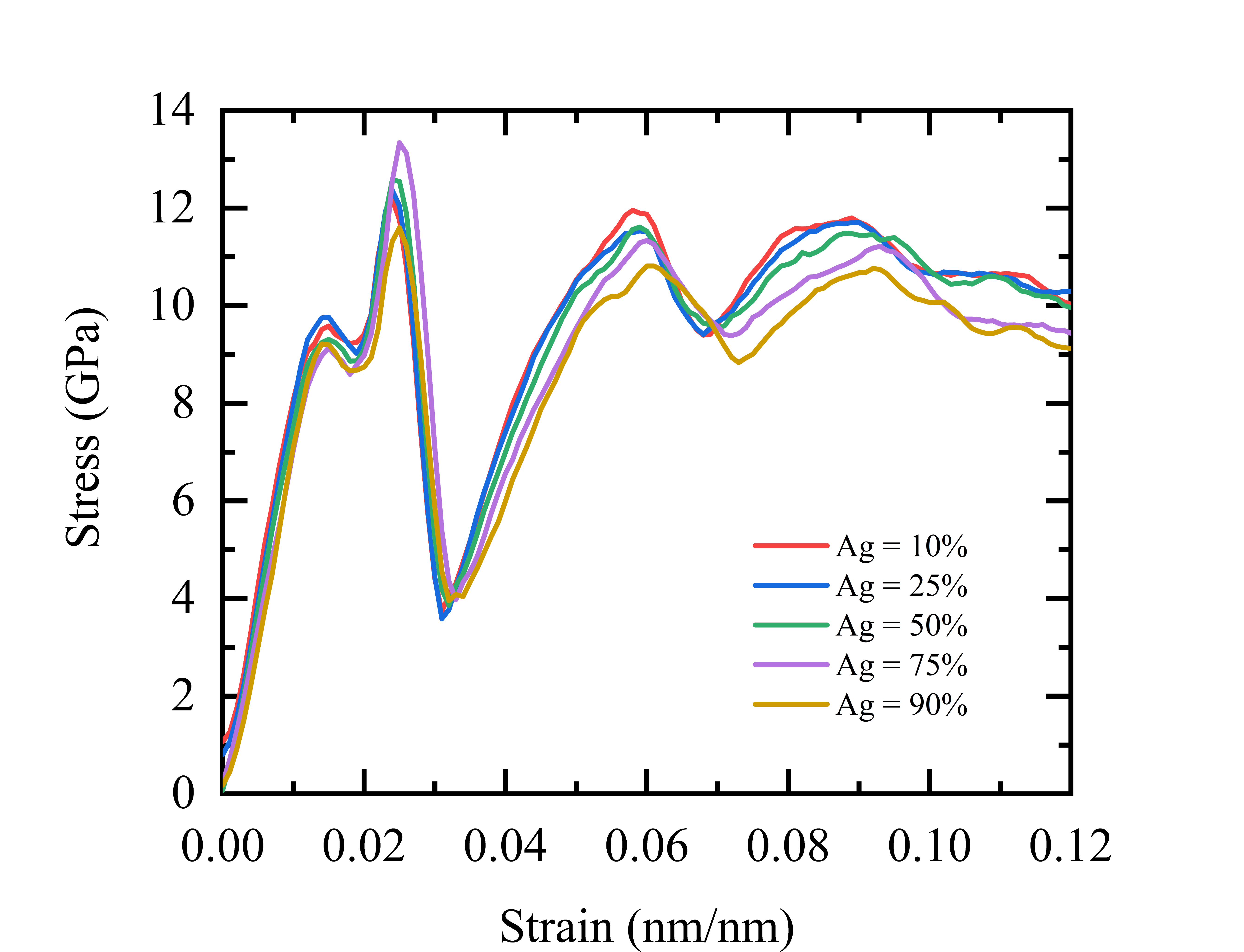}}
		\caption{ }
		\label{fig4b}
	\end{subfigure}
	\caption{Stress-Strain Curve of (a) FGM and (b) Core Shell nanosphere for different percentages of Ag.}
	\label{fig4}
\end{figure}
\begin{figure}[htbp]
	\centering
	\begin{subfigure}{0.7\textwidth}
		\centerline{\includegraphics[width=1\textwidth]{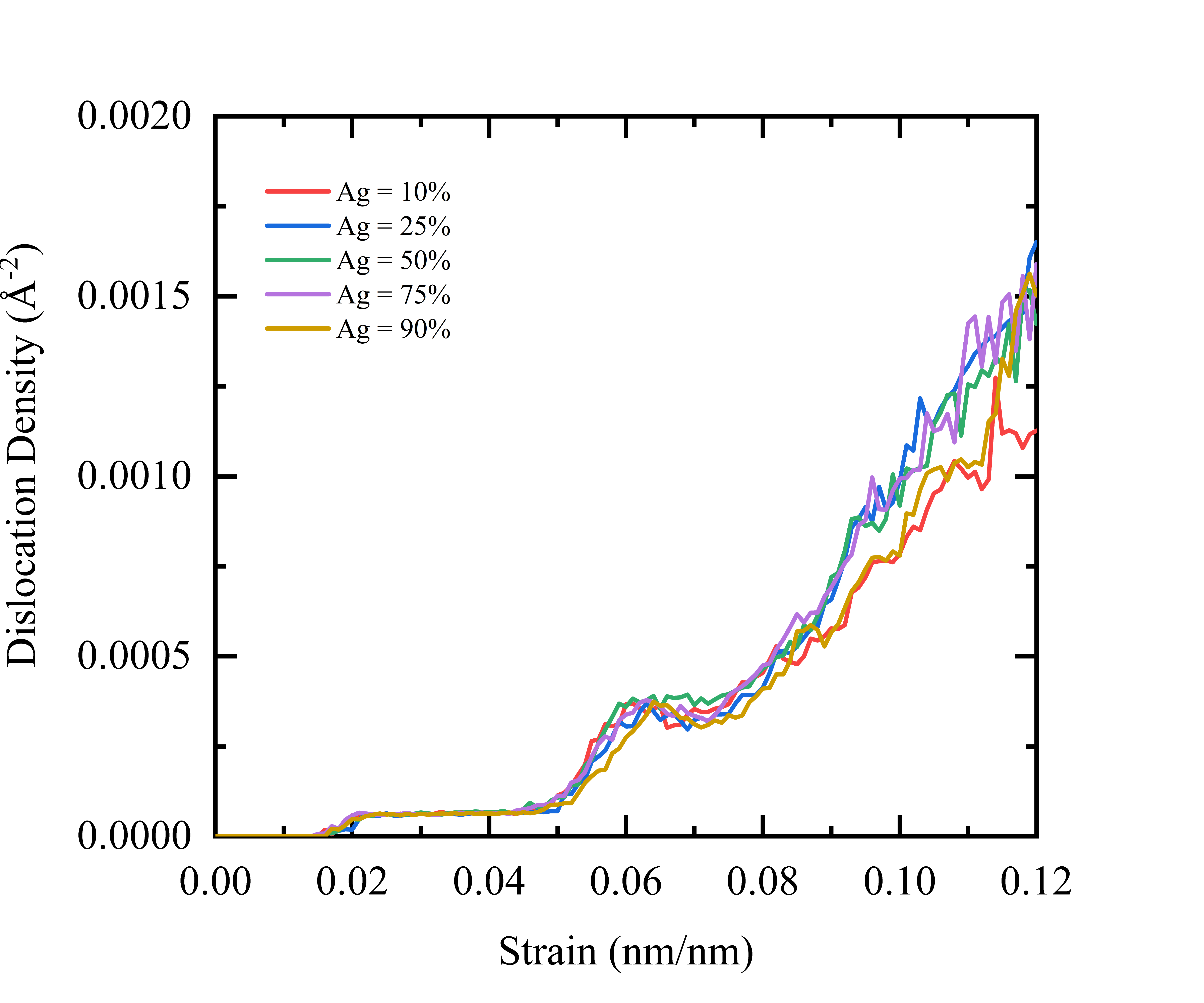}}
		\caption{ }
		\label{fig5a}
	\end{subfigure}
	\begin{subfigure}{0.7\textwidth}
		\centerline{\includegraphics[width=1\textwidth]{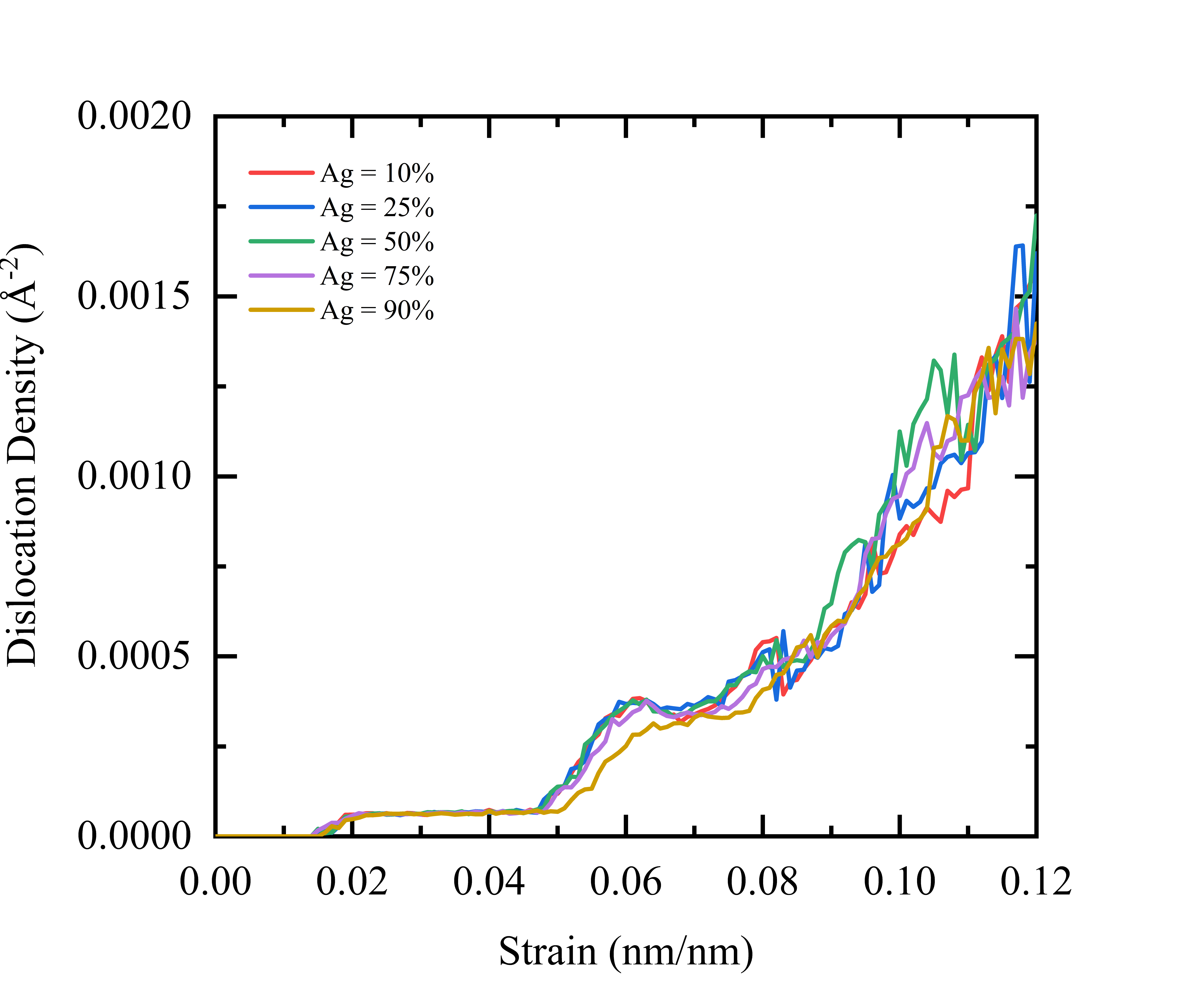}}
		\caption{ }
		\label{fig5b}
	\end{subfigure}
	\caption{Total dislocation density-Strain Curve of (a) FGM and (b) Coreshell nanosphere for different percentages of Ag.}
	\label{fig5}
\end{figure}
\begin{figure}[htbp]
	\centering
	\begin{subfigure}{0.8\textwidth}
		\centerline{\includegraphics[width=1\textwidth]{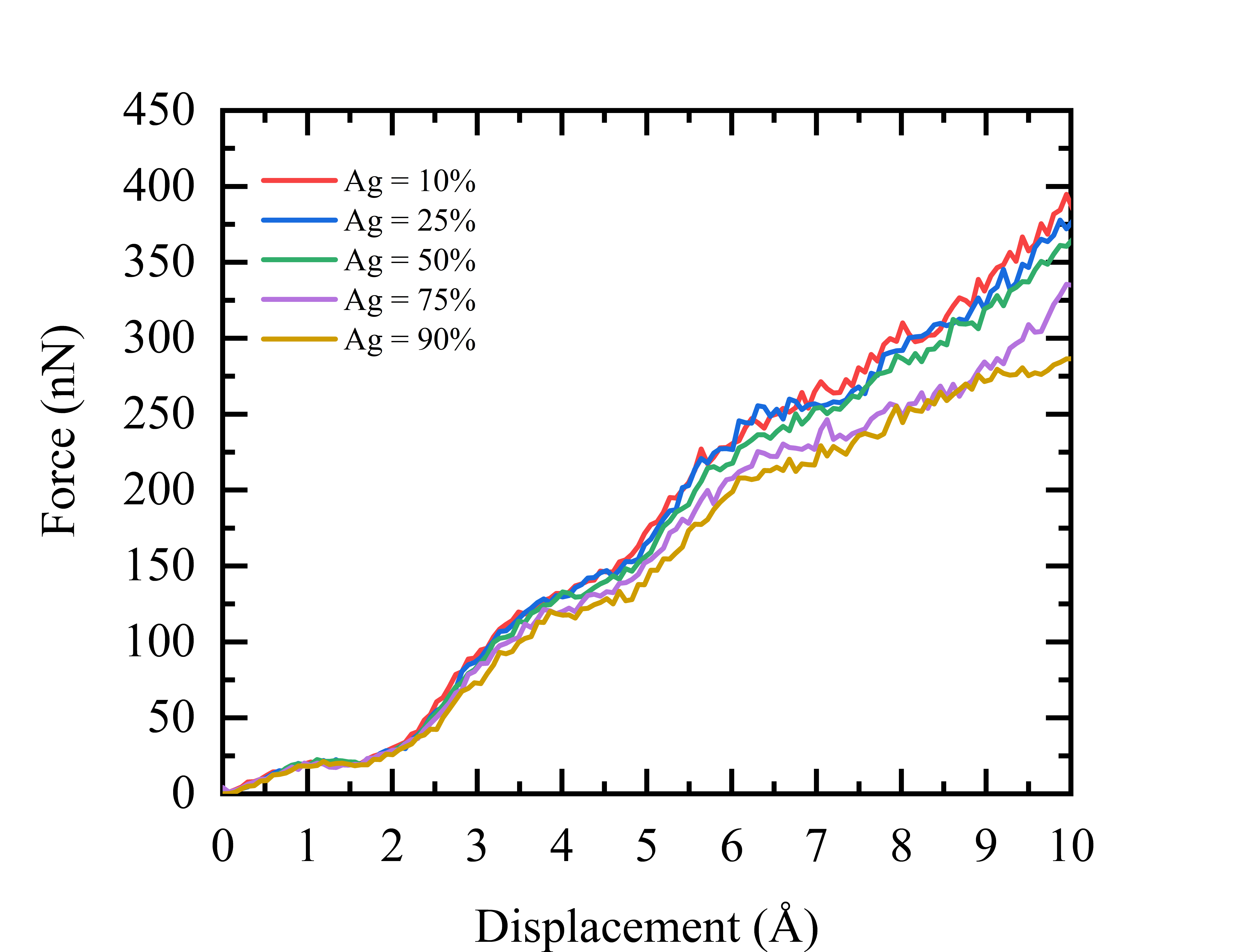}}
		\caption{ }
		\label{fig6a}
	\end{subfigure}
	\begin{subfigure}{0.8\textwidth}
		\centerline{\includegraphics[width=1\textwidth]{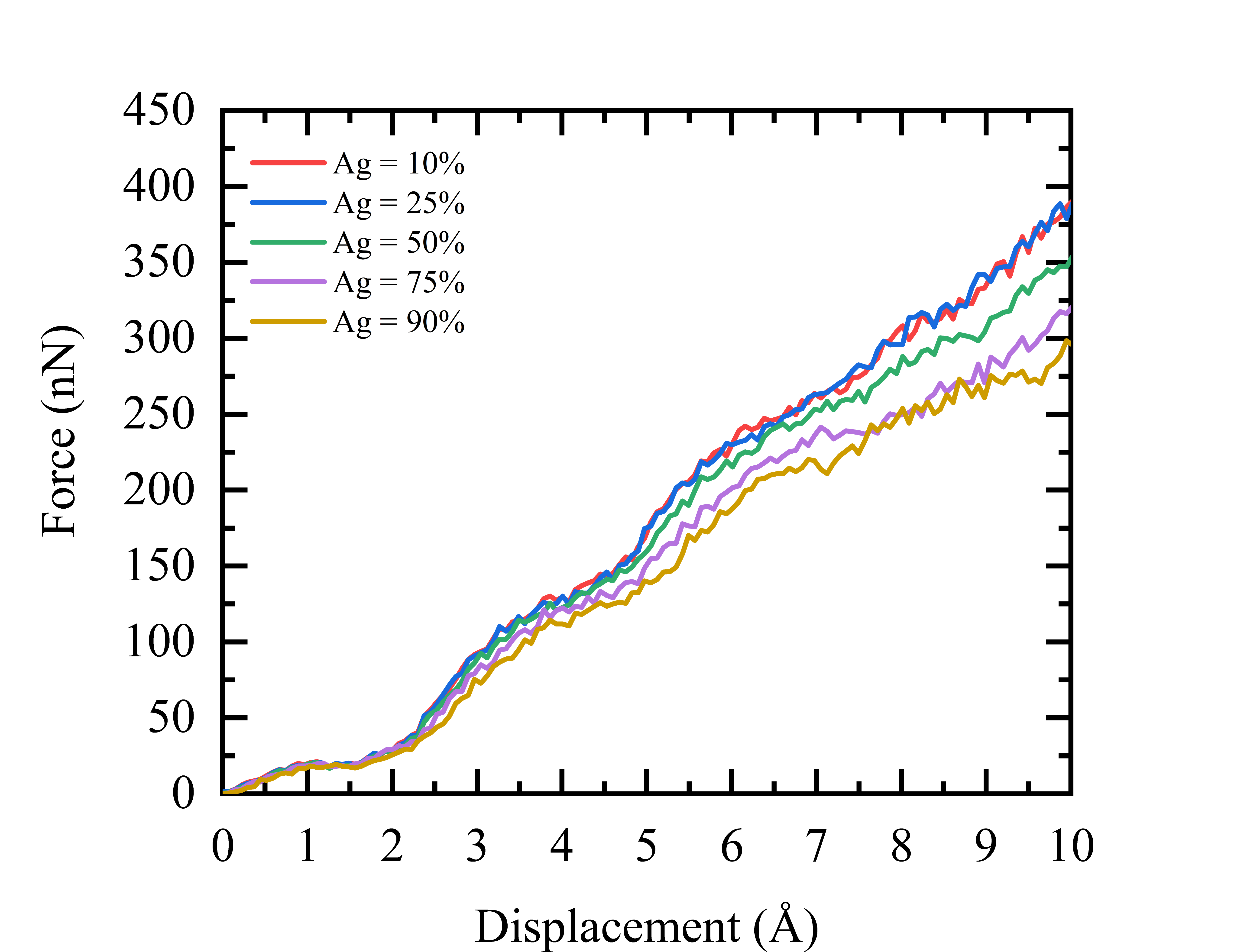}}
		\caption{ }
		\label{fig6b}
	\end{subfigure}
	\caption{Force-Displacement Curve of (a) FGM and (b) Coreshell nanosphere for different percentages of Ag.}
	\label{fig6}
\end{figure}
In this study, the influence of alloying percentages on mechanical properties and dislocation density of Au-Ag nanospheres has been investigated using FGM and Core-shell nanospheres. Figs.\ref{fig4a} and \ref{fig4b} show stress-strain curves for FGM and core-shell nanospheres with varing alloying percentages. Fig. \ref{fig5} and \ref{fig6} depict the dislocation density-strain and force-displacement curves, respectively.

Fig. \ref{fig4} shows the stress in nanospheres grow gradually at the beginning of indentation. This propensity is due to the Ag-Au nanocrystal's elastic characteristics. Dislocation causes tiny flaws to emerge on the contact surfaces. During the early stages of compression, certain dislocation loops are formed inside the Ag-Au nanoparticle. The atomic coordination representations clearly show these minor flaws \cite{zhang2011deformation}. Furthermore, dislocation-driven plasticity is common in nanoscale Silver and Gold particles. As the strain increases from roughly 1.6 to 2.2, the contact stress begins to decrease significantly in the second stage. Contact stress is reduced at this stage due to practically constant loading as seen in the inset of Fig.\ref{fig6} and increased contact area. Following that, the contact tension rapidly builds, reaches peak, and then quickly decreases. Similar to MD studies for Ag-Au nanoparticles, this property might be due to the creation of a new structural phase. The stress varies arbitrarily with increasing strain in the third stage, indicating that the Ag-Au nanocrystal has amorphized under high pressure.

The failure phenomena occur at 2.5 percent to 3.5 percent strain for Core-shell and FGM Au-Ag nanospheres, as shown in Fig.\ref{fig4}. We can see a yielding point shortly before the failure in this stress-strain curve. For both type of alloying, Fig. \ref{fig4} shows a similar tendency for nanospheres. Following the first drop in stress, we can see a fluctuating trend in stress, indicating that the nanospheres are ductile. Because pure Ag and Au are both ductile in nature, this ductile behavior is expected. The dislocation density-strain curve in Figs. \ref{fig5a} and \ref{fig5b} explain the fluctuating pattern that follows the failure point. As demonstrated in Fig. \ref{fig5}, the formation of dislocations results permanent deformation in nanospheres. The dislocation density remains constant until the strain reaches 4 to 4.5 percent. After this moment, the density of dislocations begins to rise. The fluctuating pattern in Fig.\ref{fig4}'s stress-strain curves is due to this phenomenon. 
 
The force-displacement curve for FGM and core-shell nanospheres are shown in Fig. \ref{fig6}. It shows the relation between the displacement of surface atoms occurs due to the forces exerted by the planar indenters and the forces. As the percentage of silver in FGM and Core Shell nanospheres increases, the force exerted by the planner indenter decreases.  

\begin{figure}[htbp]
	\centering
	\begin{subfigure}{0.8\textwidth}
		\centerline{\includegraphics[width=1\textwidth]{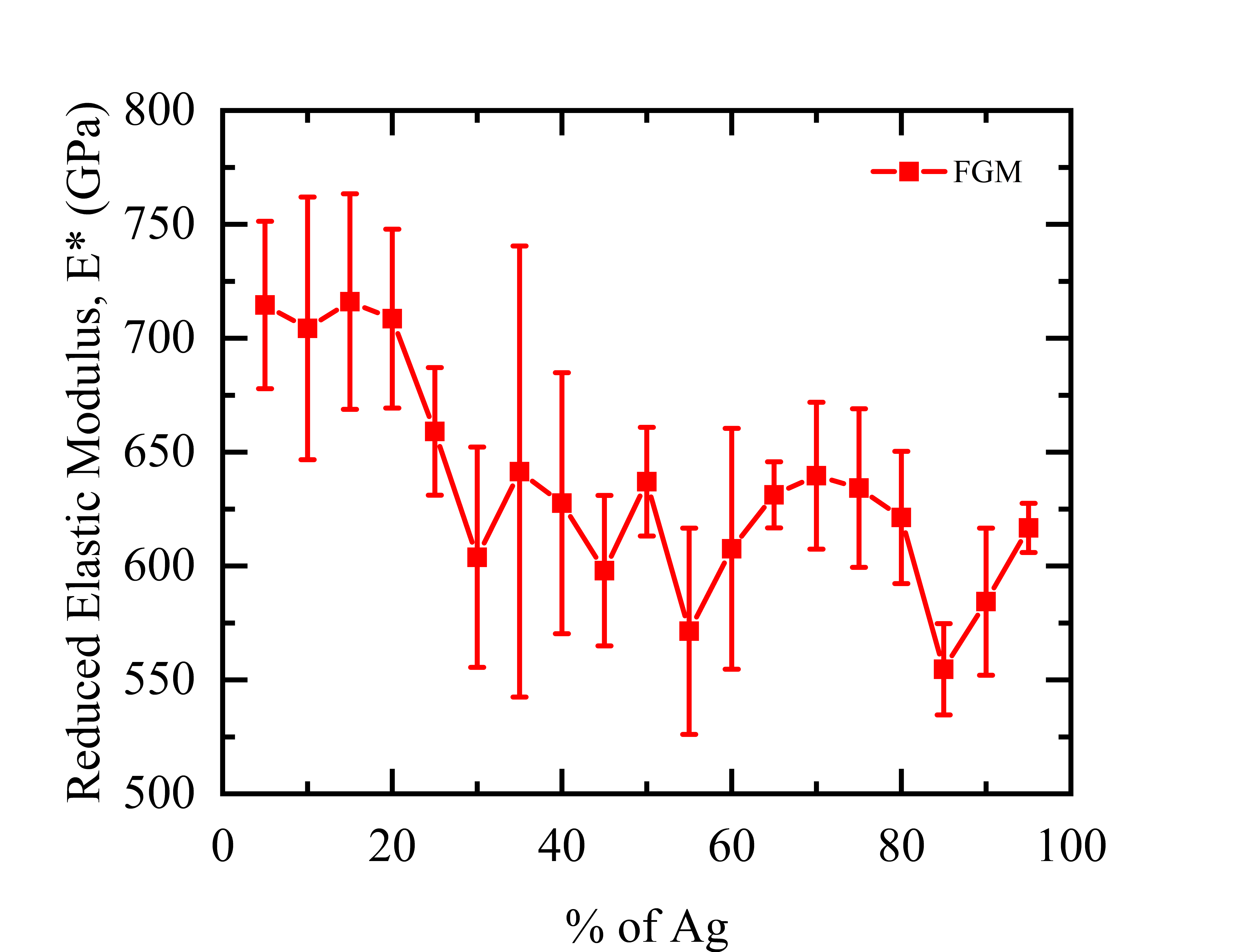}}
		\caption{ }
		\label{fig7a}
	\end{subfigure}
	\begin{subfigure}{0.8\textwidth}
		\centerline{\includegraphics[width=1\textwidth]{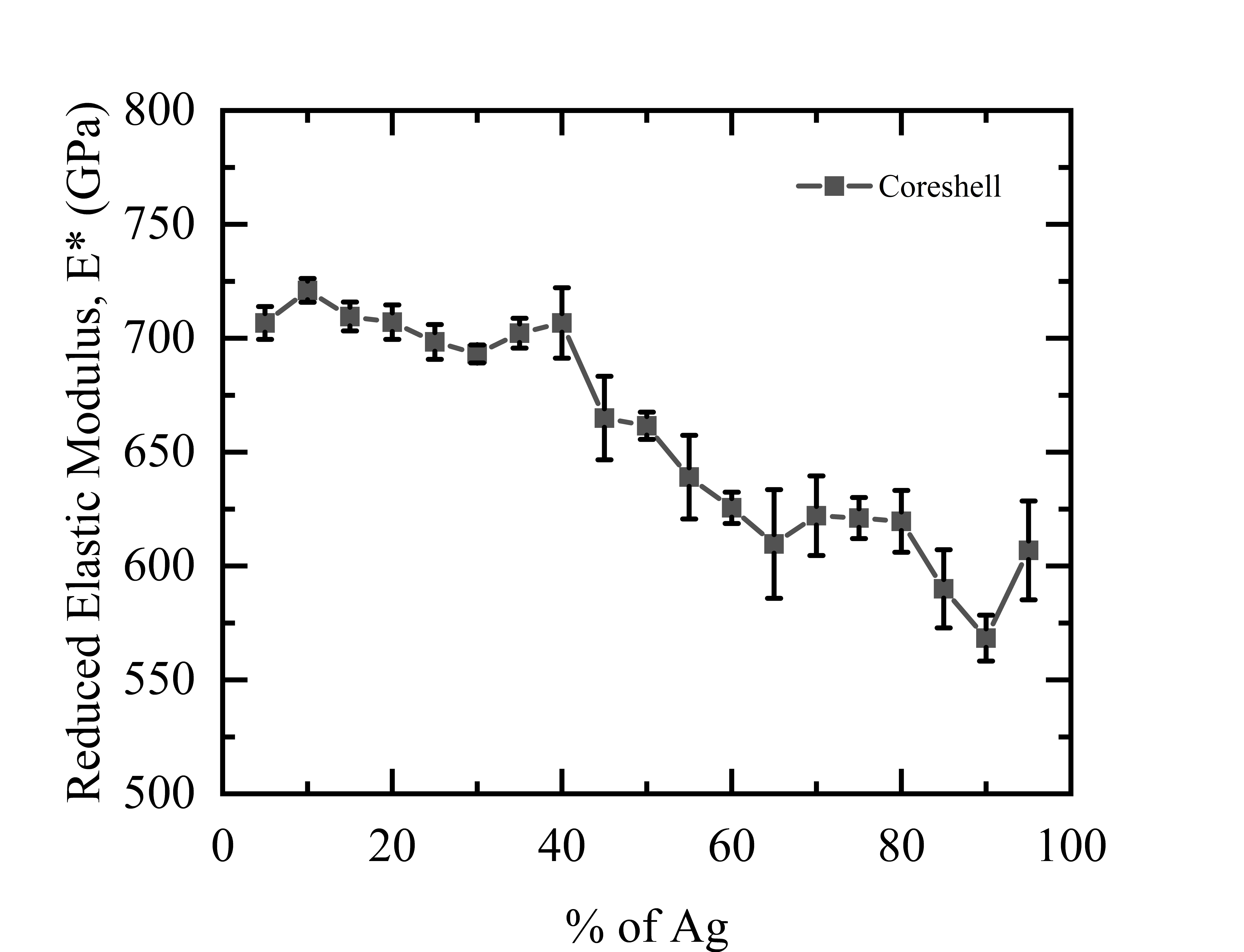}}
		\caption{ }
		\label{fig7b}
	\end{subfigure}
	\caption{Variation of Reduced Moduli of (a) FGM and (b) Coreshell nanosphere obtained from molecular dynamics simulation with different percentages of Ag.}
	\label{fig7}
\end{figure}

The nanospheres' Reduced Modulus is determined using a method outlined in earlier publications. The total elastic compressive response of nanoparticles will approach the Hertzian contact model \cite{valentini2007phase,lian2009sample}, which yields the relation between load F and compression depth $\delta$ as

\begin{equation}
	F = \frac{4}{3} E^* R^{\frac{1}{2}} \delta^{\frac{3}{2}}
\end{equation}

where $R$ is the radius of nanoparticle and $E^*$ is the reduced modulus, as Bian et al. \cite{bian2014atomistic} pointed out. The Force-displacement curve of compression is fitted using Hertzian contact theory.  Fig. \ref{fig7} depicts the fluctuation of $E^*$ with alloying percentages for FGM and Core Shell nanospheres. As seen in the graph, $E^*$ decreases as the amount of Ag in FGM and Core Shell increases. The $E^*$ fluctuation pattern may be explained by the total alloying percentages of Ag in Au, as well as the kind of alloying (FGM or Core Shell). The values of $E^*$ for Core Shell nanospheres with a high Au constituent will be larger than those for Core Shell nanospheres with a low Au constituent. $E^*$ of FGM nanospheres, on the other hand, follows a similar pattern to Core Shell nanospheres, with the exception of containing more fluctuating behavior. Some fundamental characteristics of FGM architecture will help explain this discrepancy. As previously stated, FGM nanospheres are made by replacing Au atoms with Ag atoms; the error bar depicts the randomness of the formations. 

\begin{figure}[htbp]
	\centering
	\begin{subfigure}{0.8\textwidth}
		\centerline{\includegraphics[width=1\textwidth]{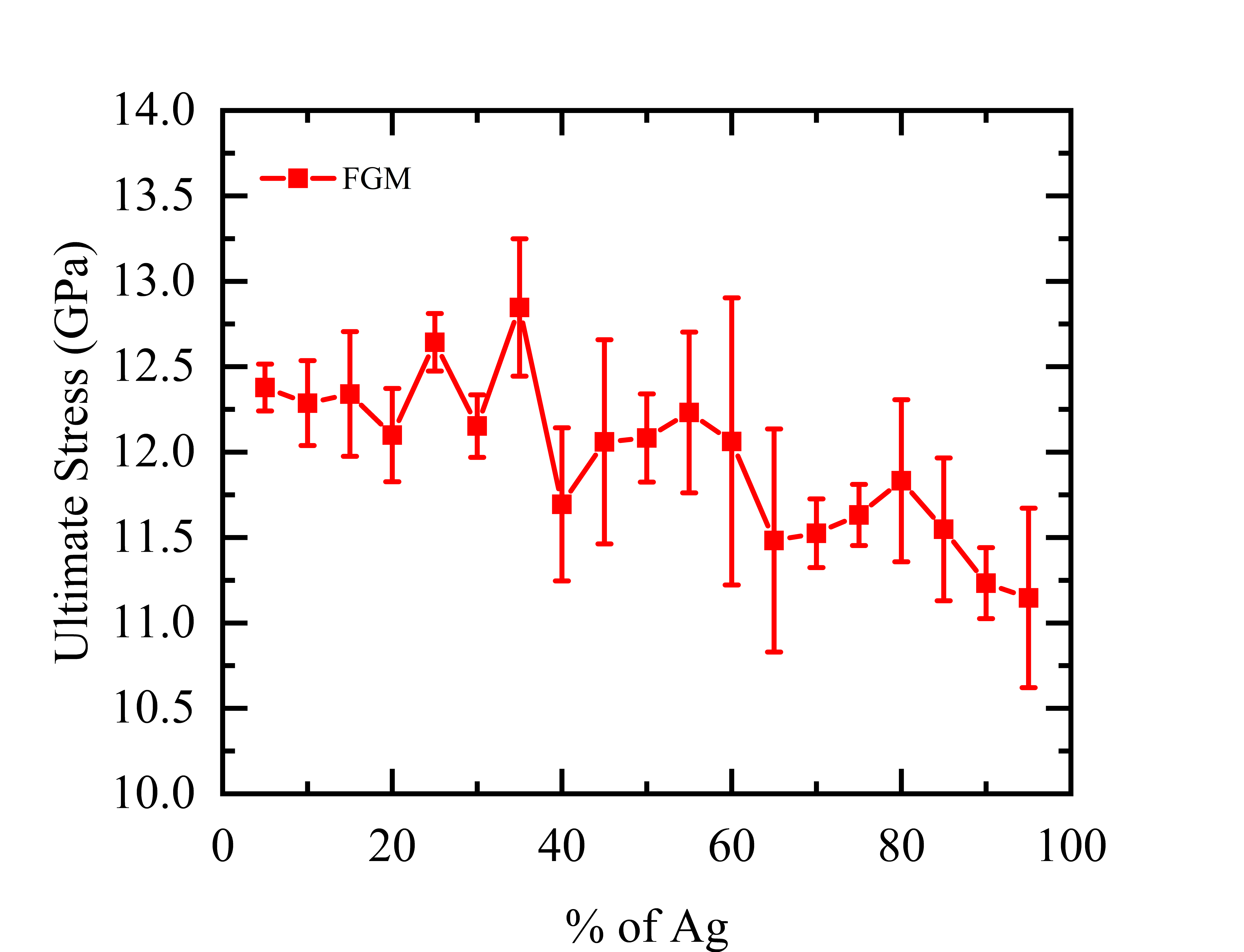}}
		\caption{ }
		\label{fig8a}
	\end{subfigure}
	\begin{subfigure}{0.8\textwidth}
		\centerline{\includegraphics[width=1\textwidth]{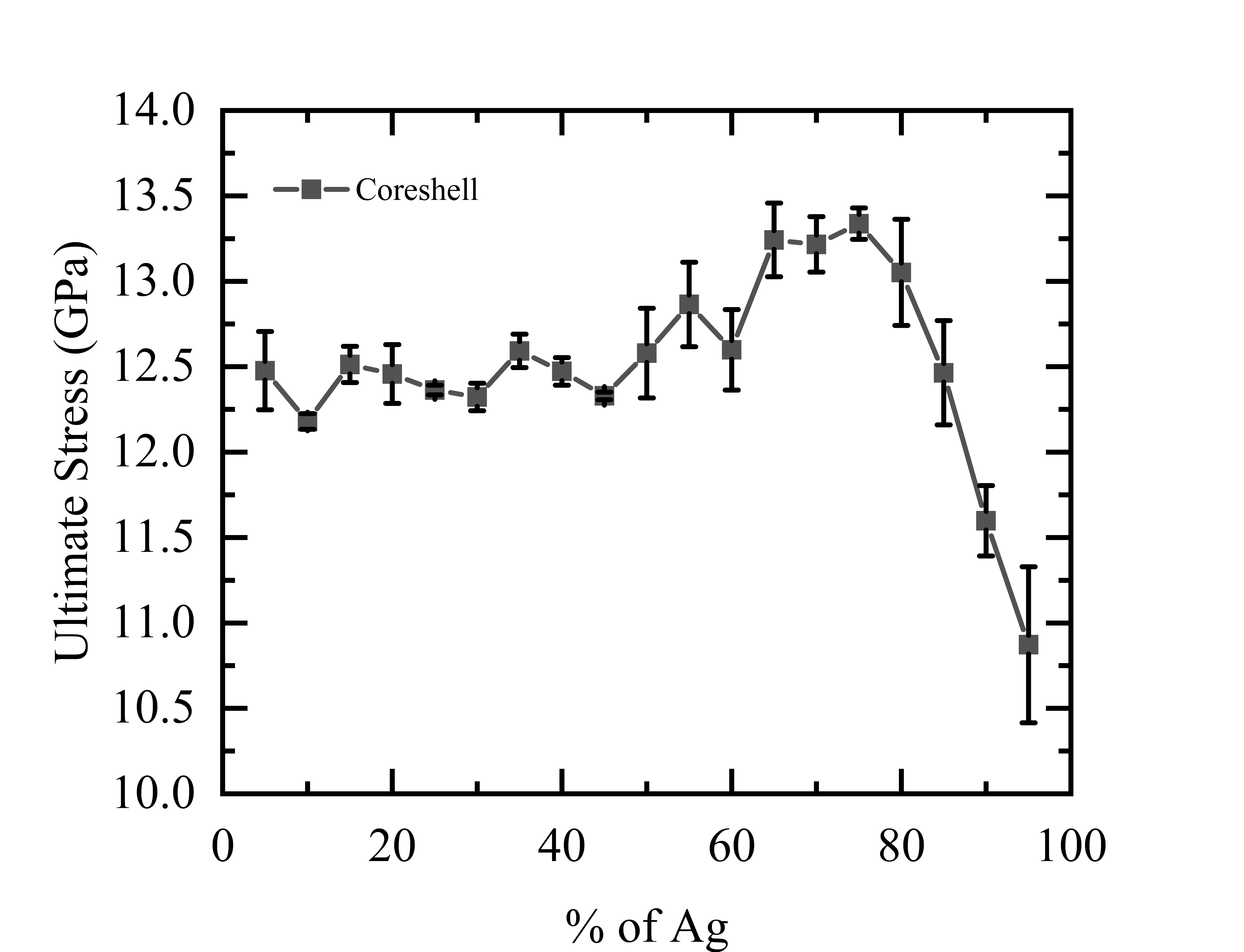}}
		\caption{ }
		\label{fig8b}
	\end{subfigure}
	\caption{Variation of Ultimate Stress of (a) FGM and (b) Coreshell nanosphere obtained from molecular dynamics simulation with different percentages of Ag.}
	\label{fig8}
\end{figure}

Fig. \ref{fig8} depicts the ultimate stress variation for FGM and Core Shell nanospheres. As the proportion of Ag in FGM nanospheres rises, the curve in Fig. \ref{fig8a} displays a downward trend. Ultimate stress created in Core Shell nanospheres, on the other hand, differed from that in FGM nanospheres in Fig. \ref{fig8b}. The ultimate stress in core shell nanospheres grows as the proportion of Ag increases until it reaches 75 percent. Ultimate stress begins to decrease after this range. This may happen due to the Core Shell effect on the nanospheres. We may conclude that until the proportion of Ag in nanospheres reaches 75\%, the ultimate stress of Core Shell and FGM nanospheres behaves in opposite ways, and after that they follow a similar pattern.

\begin{figure}[htbp]
	\centering
	\begin{subfigure}{0.7\textwidth}
		\centerline{\includegraphics[width=1\textwidth]{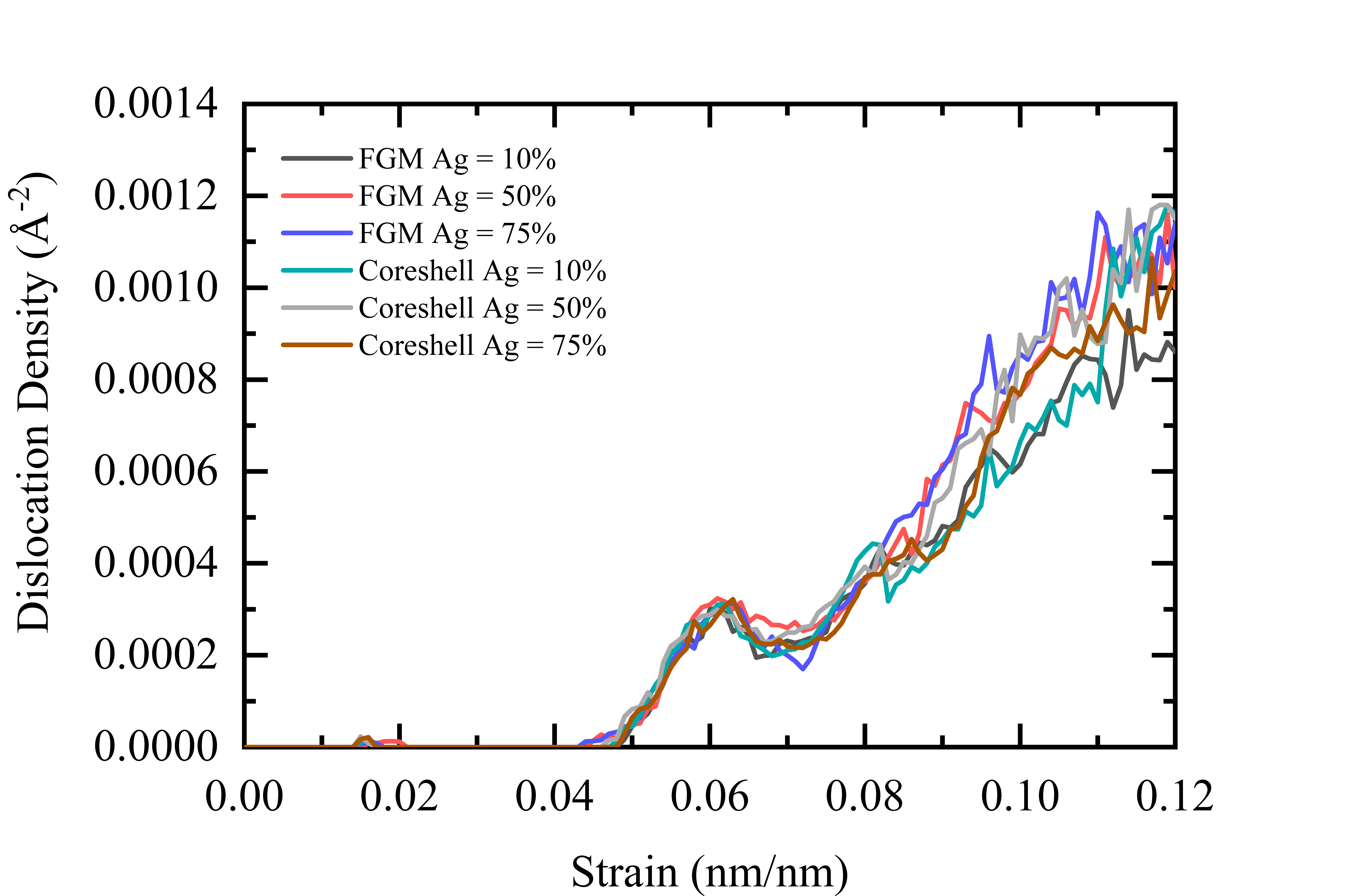}}
		\caption{ }
		\label{fig9a}
	\end{subfigure}
	\begin{subfigure}{0.7\textwidth}
		\centerline{\includegraphics[width=1\textwidth]{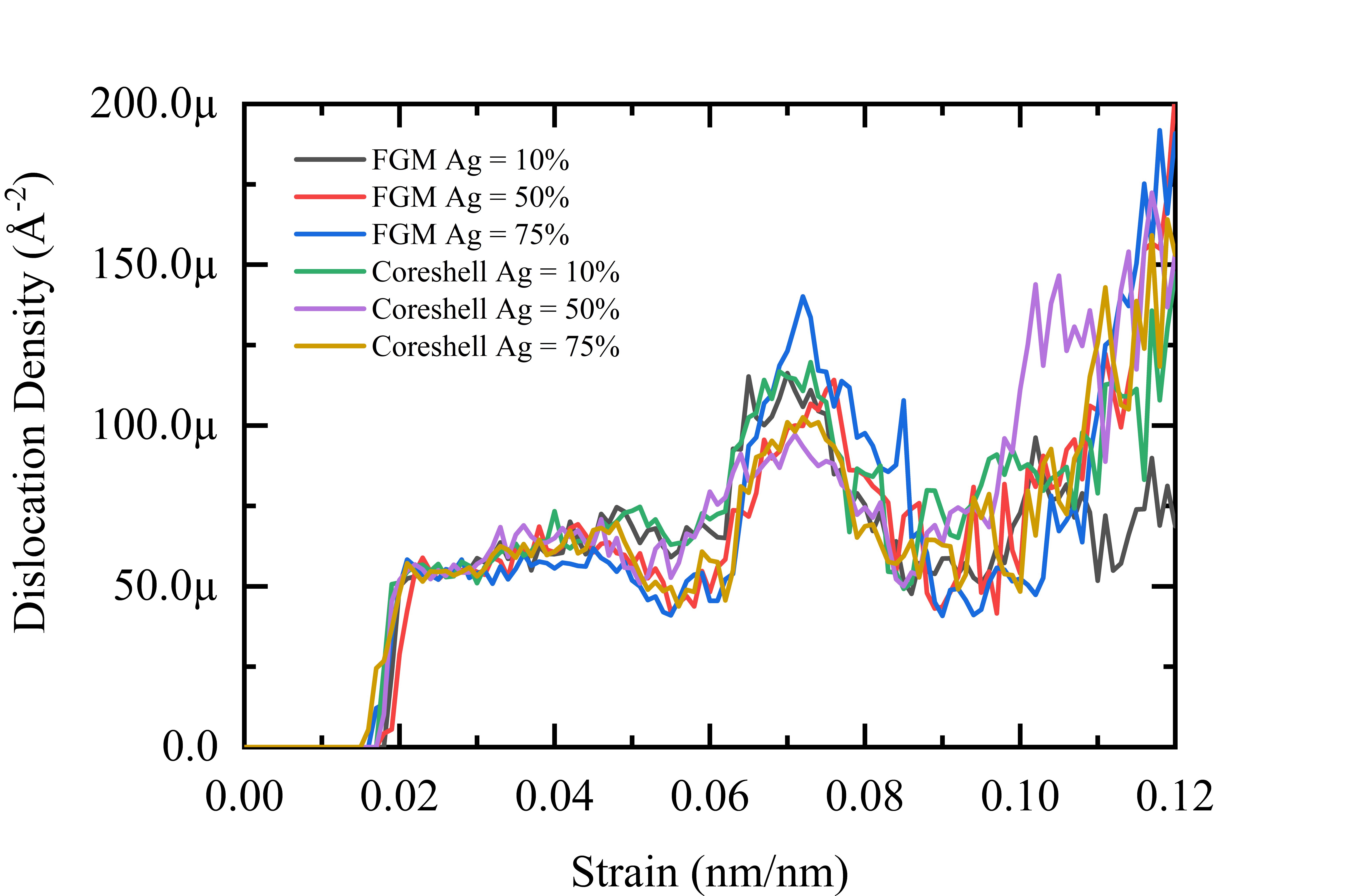}}
		\caption{ }
		\label{fig9b}
	\end{subfigure}
	\caption{Dislocation Density vs. Strain Curves of FGM and Coreshell nanosphere for different percentages of Ag. (a) Shockley and (b) Hirth dislocation density are showed in different graphs.}
	\label{fig9}
\end{figure}

\subsection{Deformation Mechanism}
We identify and describe the compression test, using the case of a 15 nm nanoparticle as an example. As illustrated on the stress vs. deformation curves in Fig.\ref{fig10}, elastic behavior is first obtained. When two Shockley partial dislocations are generated from the top and propagate in the four possible {111} slip planes, the first indication of plastic deformation appears in FGM nanospheres at a compression depth of $\delta|=1.19$\AA (i.e., 1.6\% deformation), providing the slight stress drop\cite{salah2017influence}. In all nanoparticles studied, we discovered that plasticity is initiated by the nucleation of these partial dislocations from the edges of the surface in contact with indenters. The edges form steps with the bottom layer and act as stress concentrators, a well-known phenomena in uniaxial surface deformation\cite{brochard2010elastic}. Shockley partial dislocations are a pair of dislocations that can cause stacking faults. By enabling an alternate path for atomic mobility, this pair of partial dislocations can permit dislocation motion. The Burgers vector of the Shockley partial dislocation lies in the plane of the fault, making it glissile. The term glissile refers to the mobile dislocations. A new Hirth dislocation is formed from the contact surface when  $\delta=1.33$\AA (1.8\%) and merges with Shockley partial dislocation. Hirth dislocation is sessile known as Hirth lock. When $\delta=1.48$\AA (2\%), five Hirth and two Shockley dislocations are visible. All Shockley dislocations disappear at $\delta=1.56$\AA (2.1\%); only eight Hirth dislocations are visible. Four Hirth dislocations unite in a point at $\delta=1.63$\AA (2.2\%), forming a pyramid-like structure that vanishes immediately. Again, four Hirth dislocations from each surface merge in a point and form two pyramid-like structures in the upper and lower contact surfaces during $\delta=2.82$\AA (3.8 \%). Two fresh Shockley partial dislocations are nucleated from the pyramid's pin-point as compression proceeds, and they combine with Hirth's dislocation at $\delta=3.33$\AA (4.5\% ). One of the newly formed Shockley dislocations is fixed by a Hirth dislocation, while the other propagates toward the contact surface. When $\delta=4.075$\AA (5.5\%), all Hirth and Shockley dislocations converge in a point and start to propagate towards the opposite surface. As the compression goes on, all dislocations merge and create another new pin-point and begin to propagate sidewise towards the opposing surface at $\delta=4.74$\AA (6.4\%). During $\delta=6.59 $\AA (8.9\%), the intersection of the Shockley partial dislocations forms two stair-rod $\frac{1}{6}[110]$ dislocations according to the reaction:
\begin{equation}
	\frac{1}{6}[\bar{1}21] +\frac{1}{6}[2\bar{1}\bar{1}] = \frac{1}{6}[110]
\end{equation}

This stair-rod dislocation is sessile known as Lomer-Cottrell lock - it locks slip in the two slip planes. All three types of dislocation merge in a point and propagate at $\delta=6.89$\AA (9.3\%). When $\delta=7.41$\AA (10\%), one new dislocation is introduced, known as Frank dislocation, it vanishes instantly. As its Burgers vector is not included in one of the {111} planes, the Frank partial is an edge dislocation that cannot glide and expand conservatively under the influence of applied stress. Unlike the glissile Shockley partial, this dislocation is stated to be sessile. But by climbing it can move \cite{smallman2014introduction}. A perfect dislocation results from the interaction of a Shockley partial dislocation with a Frank dislocation at $\delta=8.37$\AA (11.3\%), and it disappears quickly. The reaction is:
\begin{equation}
	\frac{1}{6}[\bar{1}21] +\frac{1}{6}[2\bar{1}\bar{1}] = \frac{1}{6}[110]
\end{equation}
When $\delta=9.26$\AA (12.5\%), perfect dislocation is visible again and gains stability. Two side Shockley dislocations meet with each other in the surface area at $\delta=9.85$\AA (13.3\%) and start to propagate.

For core-shell of 15nm, Plastic deformation occurs when the compression depth is $\delta =1.11$\AA (1.5\%). The top and bottom surfaces emit four shockley partial dislocations, which propagate. All Shockley partial dislocations vanish at $\delta =1.34$\AA (1.8\%), but two Hirth dislocations remain visible. When $\delta =1.49$\AA (2\%), there are eight Hirth dislocations visible four on each surface. At $\delta =2.3$\AA (3.1\%), four Hirth dislocations merge in a point and create a pyramid-like structure. Two pyramid-like structure appears on each surface at $\delta =2.45$\AA (3.3\%). As the compression goes on, two new Shockley partial dislocations are visible from the pyramid's pin-point at $\delta =3.49$\AA (4.7\%). New Shockley partial dislocation starts to propagate towards the contact surface. At $\delta =6.54$\AA (8.8\%), Stair-rod dislocation is visible and merges with Hirth dislocation and multiplies. Perfect dislocation is visual at $\delta =7.28 $\AA (9.8\%) and disappears immediately. At $\delta =7.5$\AA (10.1\%), Frank's dislocation appears and disappears quickly. As compression goes on, two surfaces' Shockley dislocation merges with each other at a compression depth of $\delta =10.10$\AA (13.6\%). 

Here We are focused on how dislocations spread from the contact surface and interact with one another. In our observation, plastic deformation begins with dislocation propagation from the contact surface, which is related to the nucleation of the initial dislocation in defect-free nanostructures. However, when the load is released, the initial nucleated dislocations, which lead to the formation of the pyramidal-like structure, spontaneously return to the surface. After the formation of a pyramid-shaped structure, the materials begin to act as amorphous materials with increasing compression depth , resulting in material failure.

\subsection{Effect of Dislocation Density on Strength}
\begin{figure}[htbp]
	\centerline{\includegraphics[width=1\textwidth]{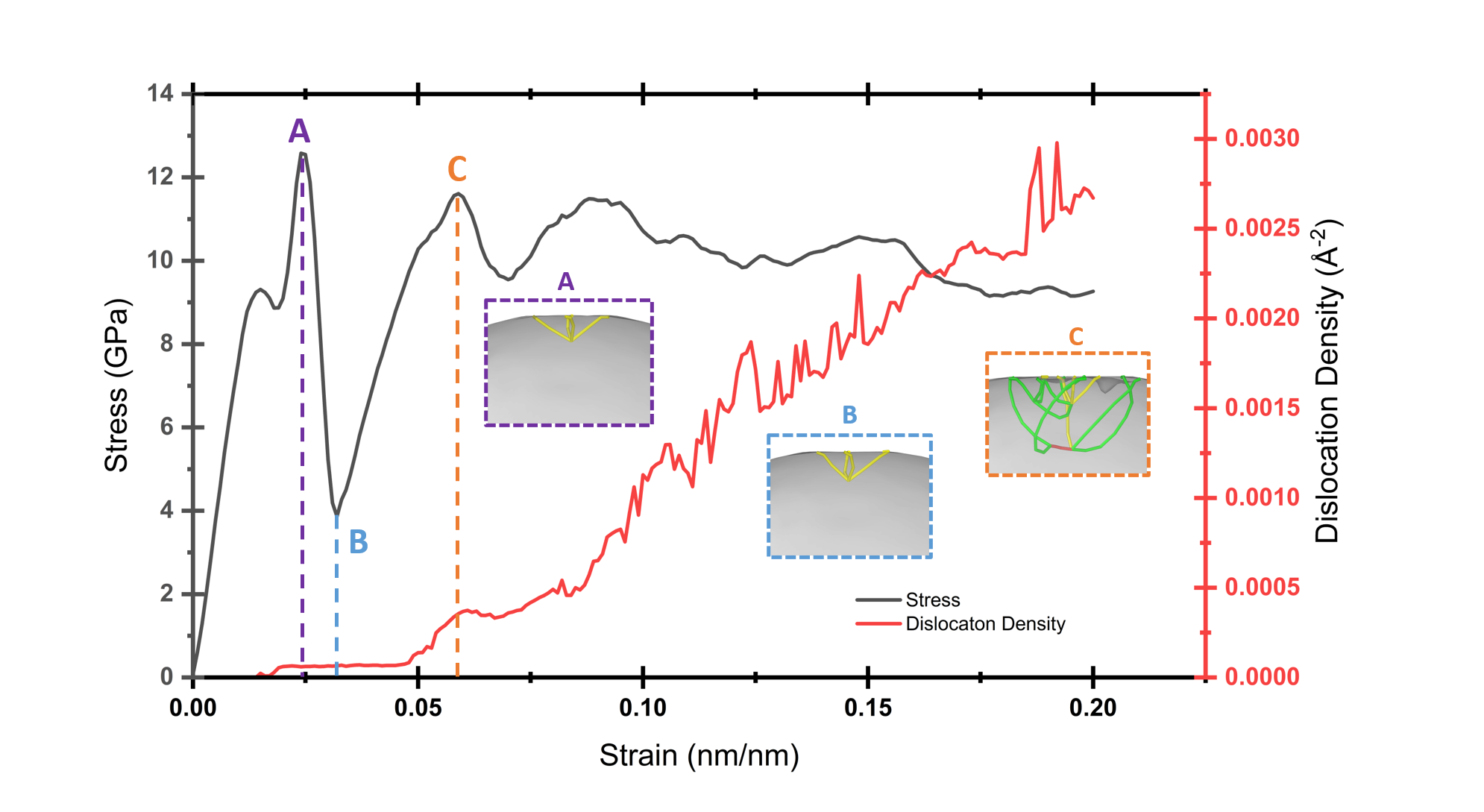}}	
	\caption{Variation of stress and Dislocation Density of Core-shell nanosphere with strain for 50\% Ag. Here, the Yellow line indicates Hirth dislocation, and the Green line means Shockley dislocation.}
	\label{fig10}
\end{figure}
There is a link between dislocation density and stress. According to Stress Vs. Strain and Dislocation density Vs. Strain curves shown in Fig.\ref{fig10}, we observed that stress is mostly regulated by dislocation density. Partial Shockley dislocation is observed at 1.5-1.6\% strain. This partial dislocation may enable dislocation motion. The term glissile refers to this form of partial dislocation. Shockley partials have the ability to glide on the glide plane, causing stacking faults. A stacking fault ribbon connects two partial dislocations as Shockley partial dislocation glides on the slip plane yielding a small stress drop\cite{salah2017influence}.

Shockley partial dislocation vanishes as compression proceeds, and Hirth dislocation emerges at 2-2.1\% strain. Hirth dislocation, also known as Hirth lock, is a sessile dislocation (a sessile dislocation cannot glide or move) having crystallographic orientations such as [100]. As these directions do not allow slip, the Hirth dislocations become locked and function as obstacles to any neighboring movable dislocations that share the same slip system, resulting in strain hardening. Because of this strain hardening, the stress curve reaches its maximum point as shown in Fig.\ref{fig10} Point A. The accumulation of dislocation barriers, which function as impediments to uninhibited dislocation movement, causes strain hardening. Immobile or sessile dislocations are the most frequent barrier in FCC metals, and they play a significant role in hardening behavior. Dislocation interactions between dislocations on various glide planes can produce these. Some dislocation interactions can cause dislocation annihilation. Dislocation annihilation can occur as a result of some dislocation interactions. Some can generate dislocation locks, which result in stable dislocation segments forming in crystallographic directions (Hirth lock or lomer lock). These are not slip directions, but rather crystallographic directions. These locks prevent slippage and create barriers to other mobile dislocations, as well as causing strain hardening\cite{hull2011introduction}.

\begin{figure}[htbp]
\centering
	\begin{subfigure}{0.7\textwidth}
		\centerline{\includegraphics[width=1\textwidth]{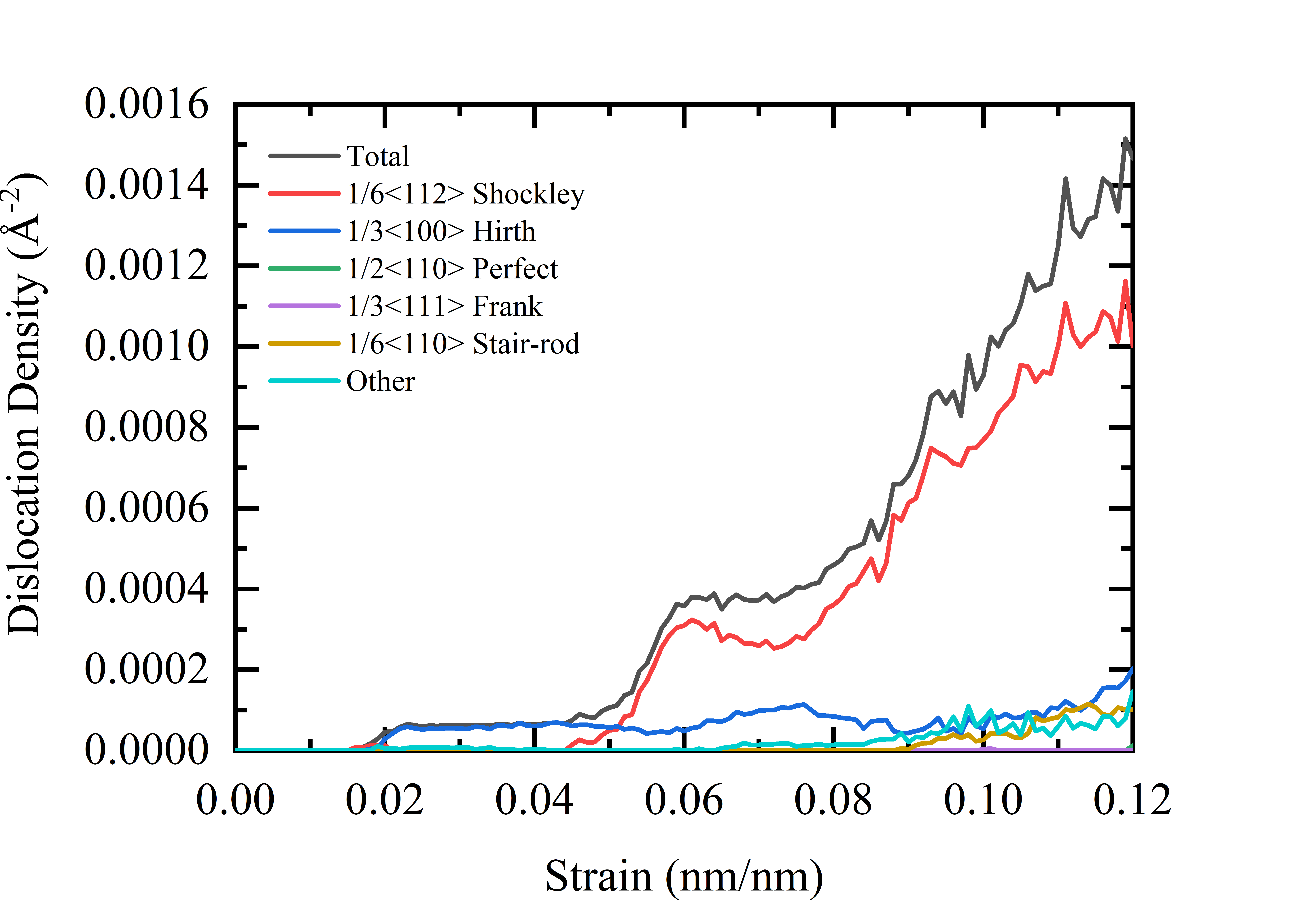}}
		\caption{ }
		\label{fig11a}
	\end{subfigure}
	\begin{subfigure}{0.7\textwidth}
		\centerline{\includegraphics[width=1\textwidth]{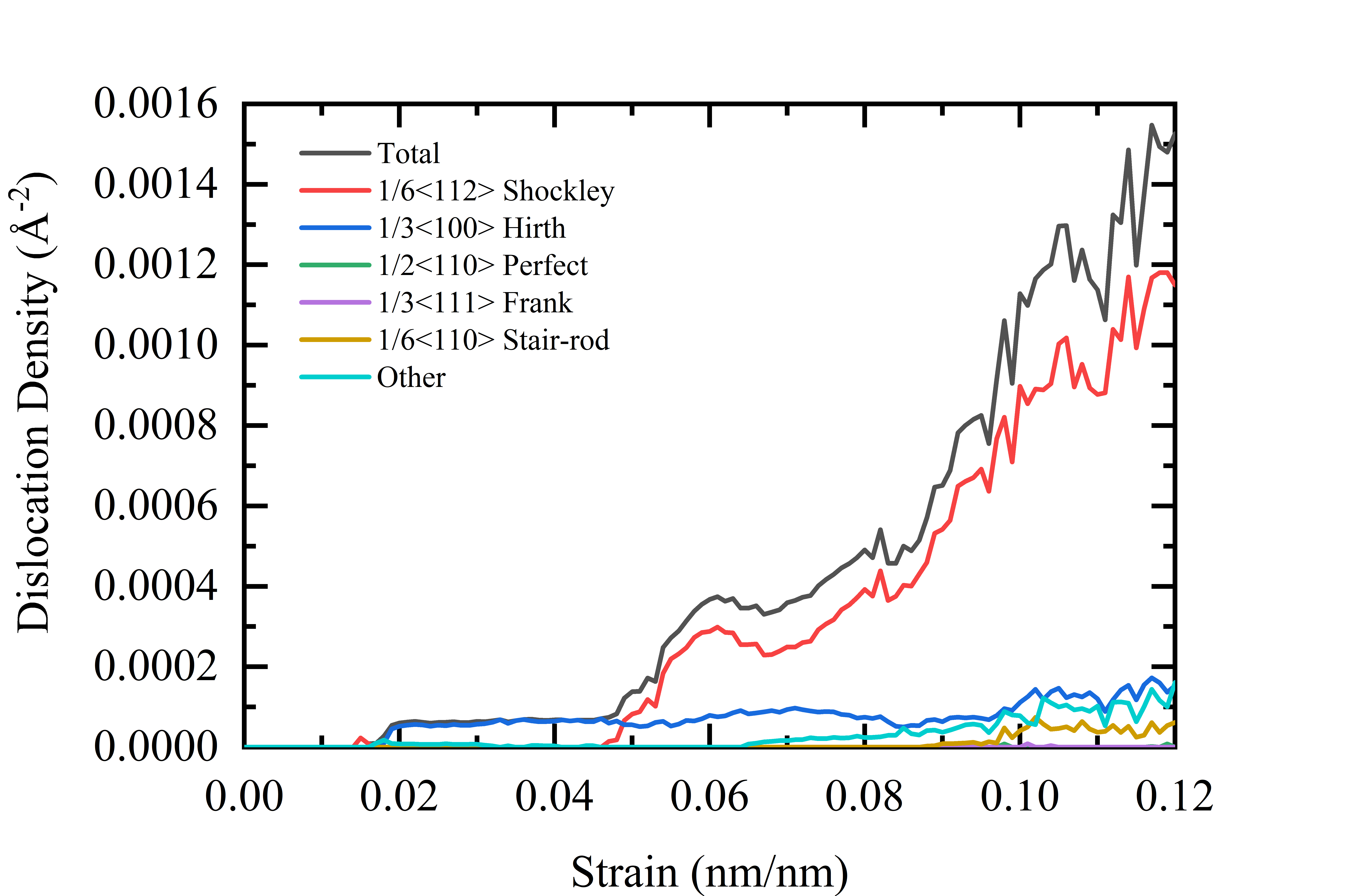}}
		\caption{ }
		\label{fig11b}
	\end{subfigure}
	\caption{Variation of dislocation density of (a) FGM and (b) Coreshell nanosphere (Ag = 50\%) with strain for different types of Partial dislocation.}
	\label{fig11}
\end{figure}
The stress curve abruptly decreases after the pick value of stress, with the growth of Hirth dislocation density, and falls to the lowest point at a 3.1-3.5 percent strain as shown in Fig. \ref{fig10} Point B. With a reduction in Hirth dislocation density from the lowest point, the stress curve begins to rise again. Shockley partial dislocation resurfaces at 4.5-5 percent strain, unites with Hirth dislocation, and propagates. Now stress largely depends on Shockley partial dislocation. The stress curve rises as the Shockley partial dislocation density increases and Hirth's dislocation density falls as shown in \ref{fig10} Point C. The stress curve is now proportional to the density of Shockley dislocations. Stair-rod dislocation is observable at 8.9\% strain, which is known as Lomer-Cottrell lock. As it locks slip in both slip planes, it is a sessile dislocation. Strain hardening happens as it prevents plane slide, and the stress curve grows as a result of this strain hardening.

We found a relationship between dislocation density and strain from the Dislocation Density Vs Strain graph, as shown in Fig 10. In the case of FGM and Core-shell, dislocation initiation begins at 1.5-1.7\% strain as shown in Fig. \ref{fig11} Shockley Partial dislocation $(\frac{1}{6}<112>)$ is observed on both surfaces at 1.5-1.7\% strain. As Shockley is a glissile type dislocation, it permits the plane to shift, lowering stress slightly. Shockley dislocations vanish from both surfaces after 2\% strain. On both surfaces, Hirth $(\frac{1}{3}<001>)$ dislocation has taken over. After 1.8\% strain, Hirth dislocation is evident. From 2.5 percent strain, Hirth dislocation density nearly stays the same. Only Hirth dislocation exists between 2\% and 4.5\% strain. Because there is just one dislocation available in that range, total dislocation density remains constant. After 4.5\% strain, a new Shockley partial dislocation forms. Hirth dislocation density remains constant from 4.5\% strain onwards, although Shockley partial dislocation density increases. Total dislocation density increases with strain starting at 4.5\% strain, as it now mostly depends on Shockley dislocation density. Other dislocations, such as Stair-rod dislocation $(\frac{1}{6}<110>)$, Frank dislocation $(\frac{1}{3}<111>)$, and Perfect dislocation $(\frac{1}{2}<110>)$, are visible after 8.9\% strain. However, with strain, their density remains low, and the material begins to behave like amorphism. As a result of their low density and lack of impact to material properties, certain forms of dislocations are neglected.

We can deduce from our findings that the stress curve is first dominated by Hirth dislocation density. After 2.7 percent strain, the relationship between Stress and Hirth Dislocation density is inversely proportional. The stress curve now depends on the density of Shockley partial dislocations as compression progresses and new Shockley partial dislocations emerge. After 4.5 percent strain, the relationship between stress and Shockley partial dislocation density is proportional.

\subsection{Comparison of Different Percentage of Alloying}
\begin{figure}[htbp]
	\centering
	\begin{subfigure}[h]{1\textwidth}
		\centering
		\begin{subfigure}[h]{0.24\textwidth}
			\centerline{\includegraphics[width=1\textwidth]{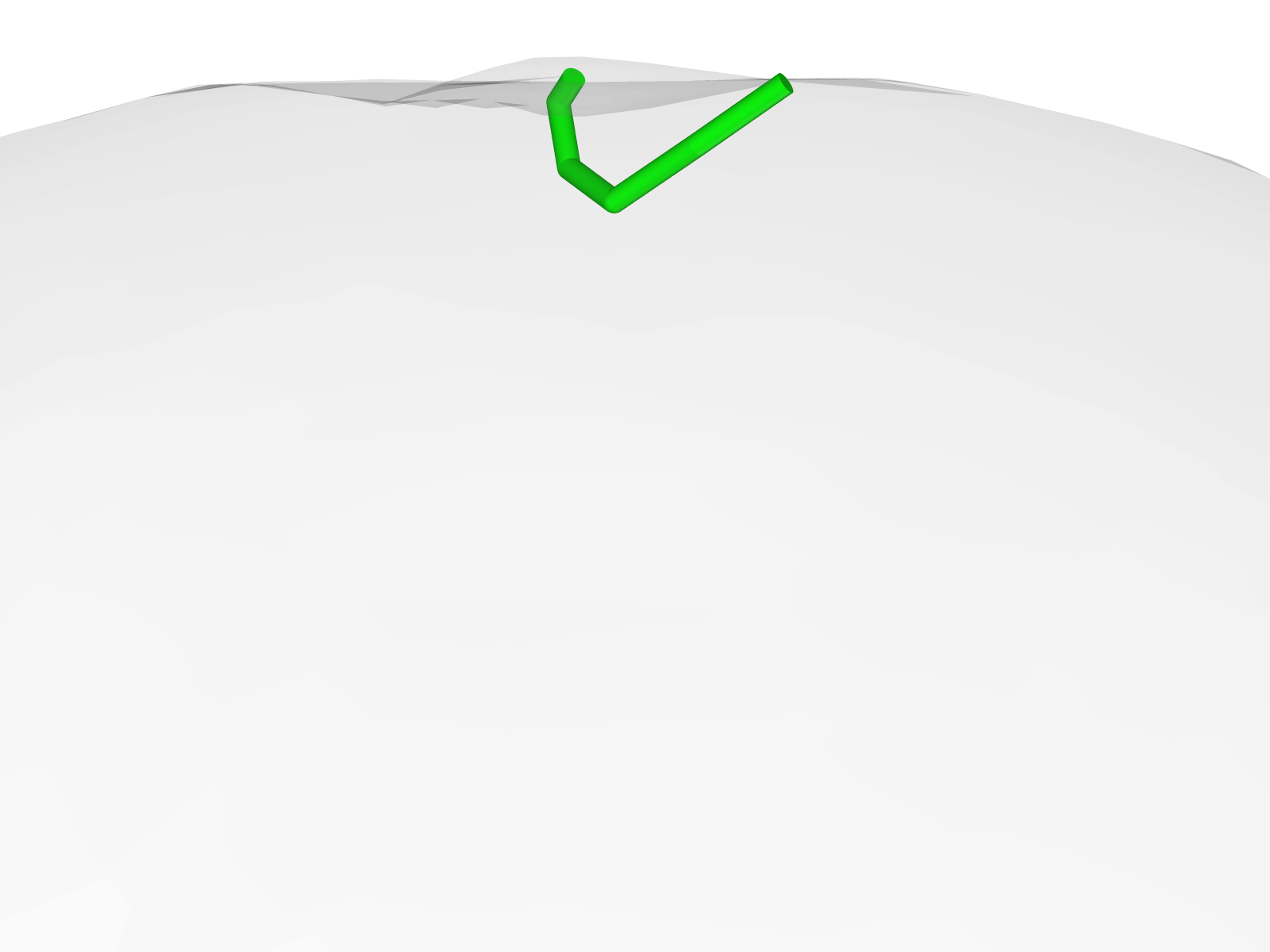}}
			\subcaption*{$\epsilon=1.6\%$}
		\end{subfigure}
		\begin{subfigure}[h]{0.24\textwidth}
			\centerline{\includegraphics[width=1\textwidth]{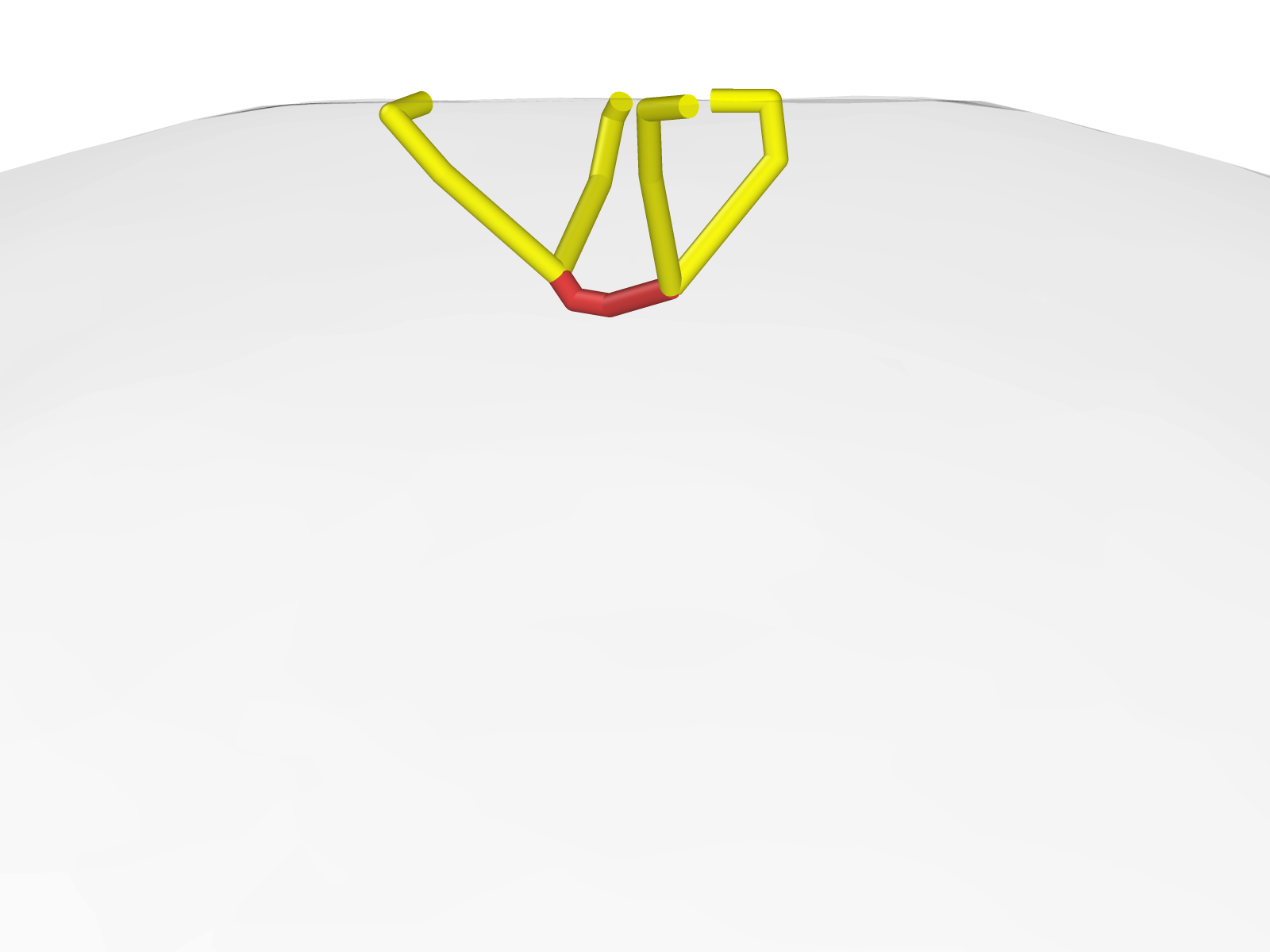}}
			\subcaption*{$\epsilon=2.6\%$}
		\end{subfigure}
		\begin{subfigure}[h]{0.24\textwidth}
			\centerline{\includegraphics[width=1\textwidth]{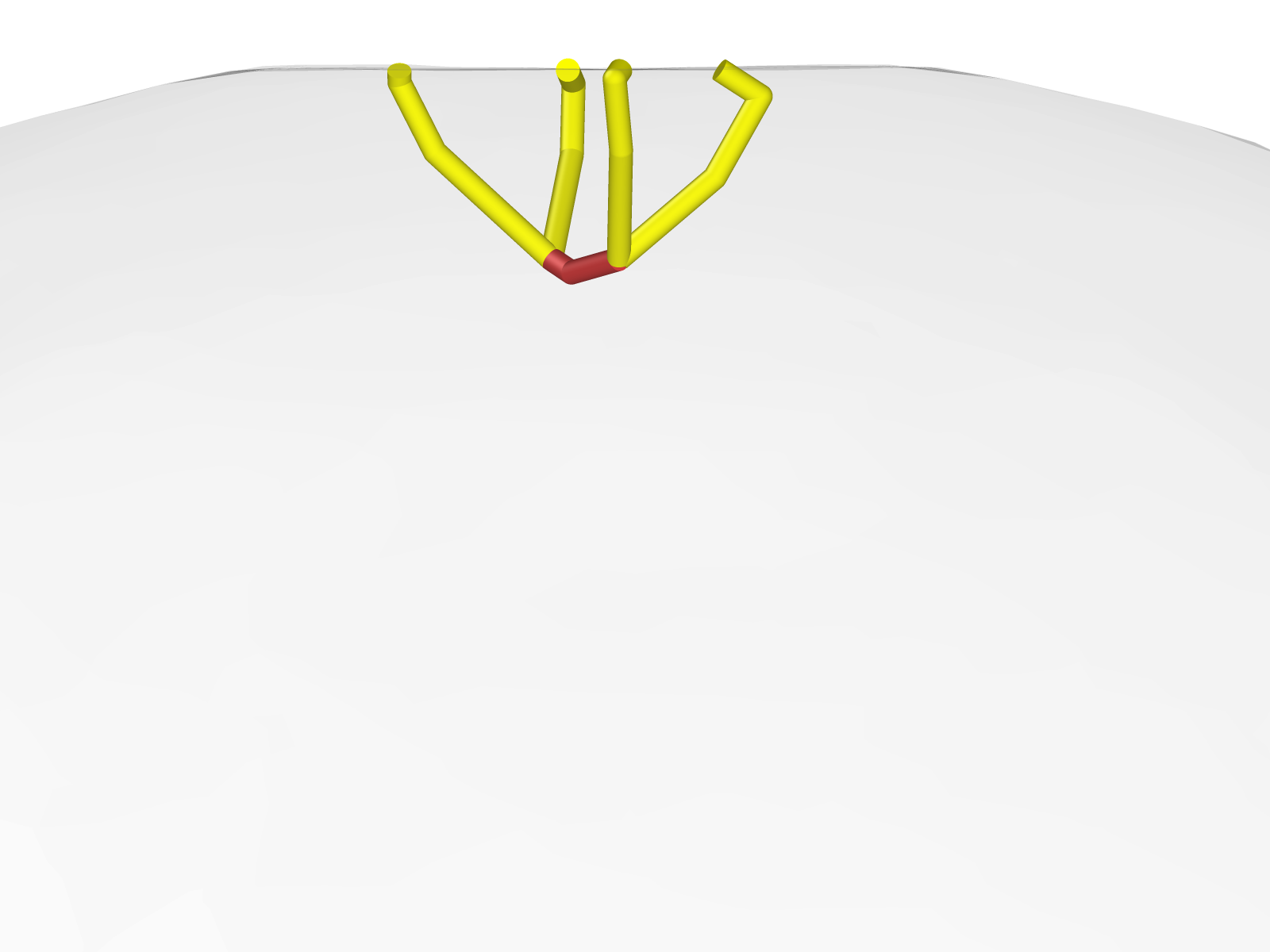}}
			\subcaption*{$\epsilon=3.1\%$}
		\end{subfigure}
		\begin{subfigure}[h]{0.24\textwidth}
			\centerline{\includegraphics[width=1\textwidth]{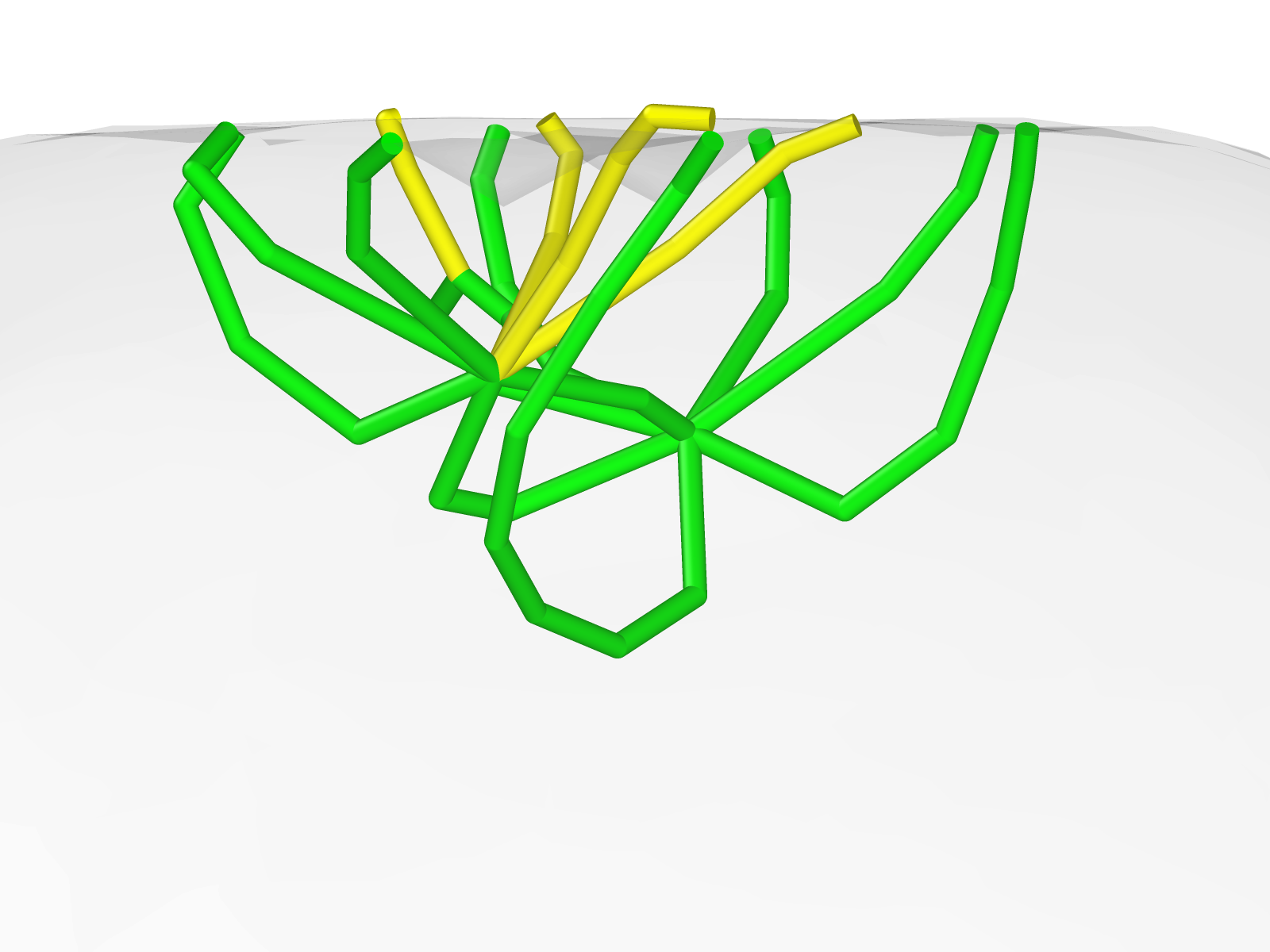}}
			\subcaption*{$\epsilon=6\%$}
		\end{subfigure}	
		\caption{Ag = 10\%}
	\end{subfigure}
	\begin{subfigure}[h]{1\textwidth}
		\centering
		\begin{subfigure}[h]{0.24\textwidth}
			\centerline{\includegraphics[width=1\textwidth]{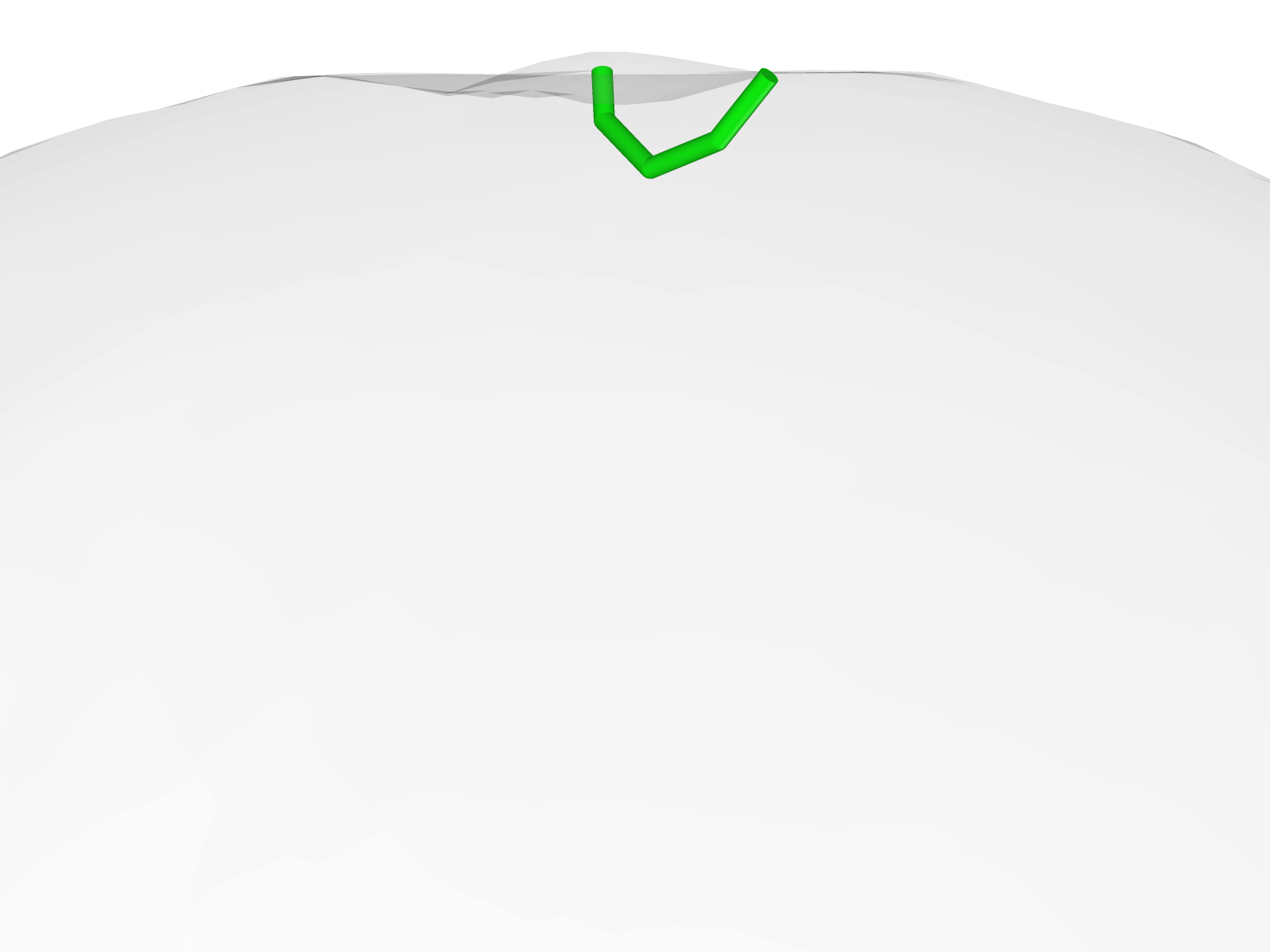}}
			\subcaption*{$\epsilon=1.6\%$}
		\end{subfigure}
		\begin{subfigure}[h]{0.24\textwidth}
			\centerline{\includegraphics[width=1\textwidth]{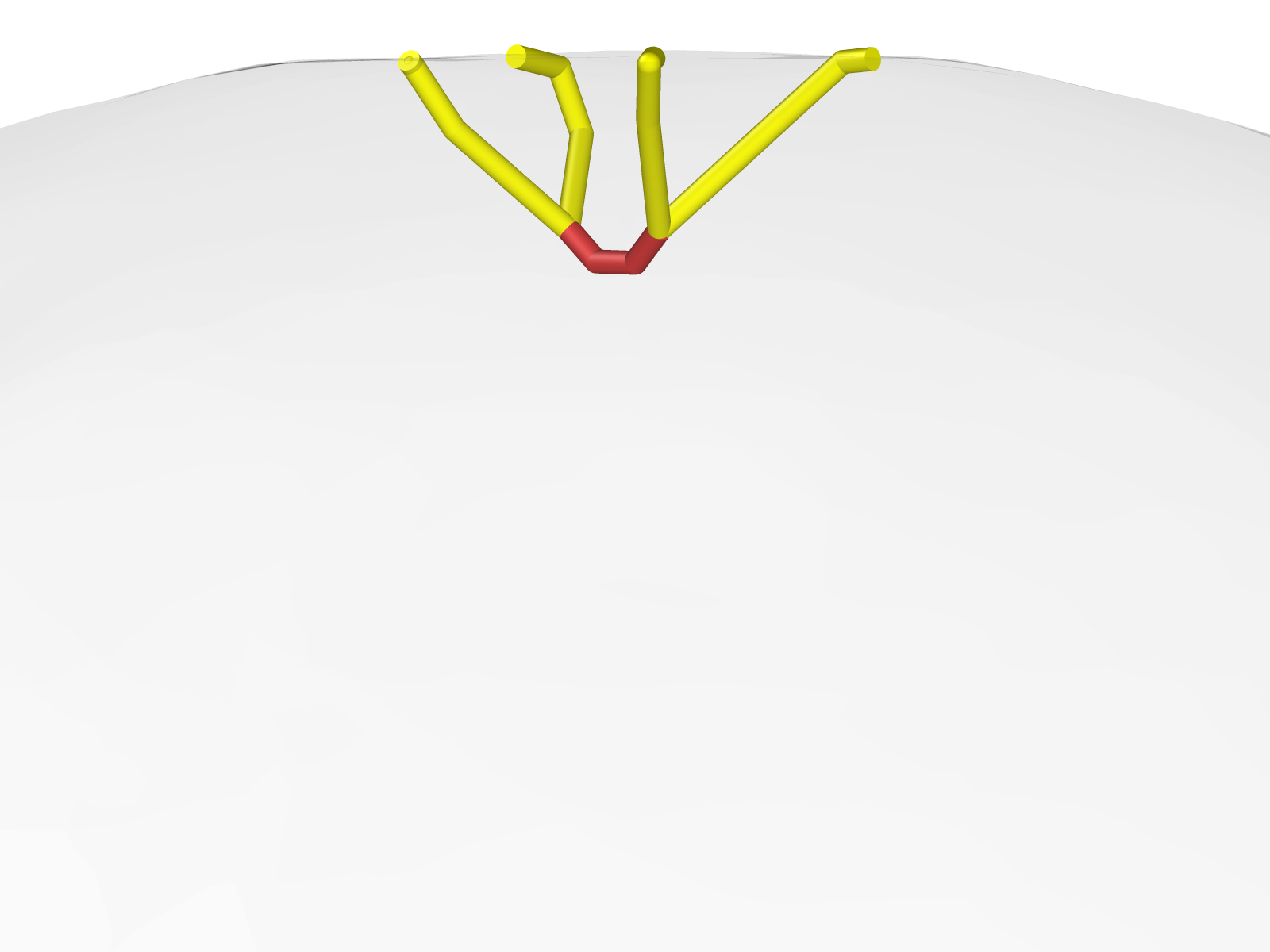}}
			\subcaption*{$\epsilon=2.6\%$}
		\end{subfigure}
		\begin{subfigure}[h]{0.24\textwidth}
			\centerline{\includegraphics[width=1\textwidth]{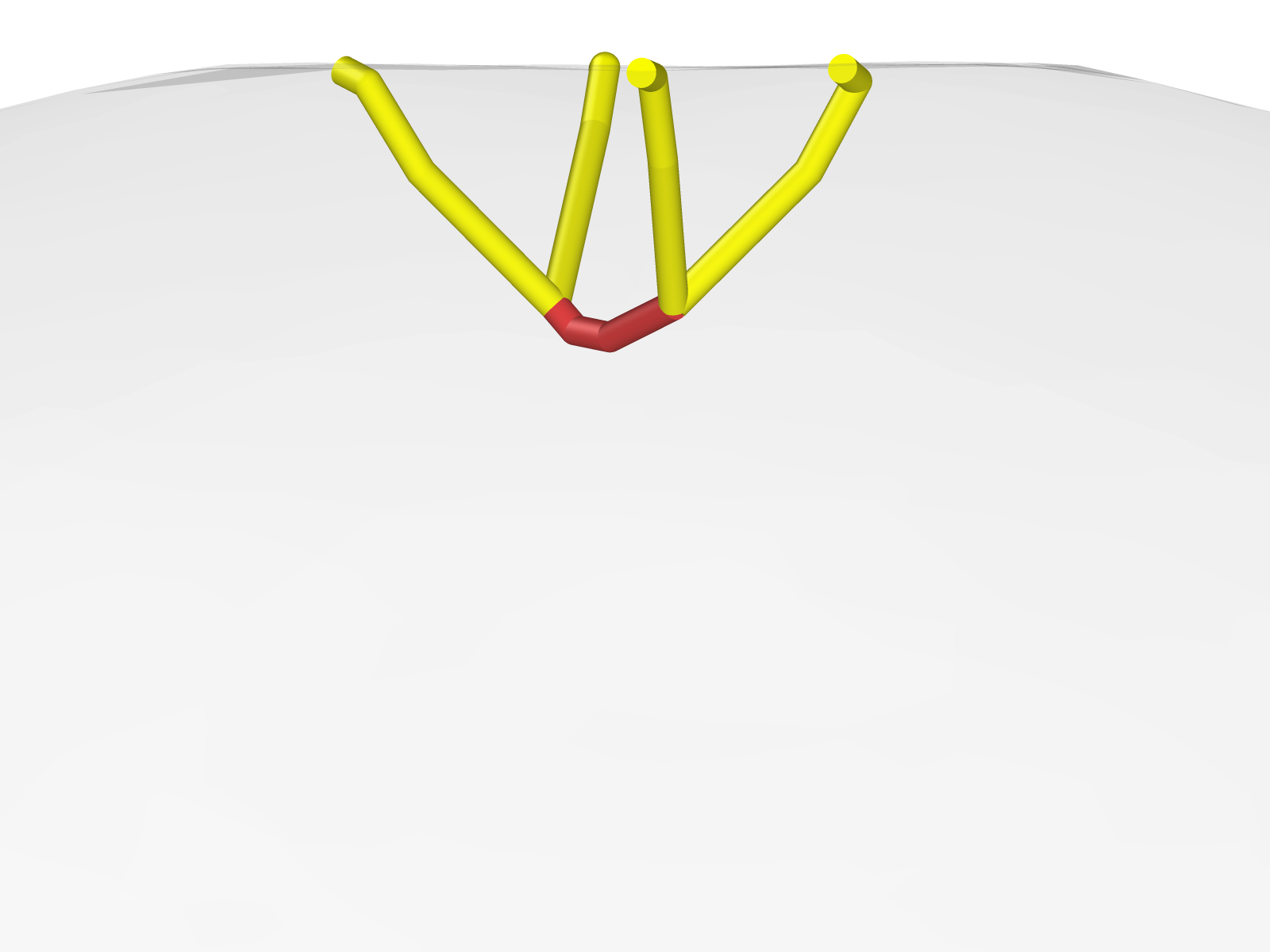}}
			\subcaption*{$\epsilon=3.2\%$}
		\end{subfigure}
		\begin{subfigure}[h]{0.24\textwidth}
			\centerline{\includegraphics[width=1\textwidth]{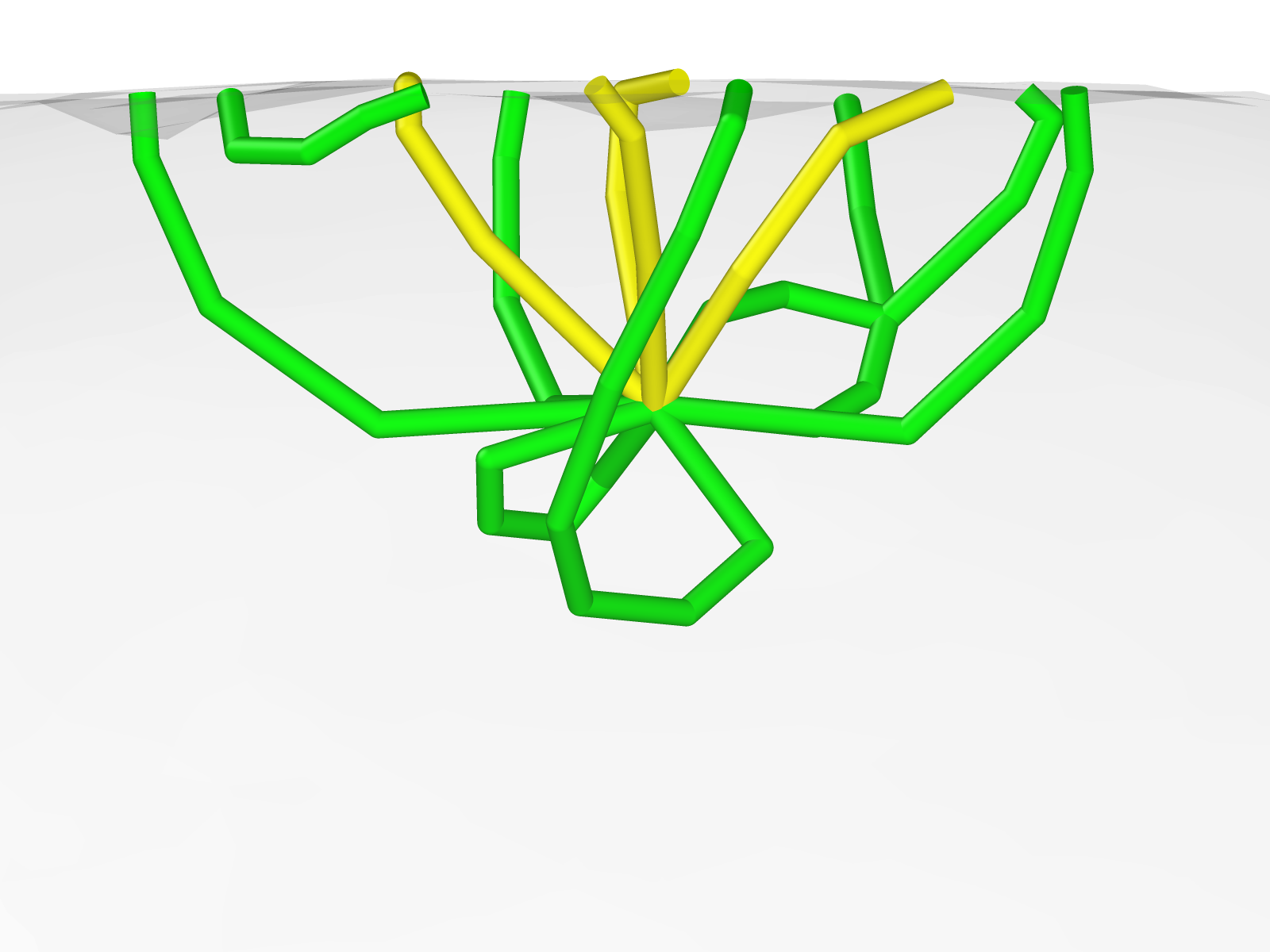}}
			\subcaption*{$\epsilon=6\%$}
		\end{subfigure}	
		\caption{Ag = 25\%}
	\end{subfigure}
	\begin{subfigure}[h]{1\textwidth}
		\centering
		\begin{subfigure}[h]{0.24\textwidth}
			\centerline{\includegraphics[width=1\textwidth]{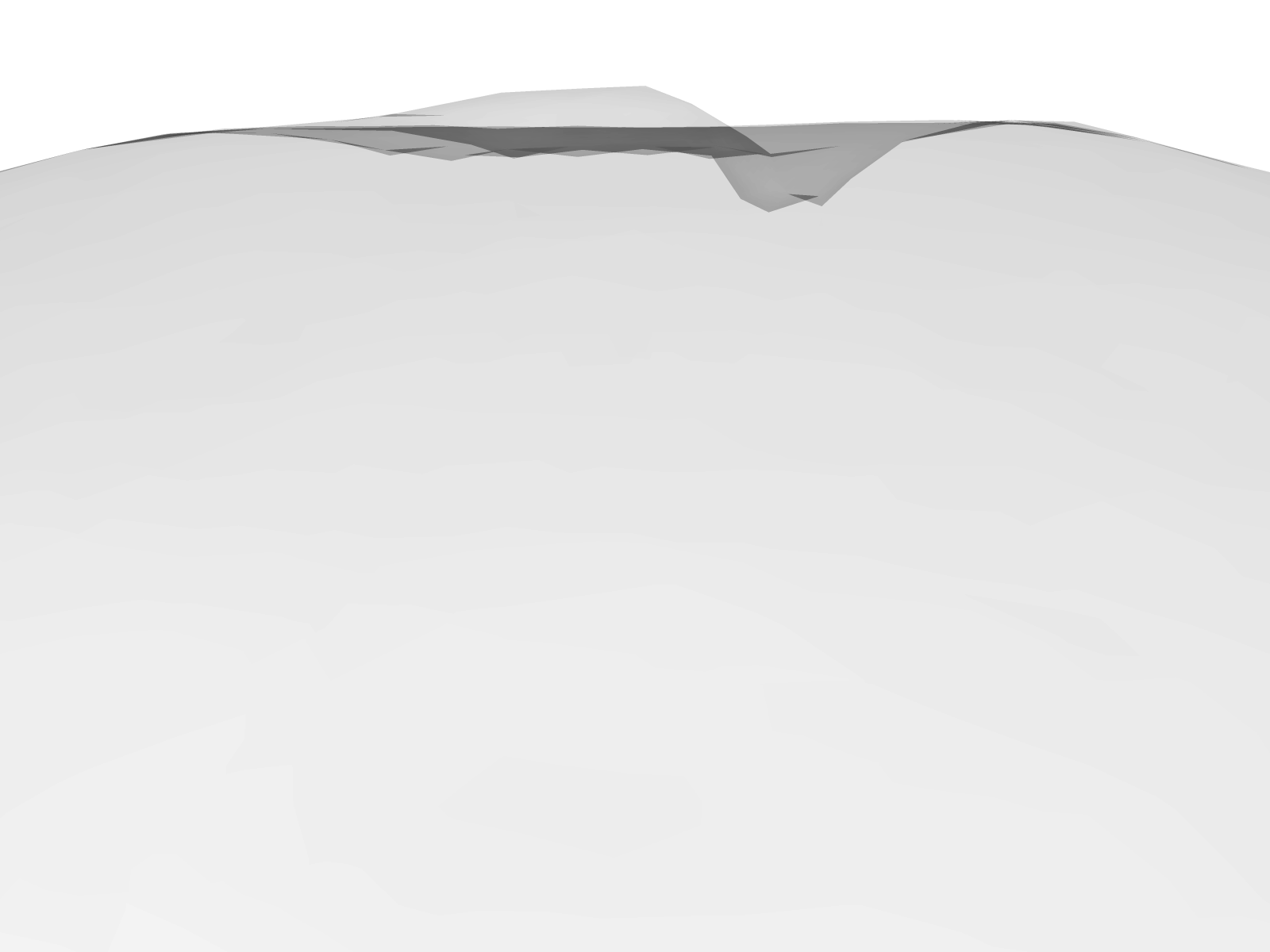}}
			\subcaption*{$\epsilon=1.6\%$}
		\end{subfigure}
		\begin{subfigure}[h]{0.24\textwidth}
			\centerline{\includegraphics[width=1\textwidth]{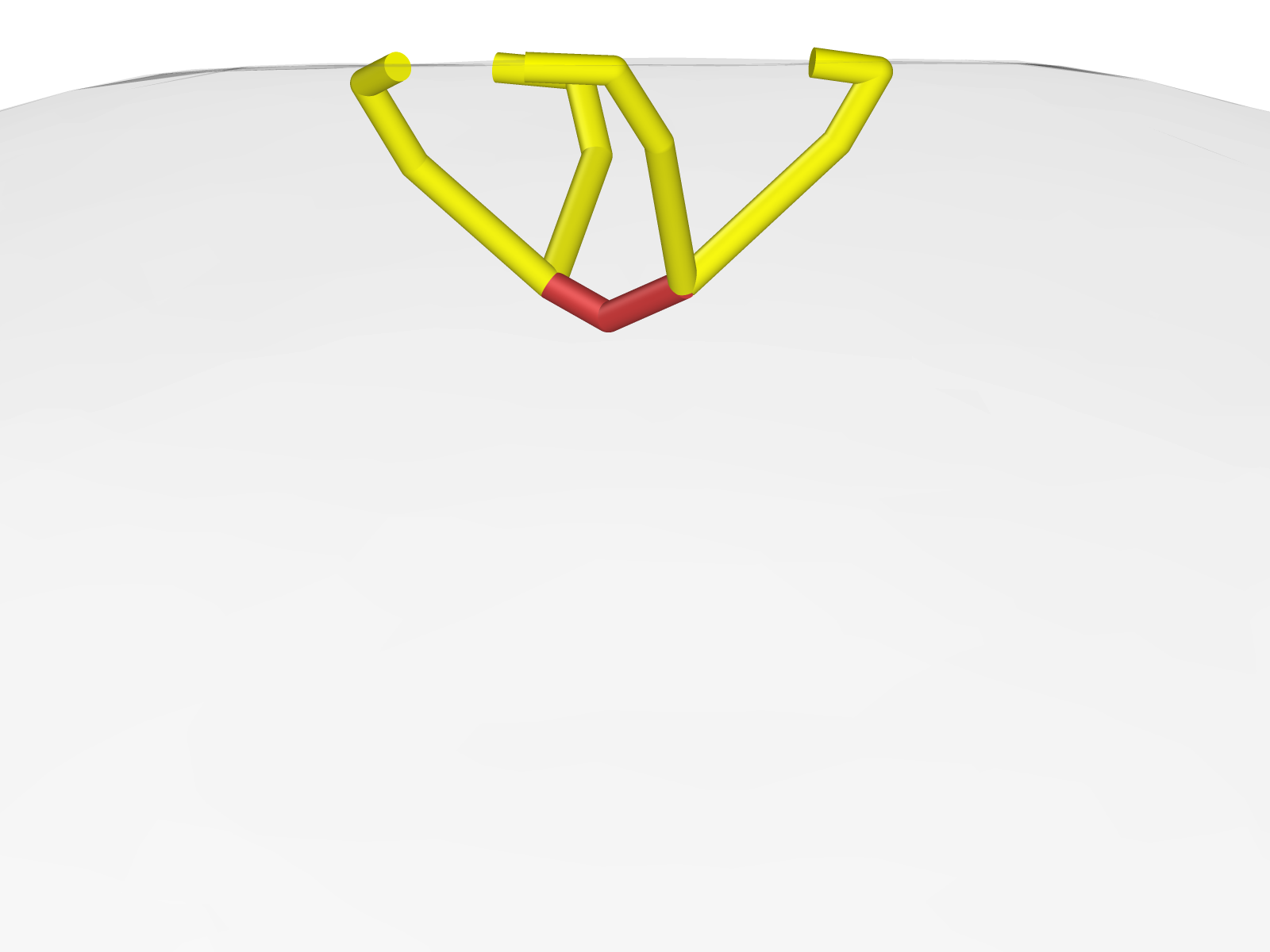}}
			\subcaption*{$\epsilon=2.7\%$}
		\end{subfigure}
		\begin{subfigure}[h]{0.24\textwidth}
			\centerline{\includegraphics[width=1\textwidth]{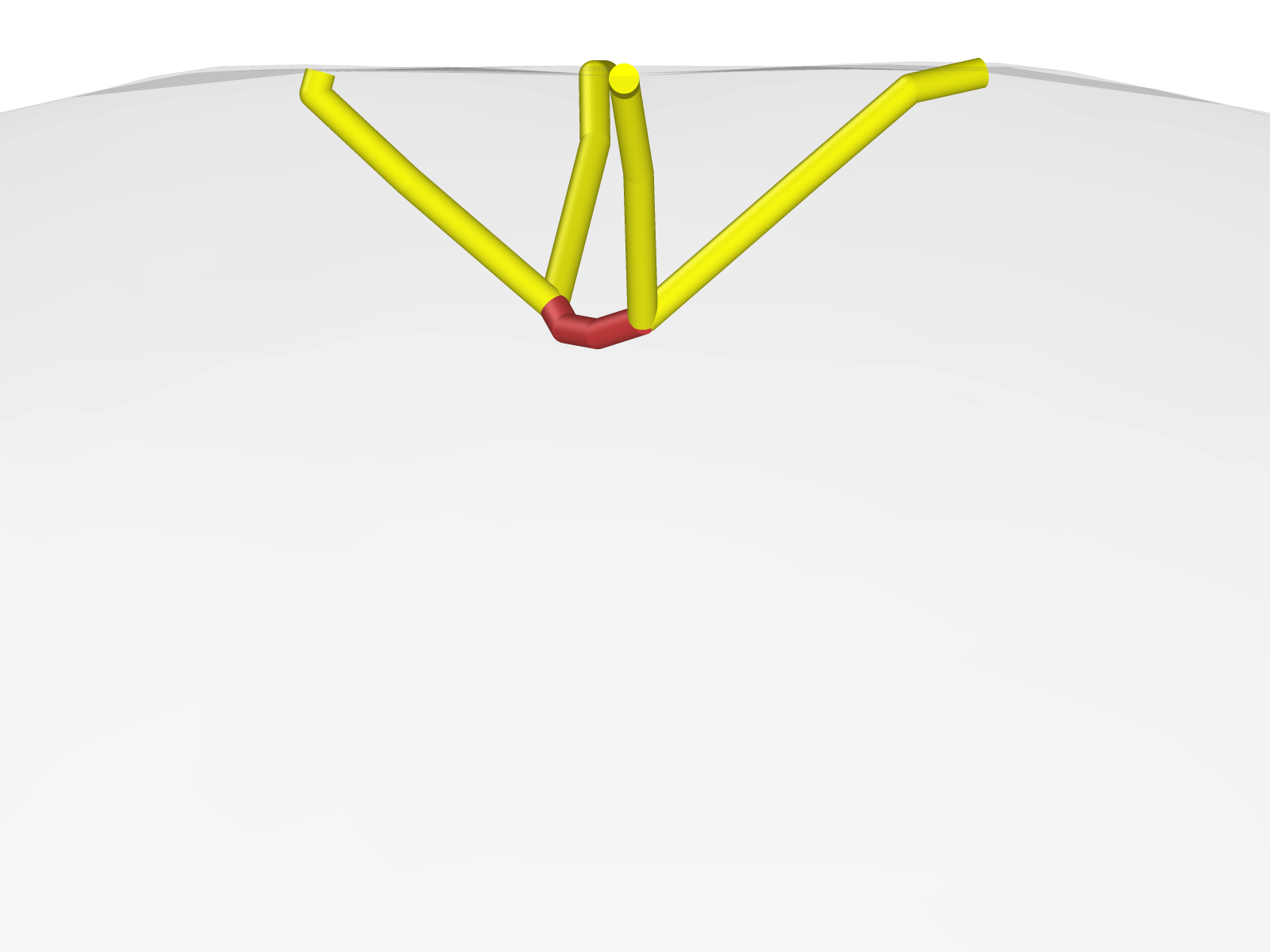}}
			\subcaption*{$\epsilon=3.1\%$}
		\end{subfigure}
		\begin{subfigure}[h]{0.24\textwidth}
			\centerline{\includegraphics[width=1\textwidth]{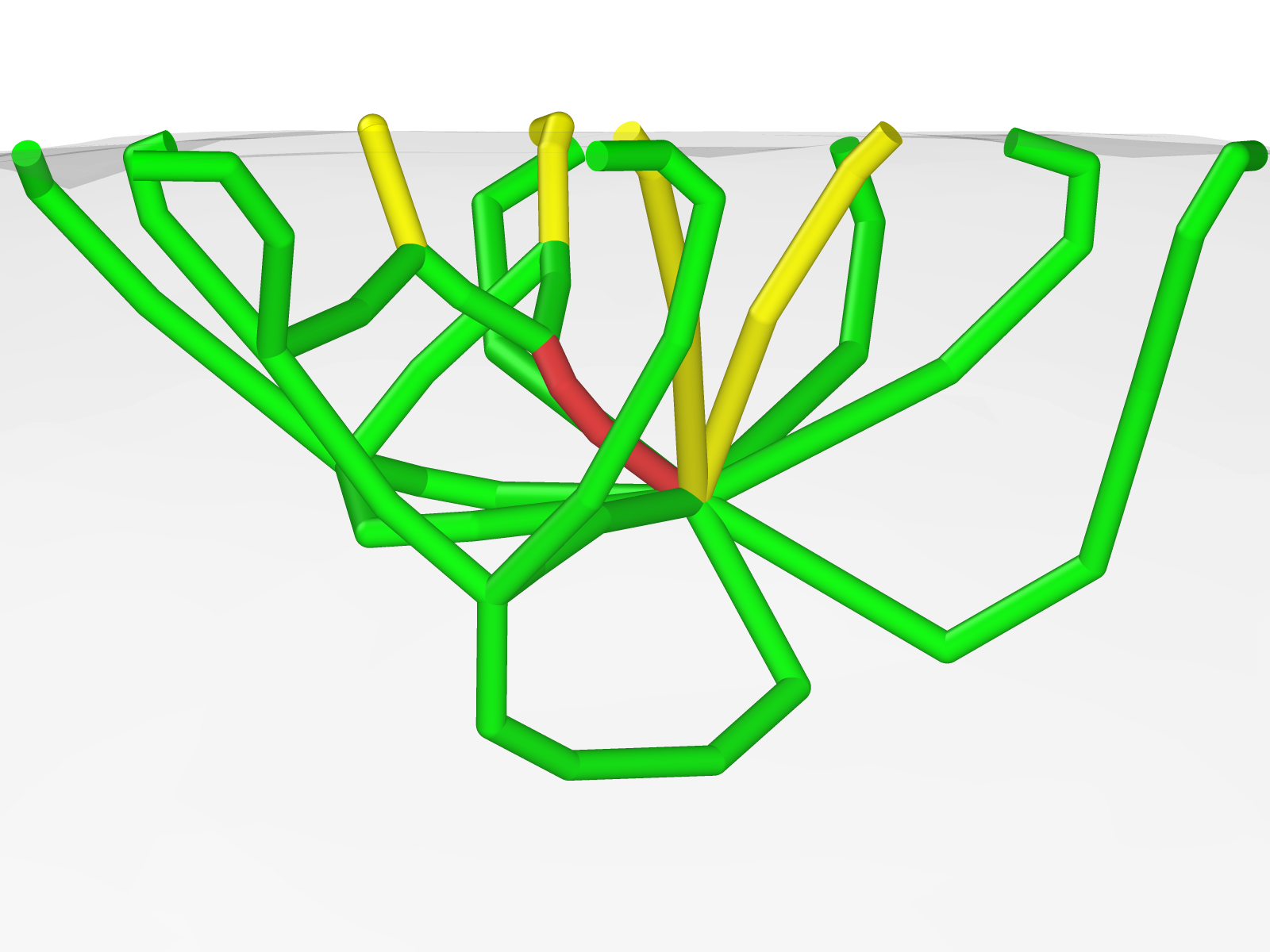}}
			\subcaption*{$\epsilon=6\%$}
		\end{subfigure}	
		\caption{Ag = 50\%}
	\end{subfigure}
	\begin{subfigure}[h]{1\textwidth}
		\centering
		\begin{subfigure}[h]{0.24\textwidth}
			\centerline{\includegraphics[width=1\textwidth]{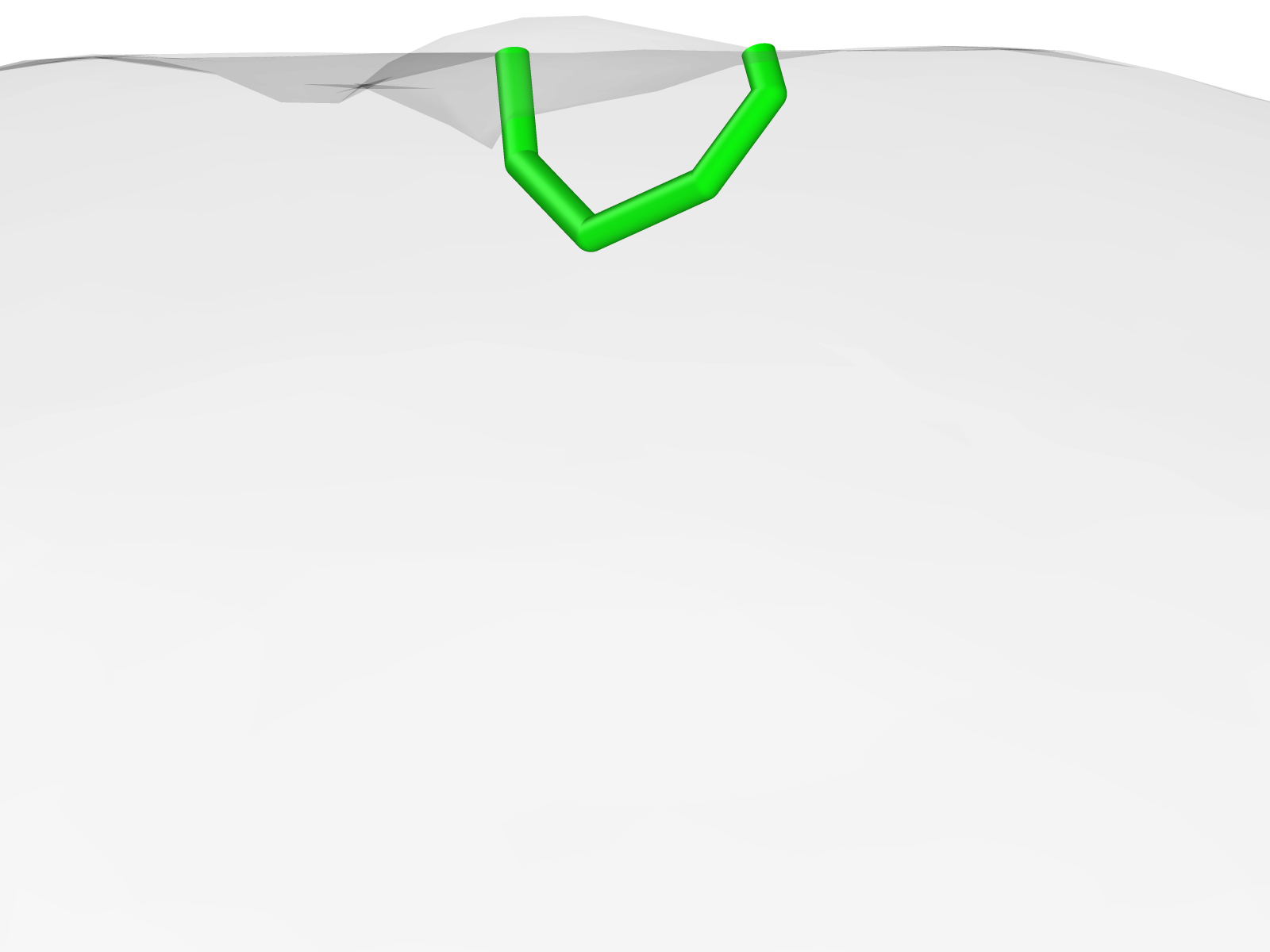}}
			\subcaption*{$\epsilon=1.7\%$}
		\end{subfigure}
		\begin{subfigure}[h]{0.24\textwidth}
			\centerline{\includegraphics[width=1\textwidth]{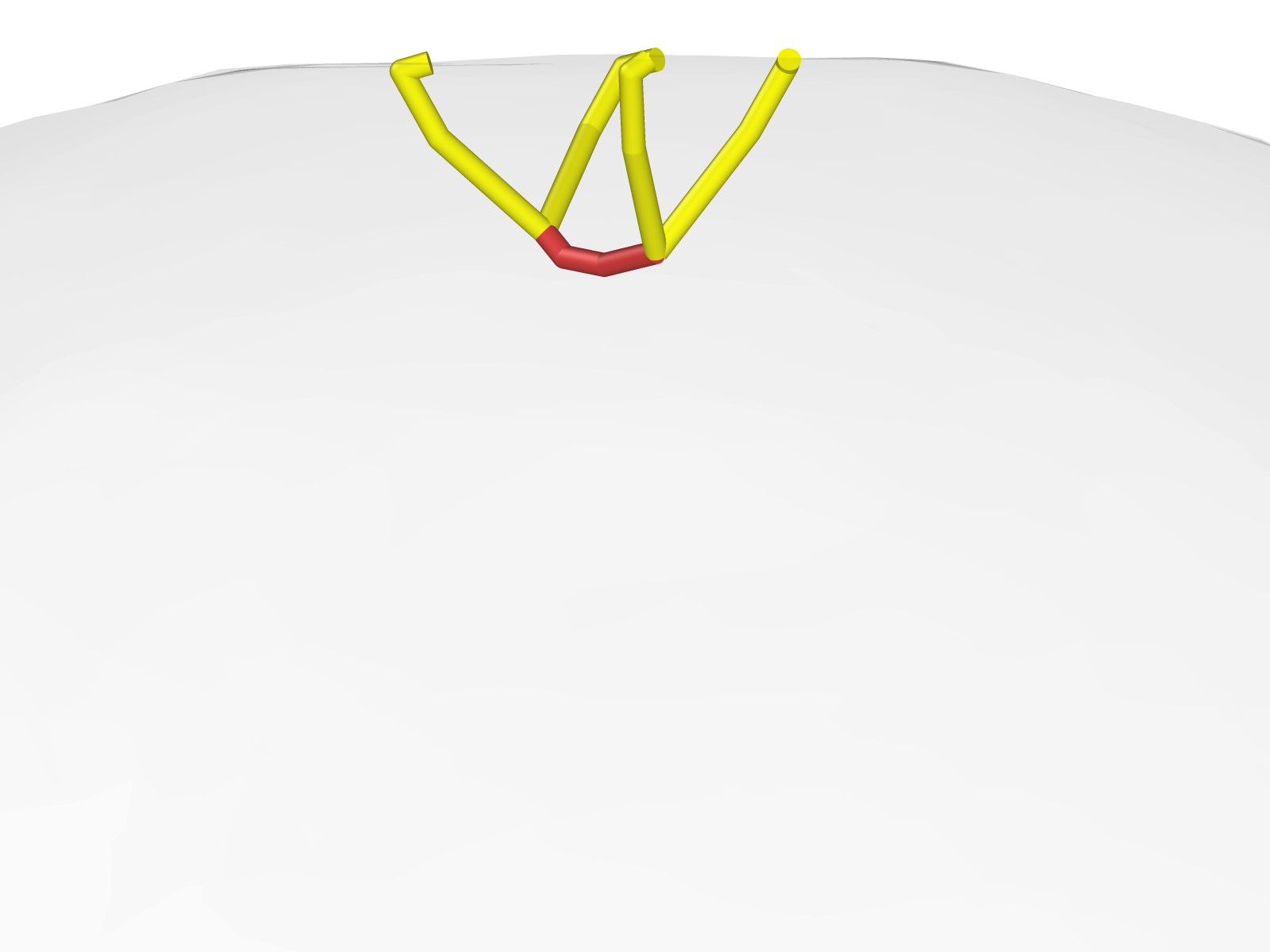}}
			\subcaption*{$\epsilon=2.5\%$}
		\end{subfigure}
		\begin{subfigure}[h]{0.24\textwidth}
			\centerline{\includegraphics[width=1\textwidth]{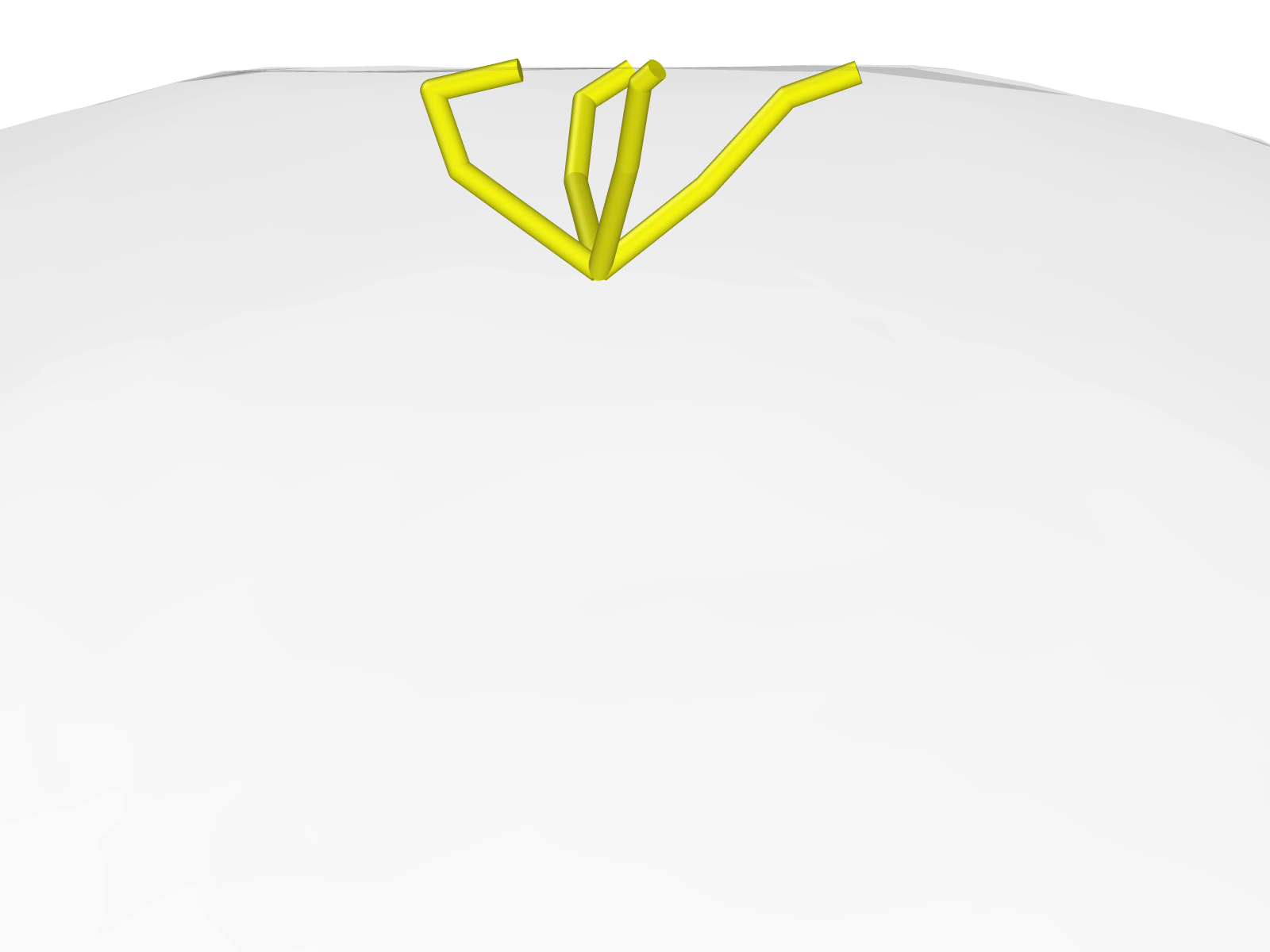}}
			\subcaption*{$\epsilon=3.3\%$}
		\end{subfigure}
		\begin{subfigure}[h]{0.24\textwidth}
			\centerline{\includegraphics[width=1\textwidth]{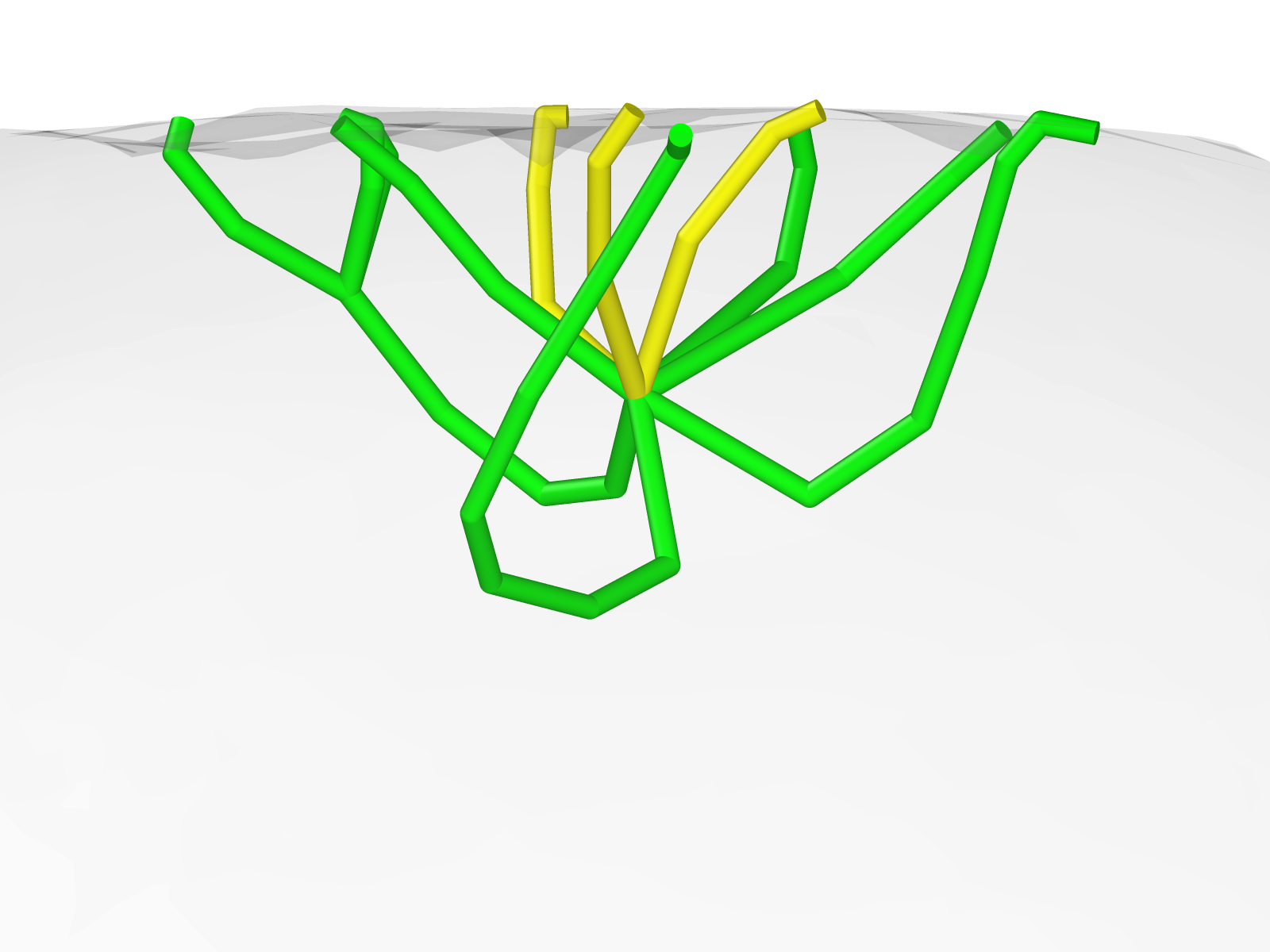}}
			\subcaption*{$\epsilon=5.9\%$}
		\end{subfigure}	
		\caption{Ag = 75\%}
	\end{subfigure}
	\begin{subfigure}[h]{1\textwidth}
		\centering
		\begin{subfigure}[h]{0.24\textwidth}
			\centerline{\includegraphics[width=1\textwidth]{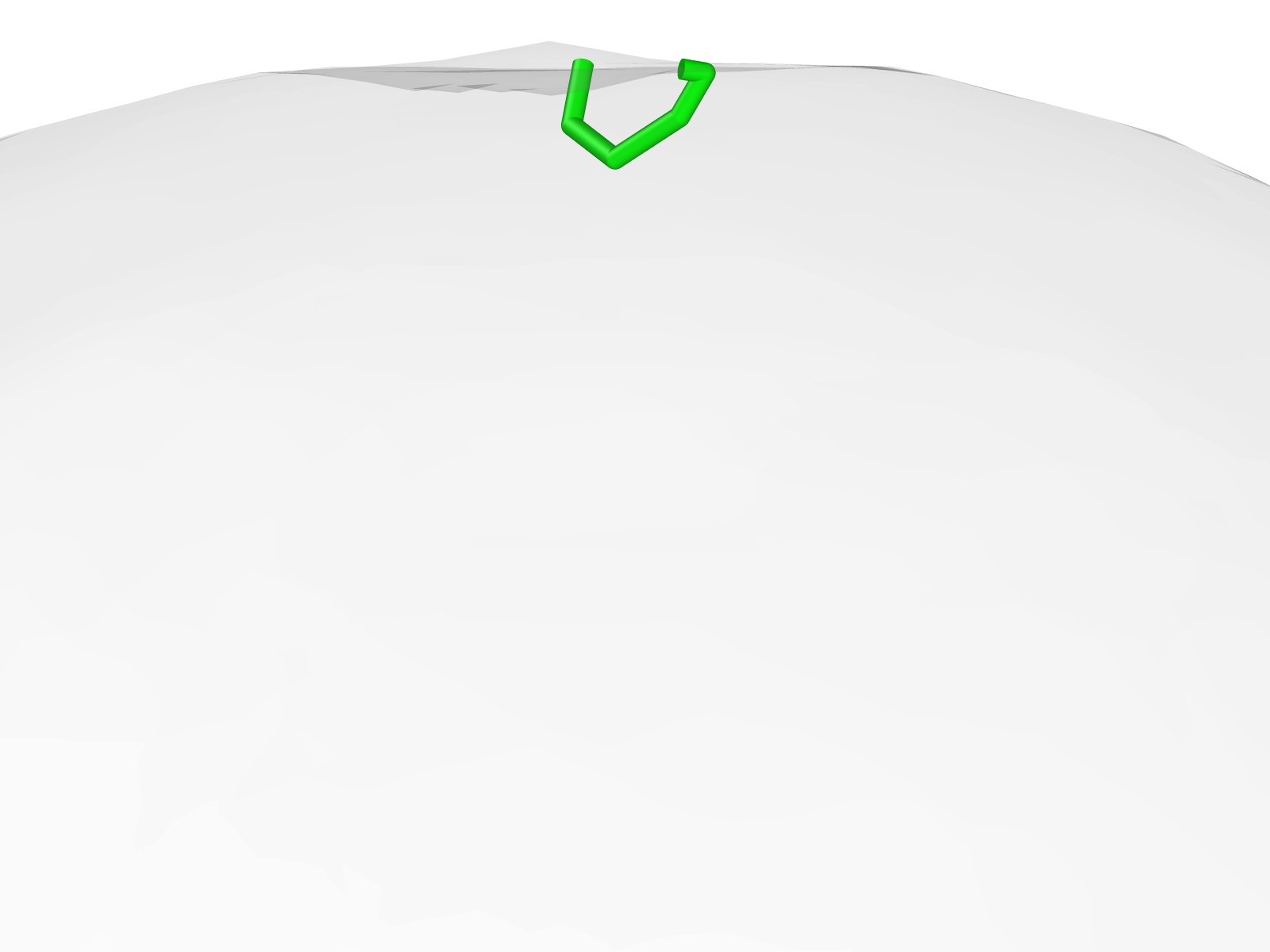}}
			\subcaption*{$\epsilon=1.75\%$}
		\end{subfigure}
		\begin{subfigure}[h]{0.24\textwidth}
			\centerline{\includegraphics[width=1\textwidth]{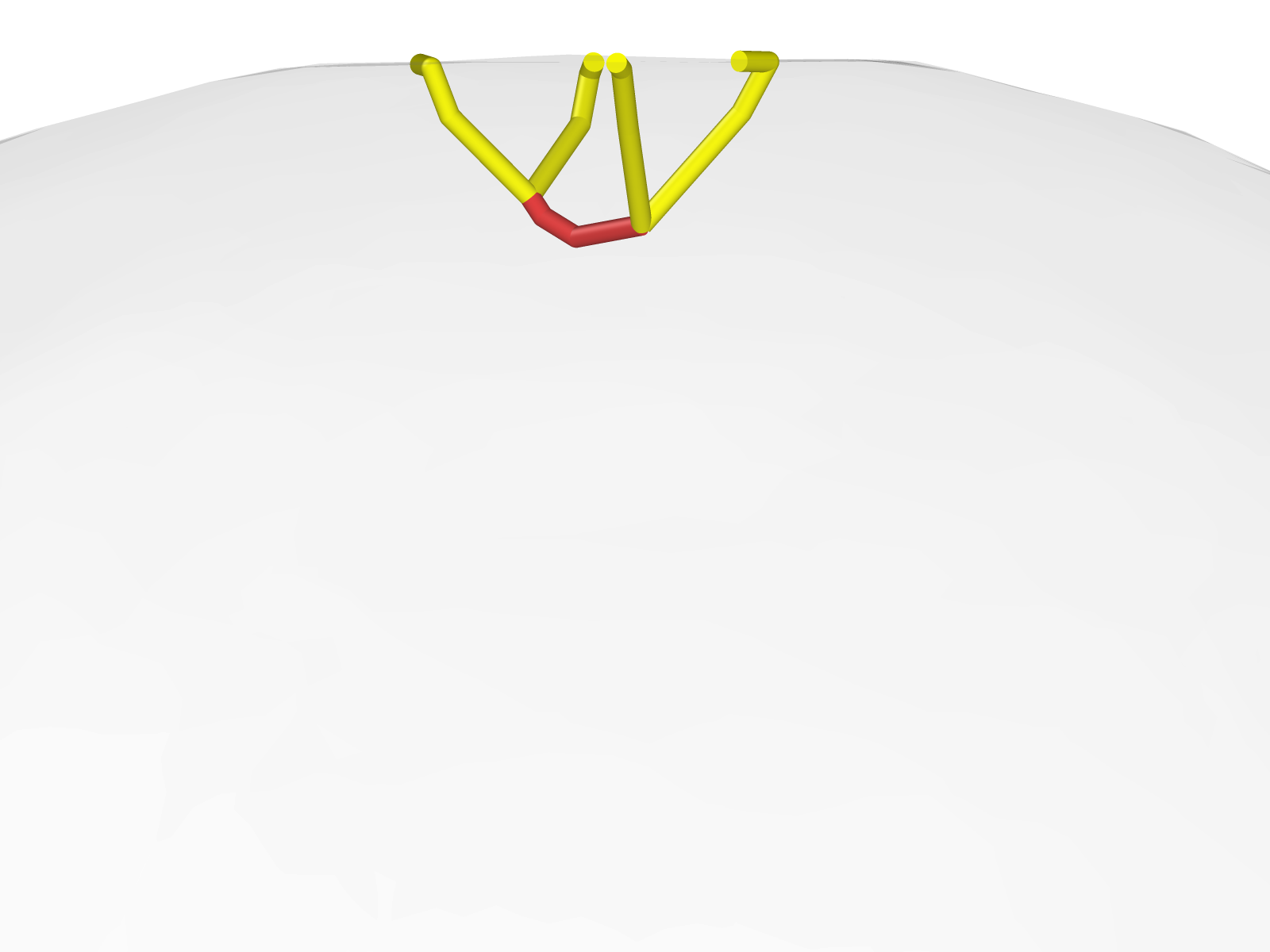}}
			\subcaption*{$\epsilon=2.6\%$}
		\end{subfigure}
		\begin{subfigure}[h]{0.24\textwidth}
			\centerline{\includegraphics[width=1\textwidth]{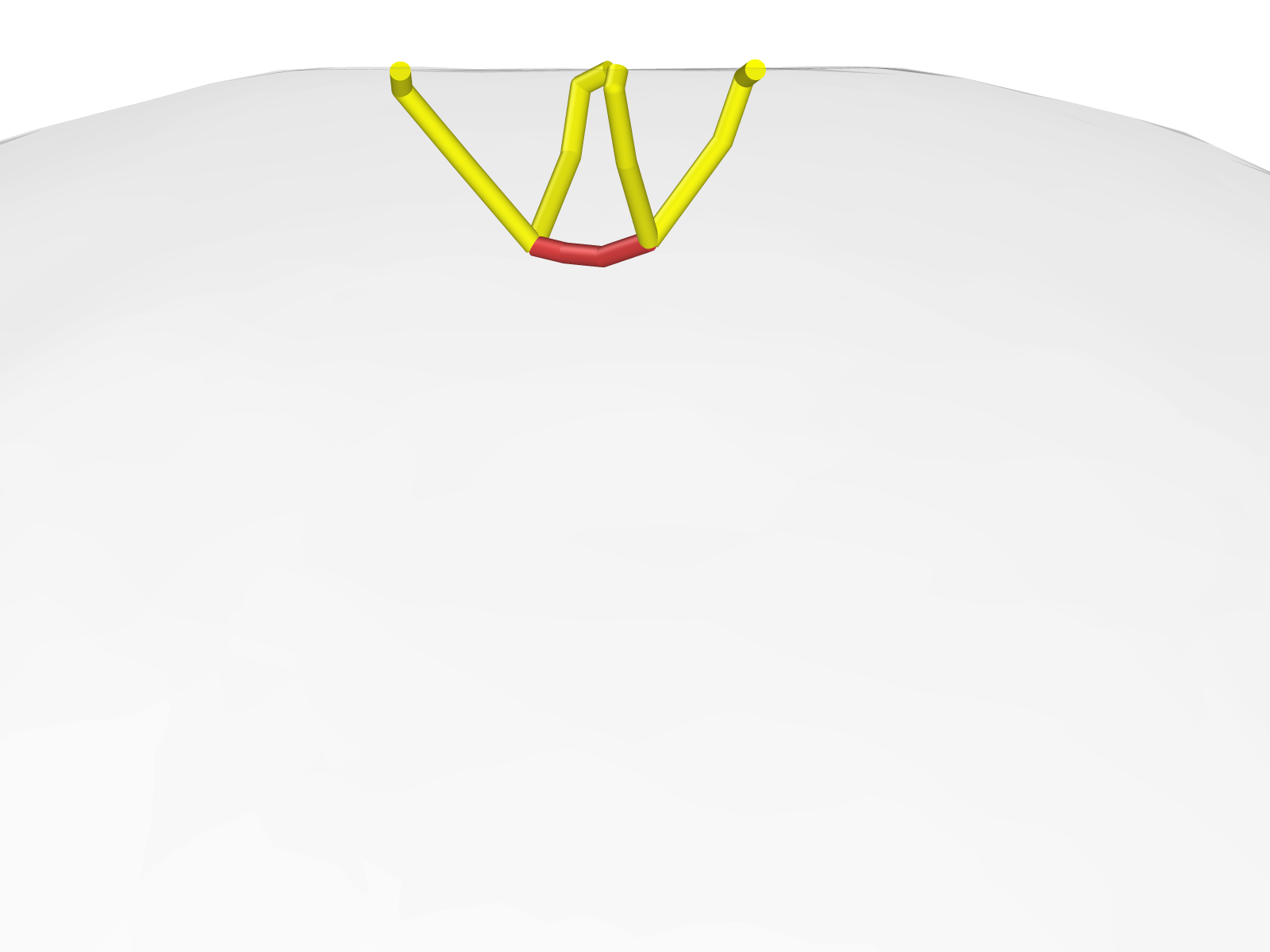}}
			\subcaption*{$\epsilon=3.4\%$}
		\end{subfigure}
		\begin{subfigure}[h]{0.24\textwidth}
			\centerline{\includegraphics[width=1\textwidth]{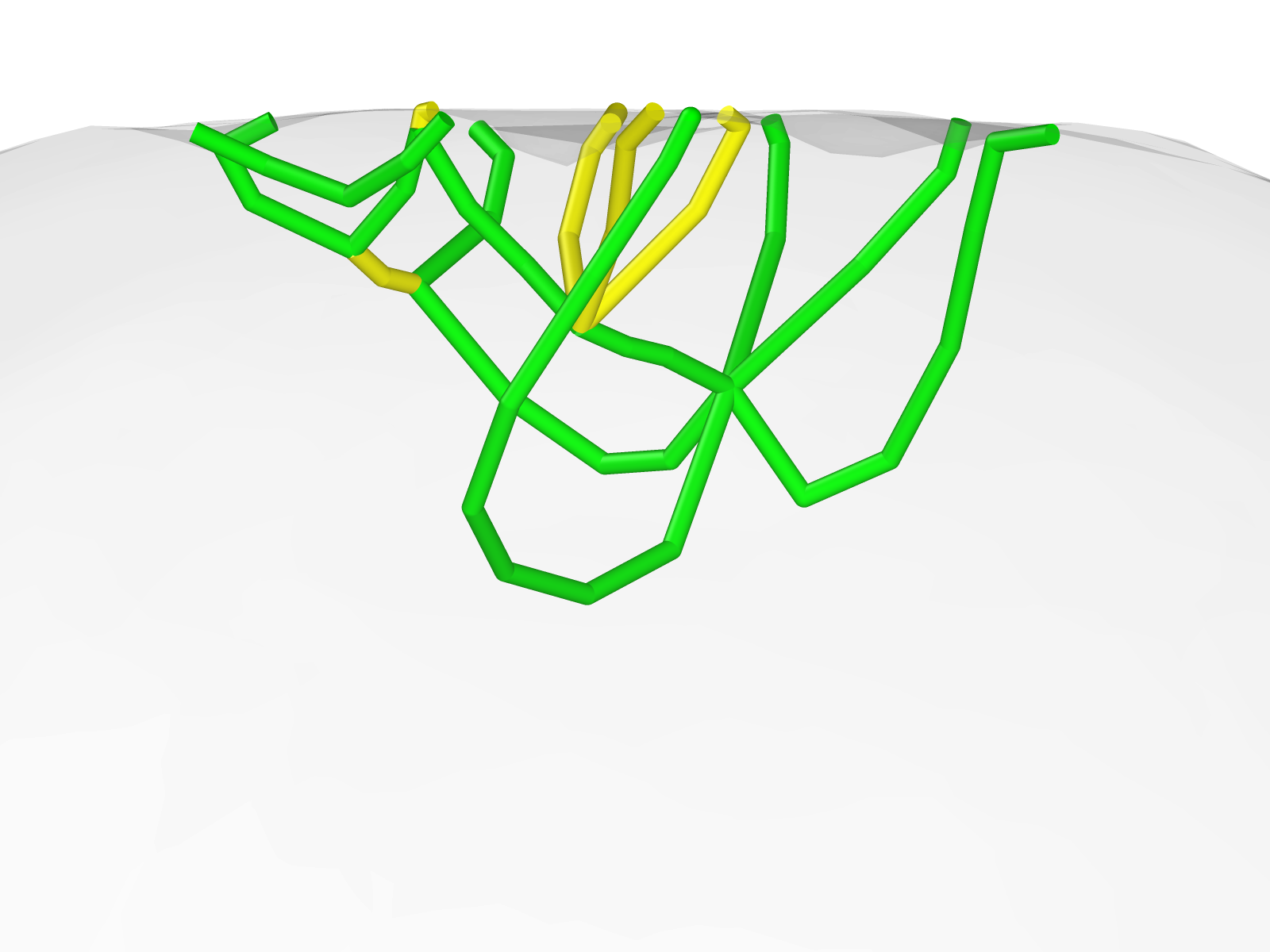}}
			\subcaption*{$\epsilon=6.1\%$}
		\end{subfigure}	
		\caption{Ag = 90\%}
	\end{subfigure}

	\caption{Dislocation structures during uniaxial compression of FGM nanosphere for different percentages of Ag. (Green line, Shockley partial dislocation with Burgers vectors $\frac{1}{6}<112>$; yellow line, hirth dislocation with Burgers vectors $\frac{1}{3}<100>$; Pink line, stair-rod dislocation with Burgers vectors $\frac{1}{6}<110>$; Red line indicates all other dislocation)}
	\label{fig12}
\end{figure}

\begin{figure}[htbp]
	\centering
	\begin{subfigure}[h]{1\textwidth}
		\centering
		\begin{subfigure}[h]{0.24\textwidth}
			\centerline{\includegraphics[width=1\textwidth]{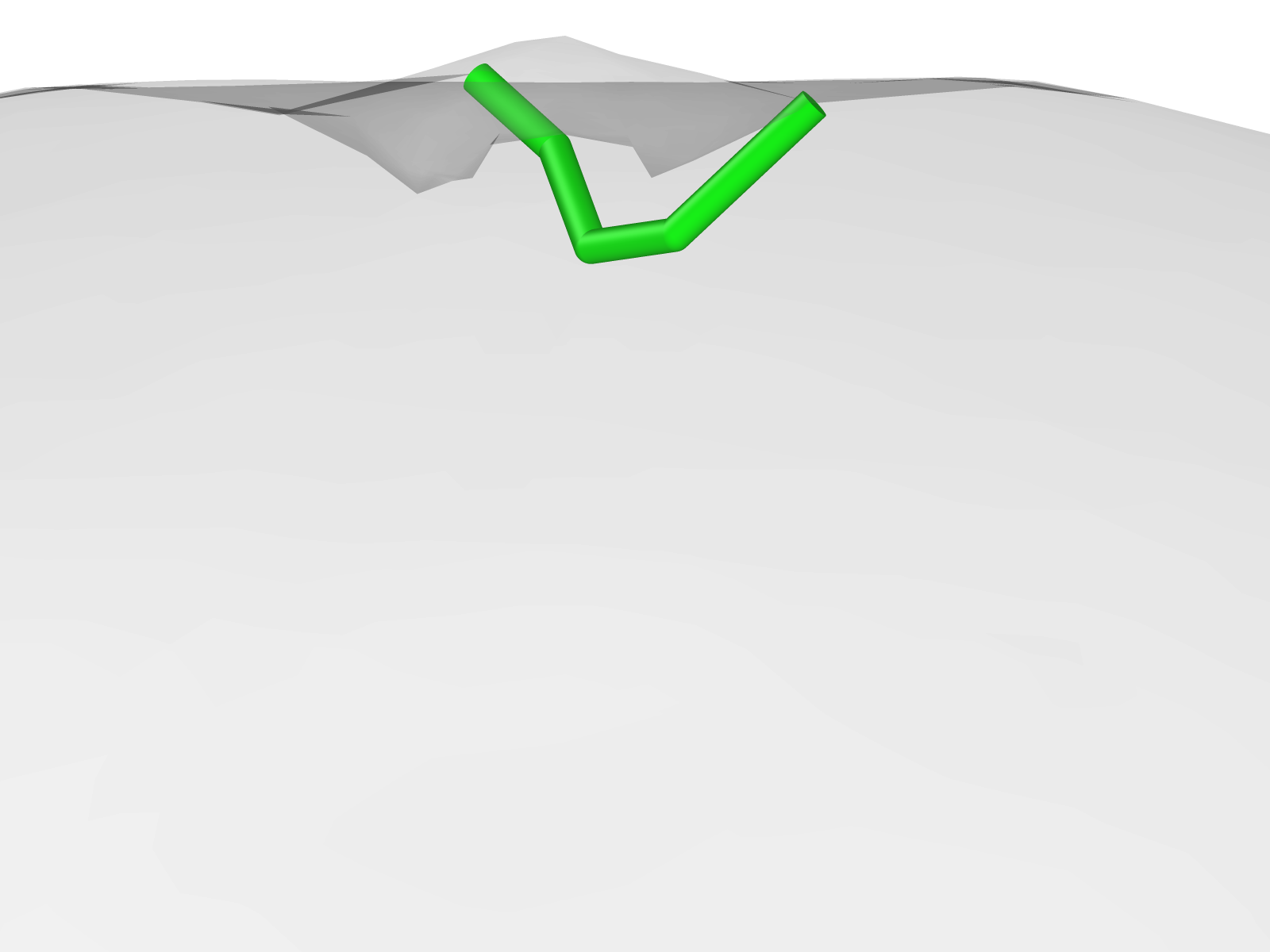}}
			\subcaption*{$\epsilon=1.5\%$}
		\end{subfigure}
		\begin{subfigure}[h]{0.24\textwidth}
			\centerline{\includegraphics[width=1\textwidth]{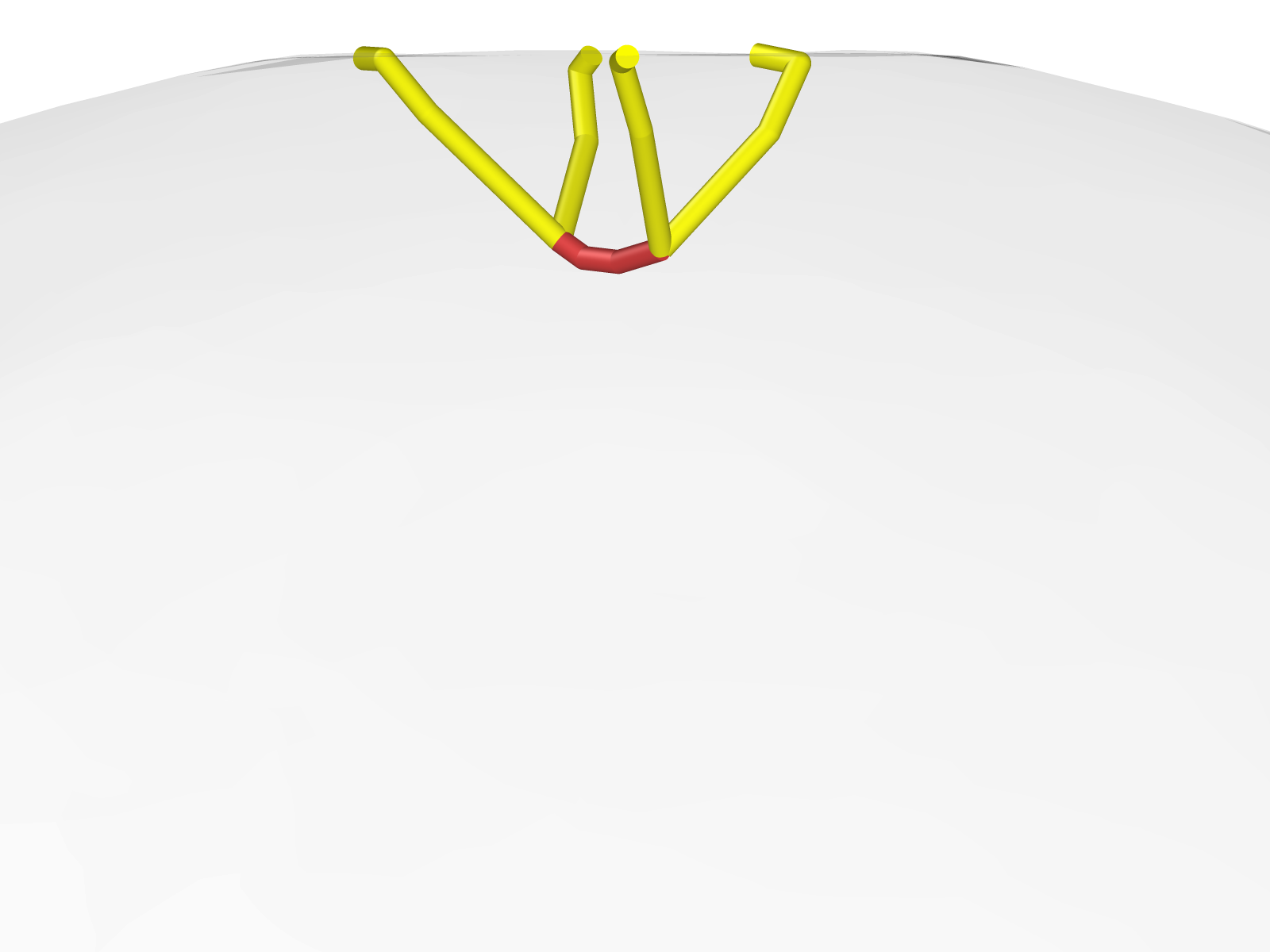}}
			\subcaption*{$\epsilon=2.6\%$}
		\end{subfigure}
		\begin{subfigure}[h]{0.24\textwidth}
			\centerline{\includegraphics[width=1\textwidth]{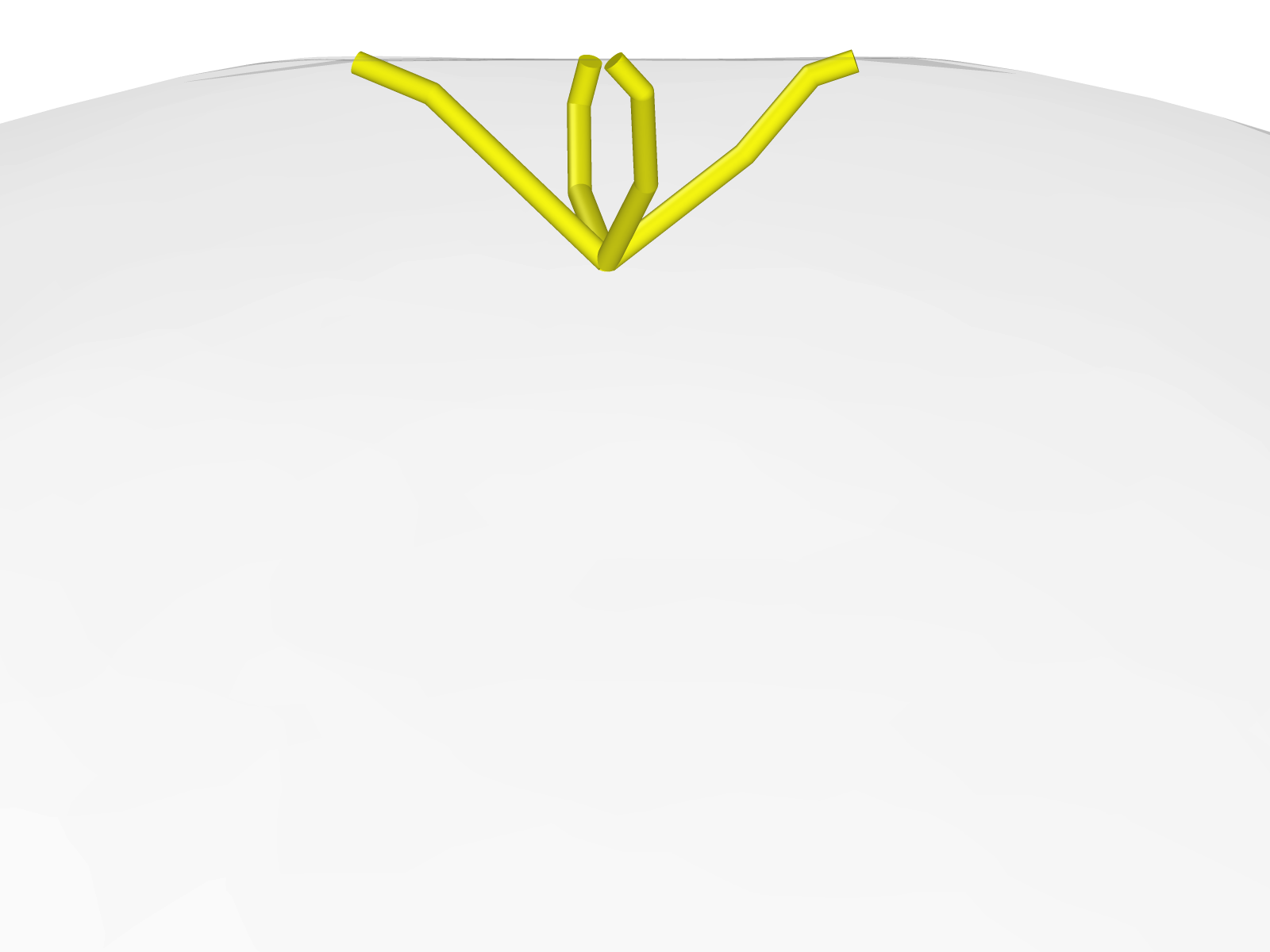}}
			\subcaption*{$\epsilon=3.1\%$}
		\end{subfigure}
		\begin{subfigure}[h]{0.24\textwidth}
			\centerline{\includegraphics[width=1\textwidth]{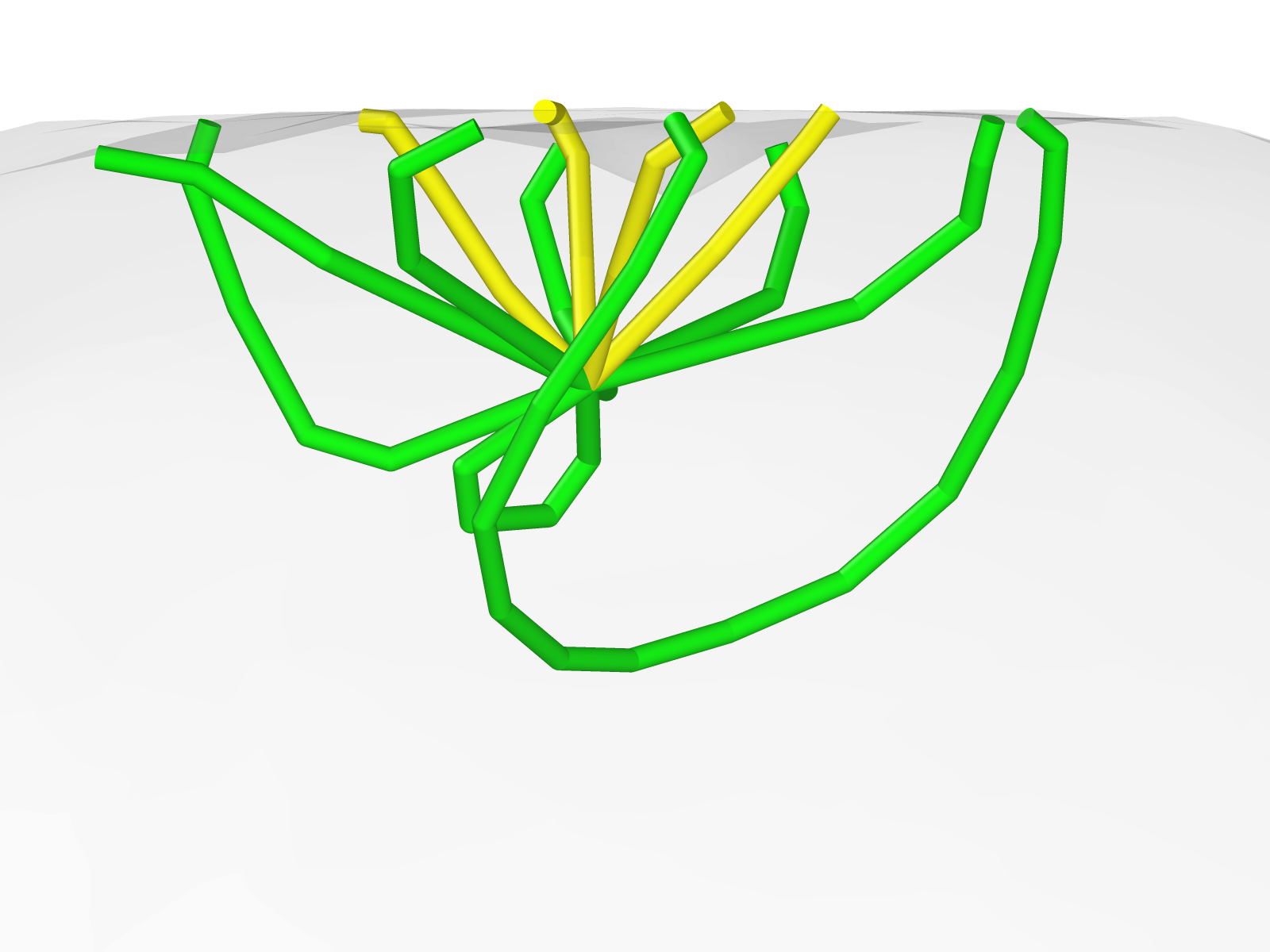}}
			\subcaption*{$\epsilon=6\%$}
		\end{subfigure}	
		\caption{Ag = 10\%}
	\end{subfigure}
	\begin{subfigure}[h]{1\textwidth}
		\centering
		\begin{subfigure}[h]{0.24\textwidth}
			\centerline{\includegraphics[width=1\textwidth]{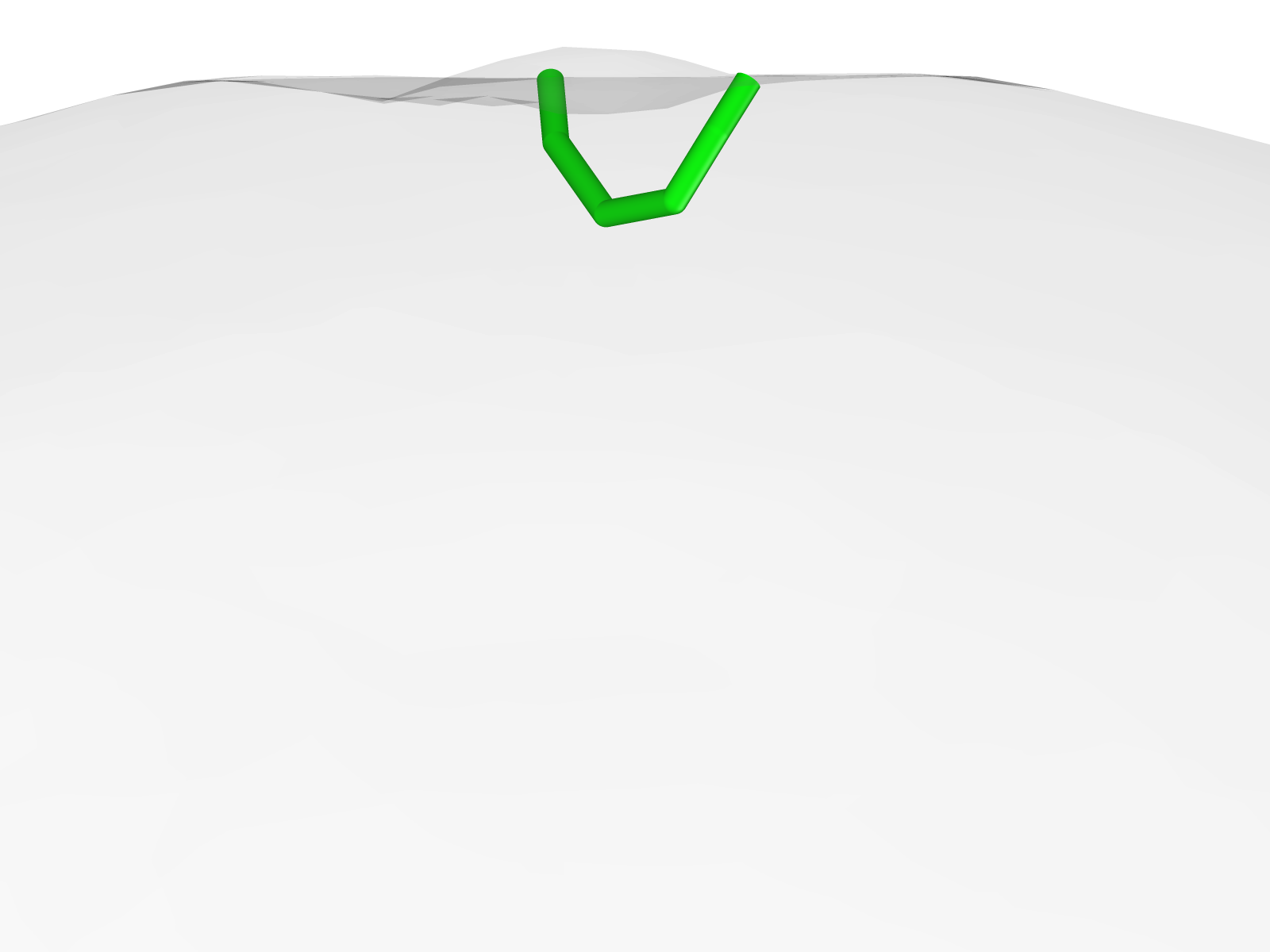}}
			\subcaption*{$\epsilon=1.5\%$}
		\end{subfigure}
		\begin{subfigure}[h]{0.24\textwidth}
			\centerline{\includegraphics[width=1\textwidth]{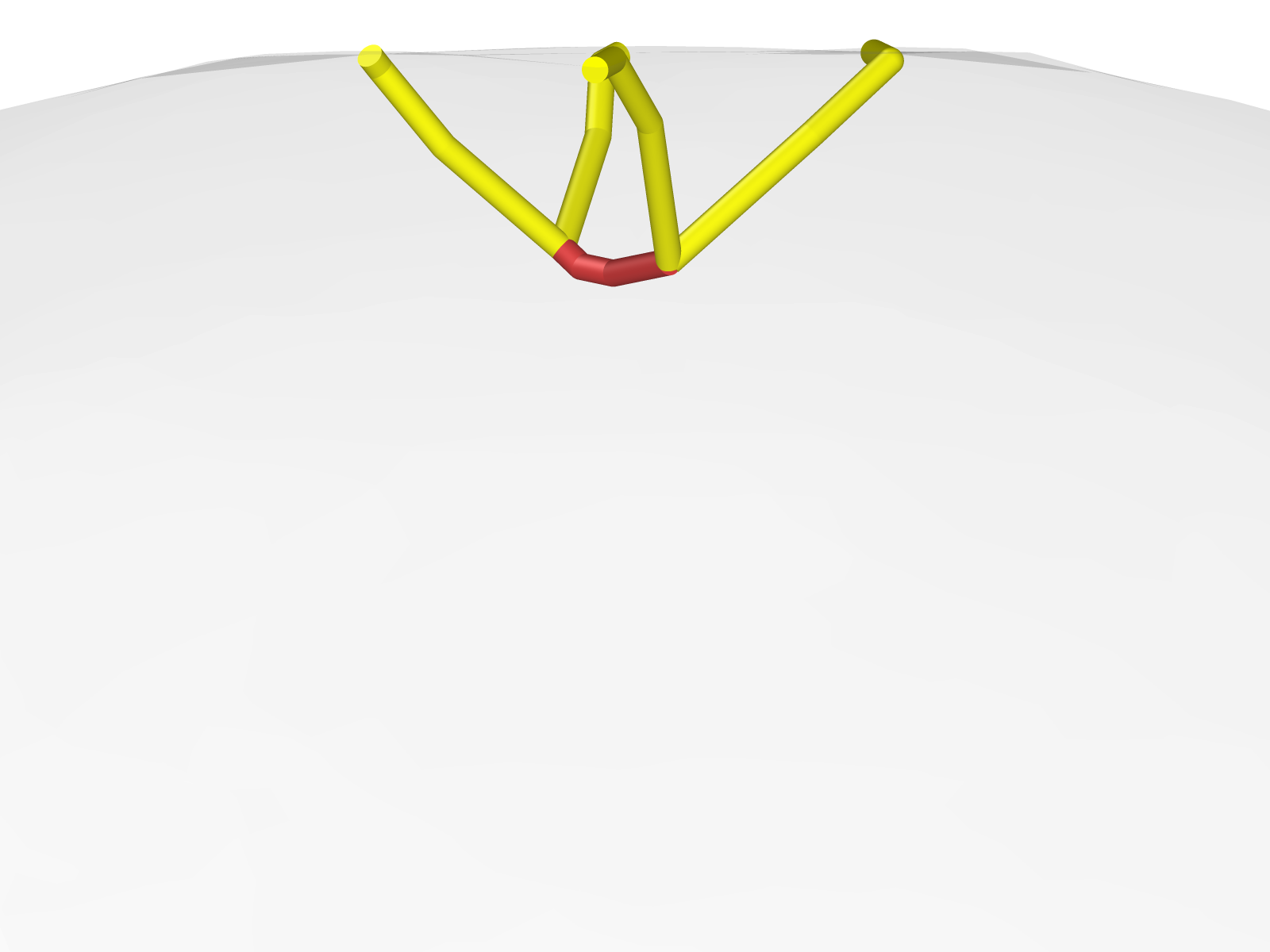}}
			\subcaption*{$\epsilon=2.6\%$}
		\end{subfigure}
		\begin{subfigure}[h]{0.24\textwidth}
			\centerline{\includegraphics[width=1\textwidth]{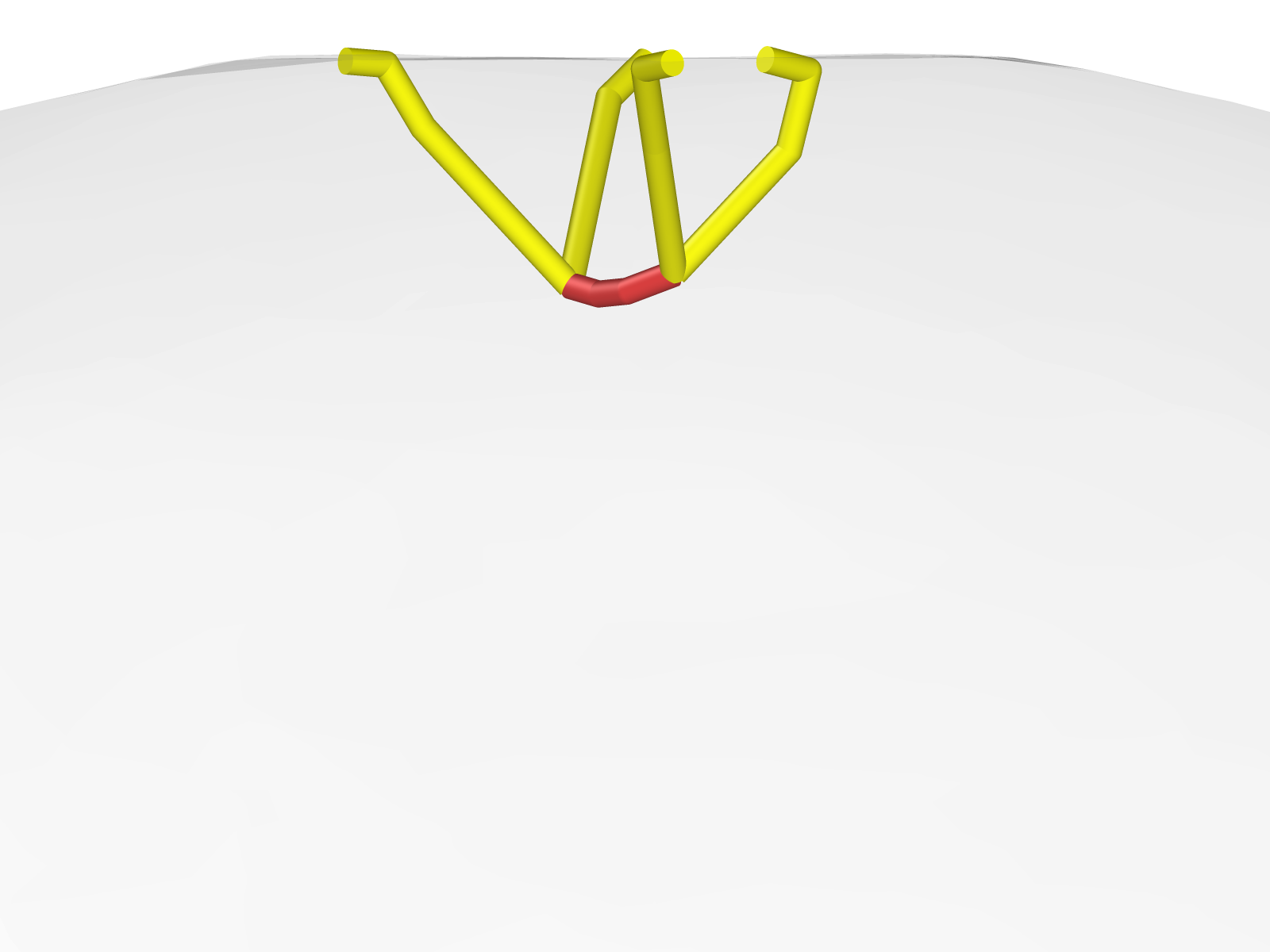}}
			\subcaption*{$\epsilon=3.1\%$}
		\end{subfigure}
		\begin{subfigure}[h]{0.24\textwidth}
			\centerline{\includegraphics[width=1\textwidth]{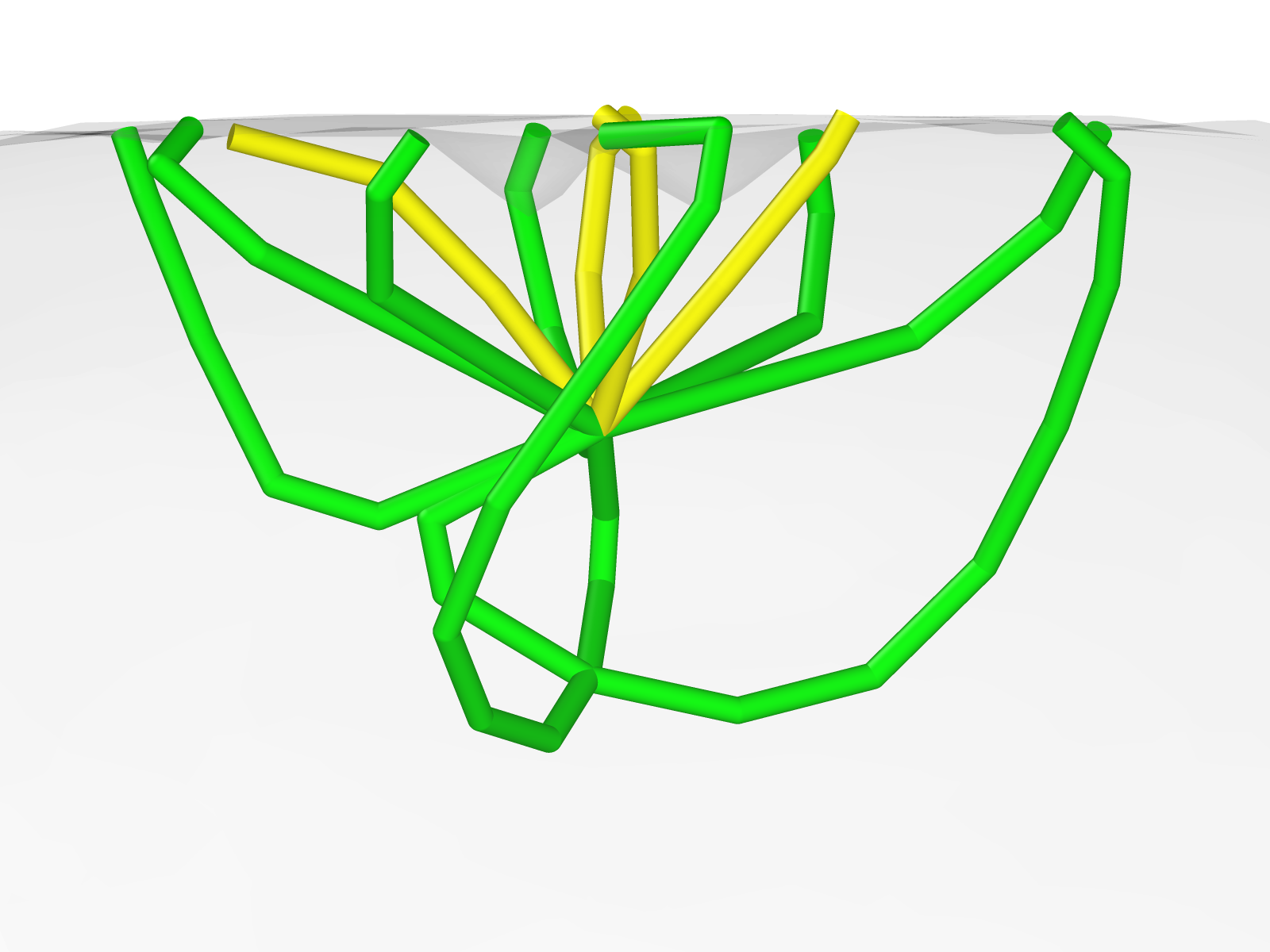}}
			\subcaption*{$\epsilon=6\%$}
		\end{subfigure}	
		\caption{Ag = 25\%}
	\end{subfigure}
	\begin{subfigure}[h]{1\textwidth}
		\centering
		\begin{subfigure}[h]{0.24\textwidth}
			\centerline{\includegraphics[width=1\textwidth]{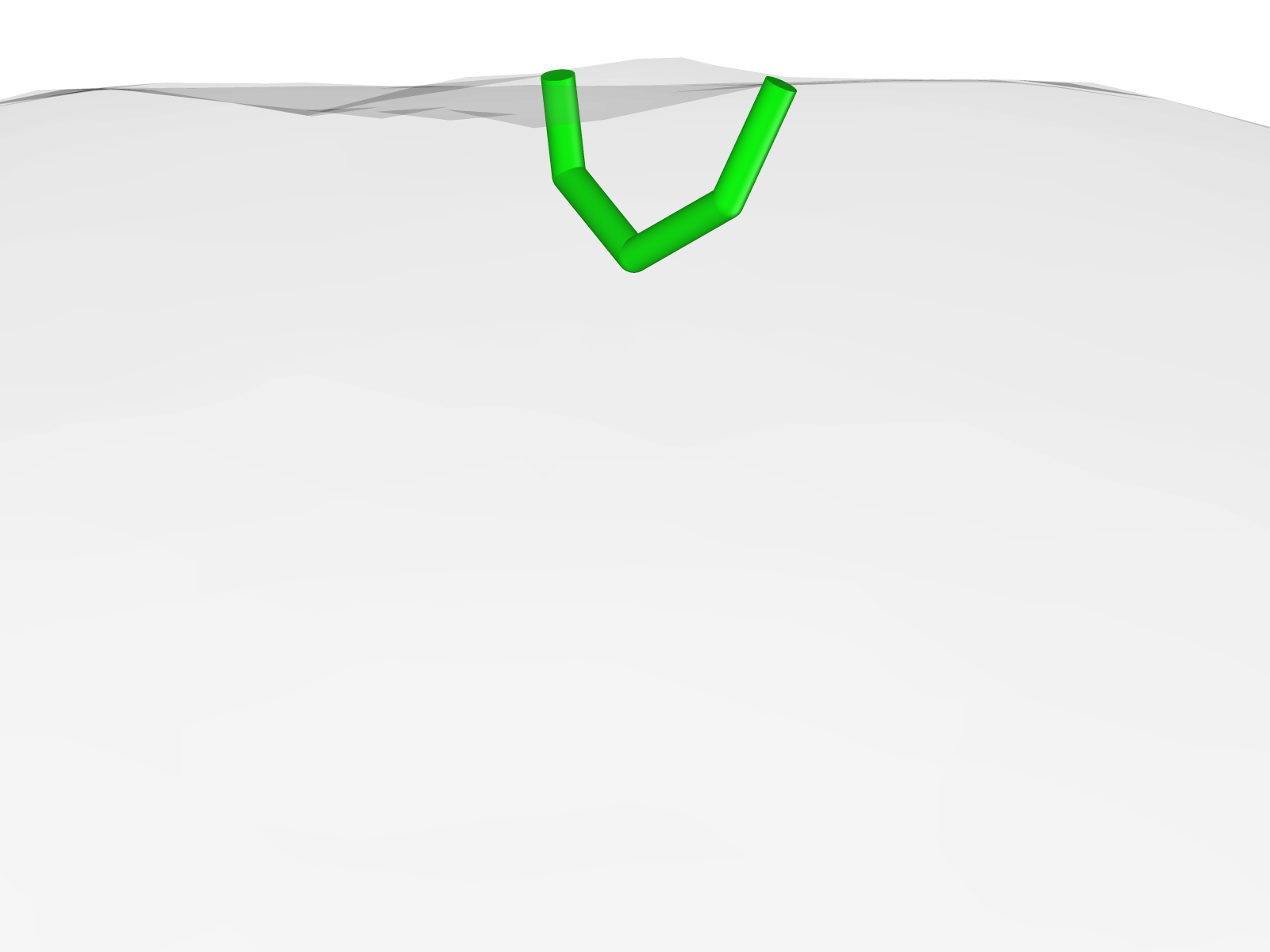}}
			\subcaption*{$\epsilon=1.5\%$}
		\end{subfigure}
		\begin{subfigure}[h]{0.24\textwidth}
			\centerline{\includegraphics[width=1\textwidth]{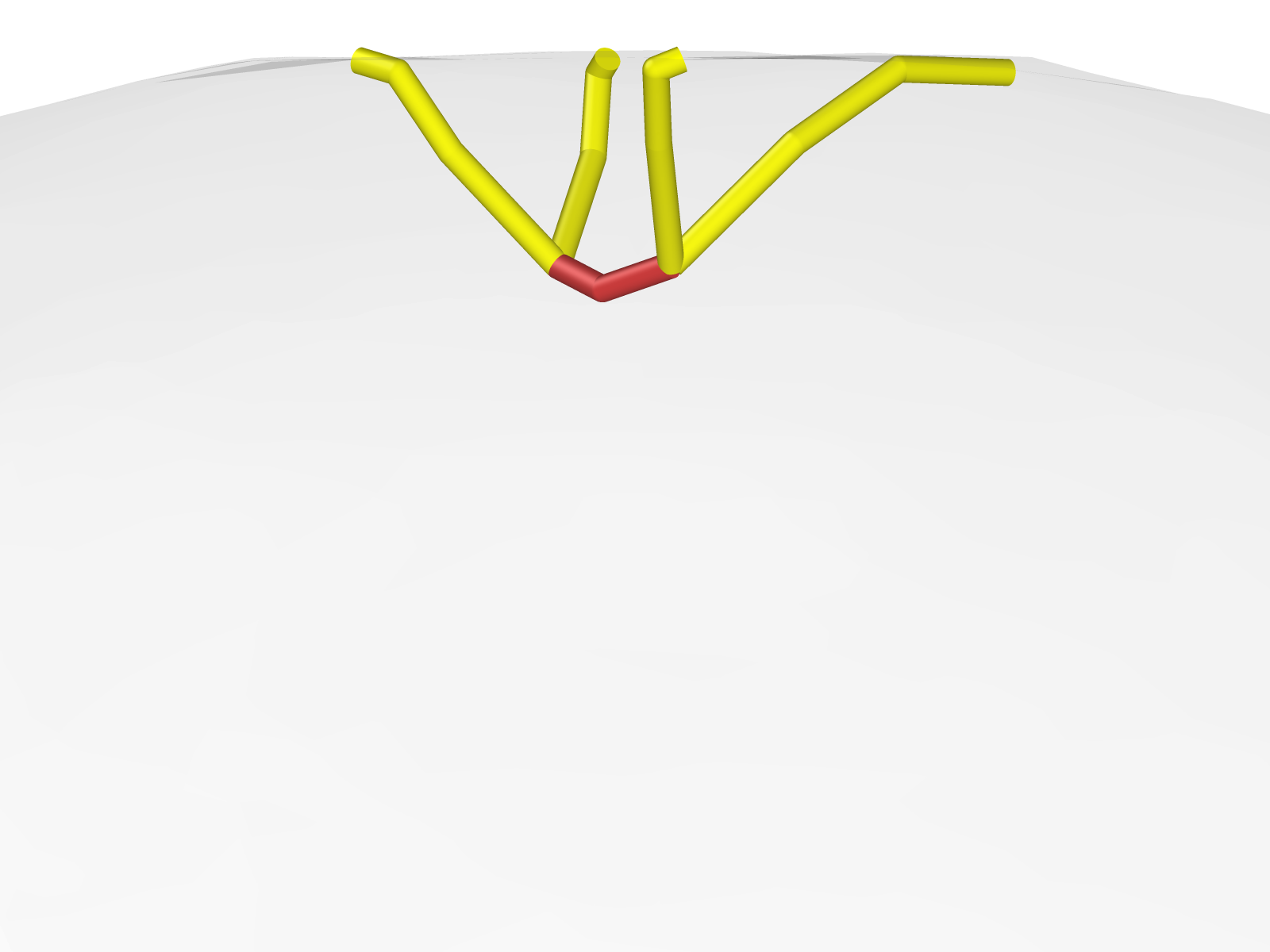}}
			\subcaption*{$\epsilon=2.6\%$}
		\end{subfigure}
		\begin{subfigure}[h]{0.24\textwidth}
			\centerline{\includegraphics[width=1\textwidth]{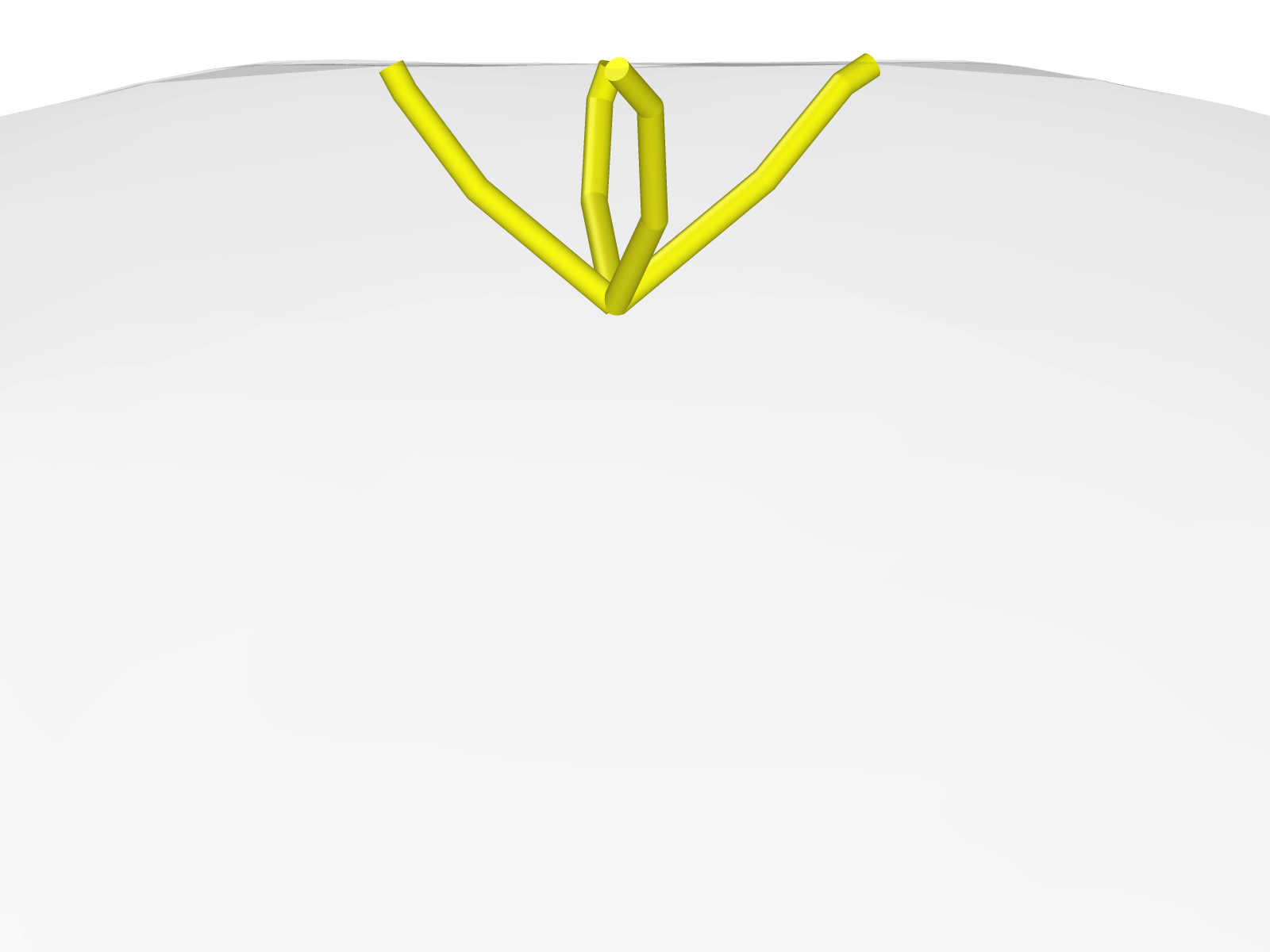}}
			\subcaption*{$\epsilon=3.2\%$}
		\end{subfigure}
		\begin{subfigure}[h]{0.24\textwidth}
			\centerline{\includegraphics[width=1\textwidth]{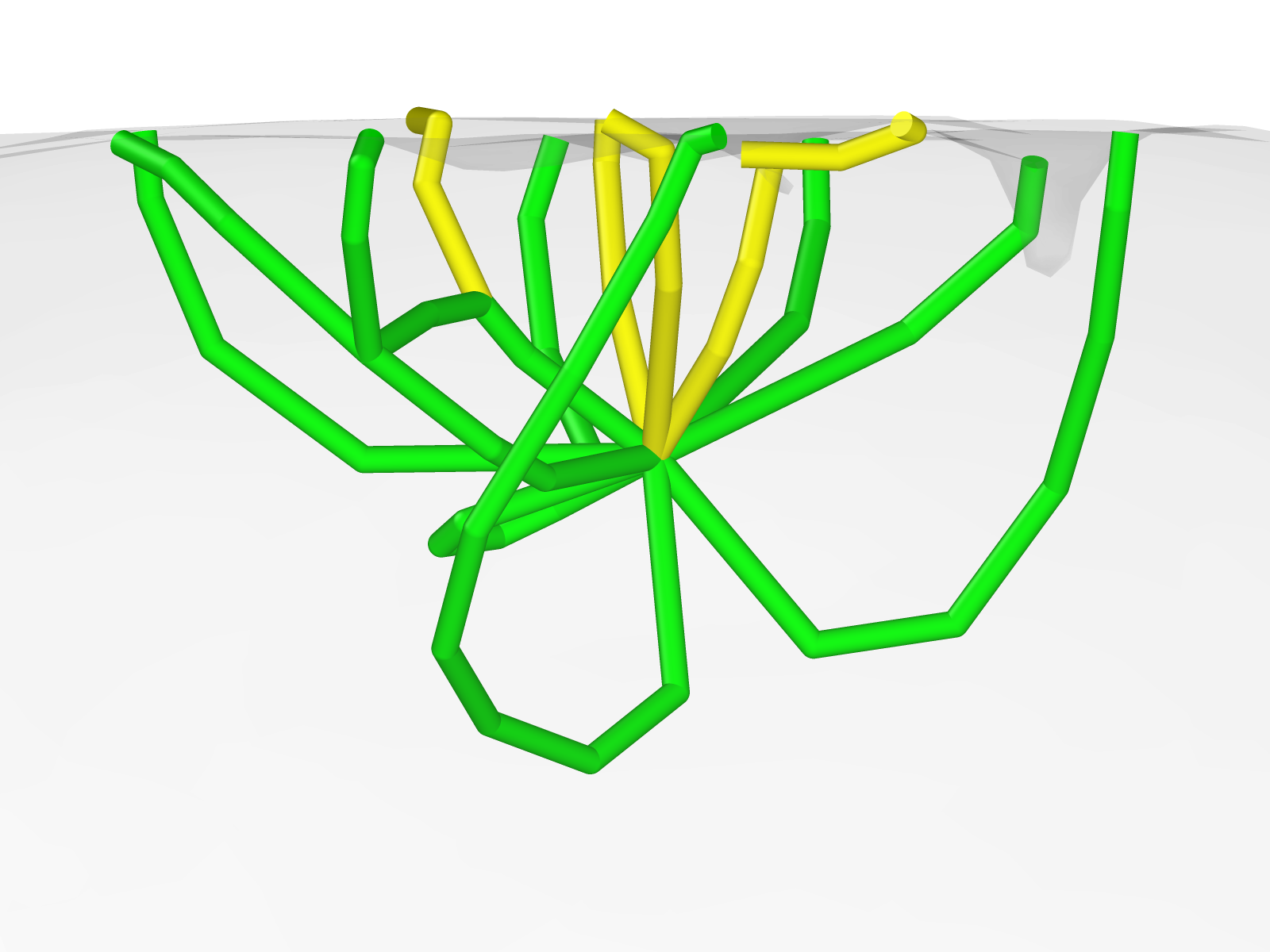}}
			\subcaption*{$\epsilon=6\%$}
		\end{subfigure}	
		\caption{Ag = 50\%}
	\end{subfigure}
	\begin{subfigure}[h]{1\textwidth}
		\centering
		\begin{subfigure}[h]{0.24\textwidth}
			\centerline{\includegraphics[width=1\textwidth]{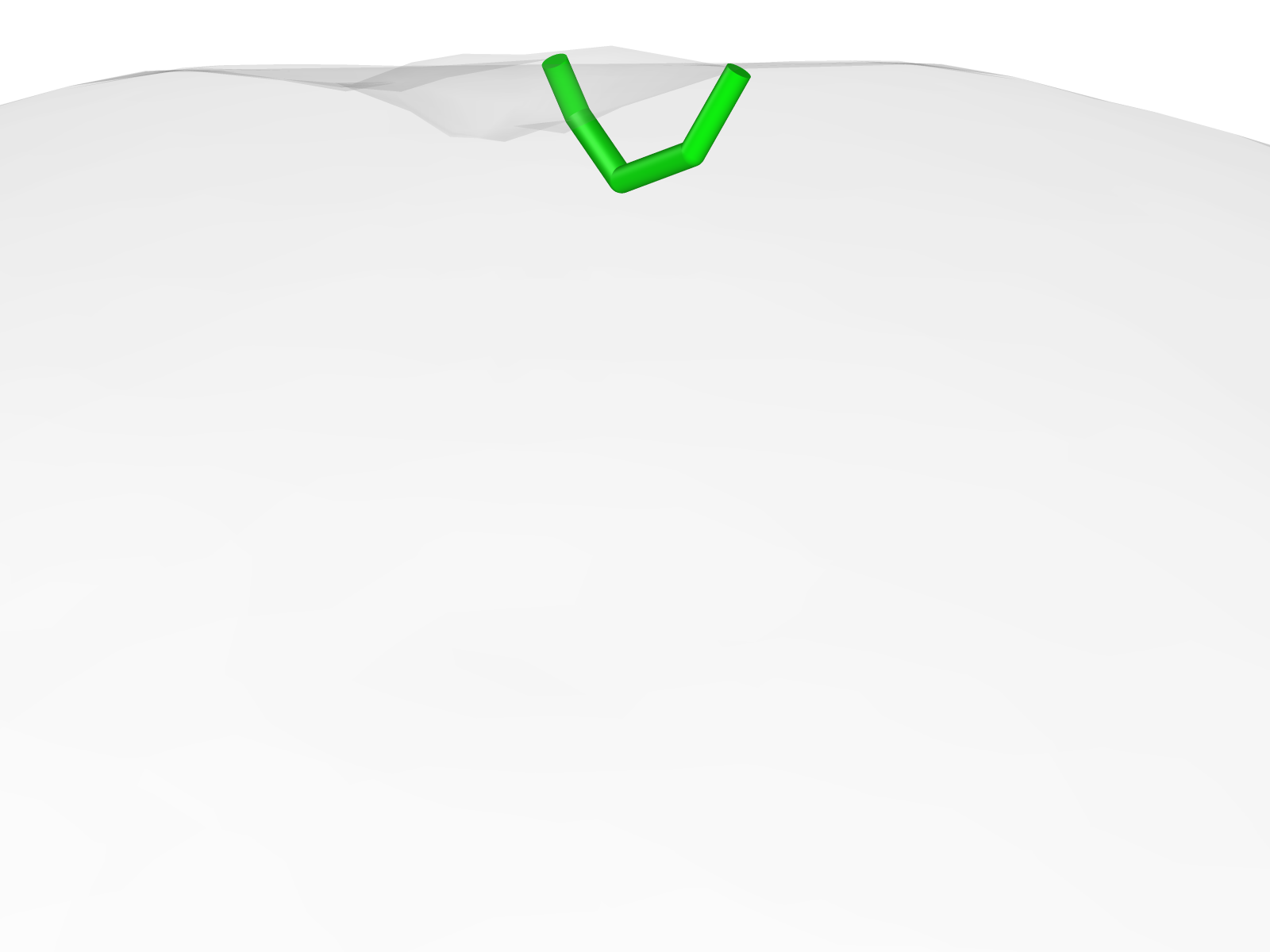}}
			\subcaption*{$\epsilon=1.5\%$}
		\end{subfigure}
		\begin{subfigure}[h]{0.24\textwidth}
			\centerline{\includegraphics[width=1\textwidth]{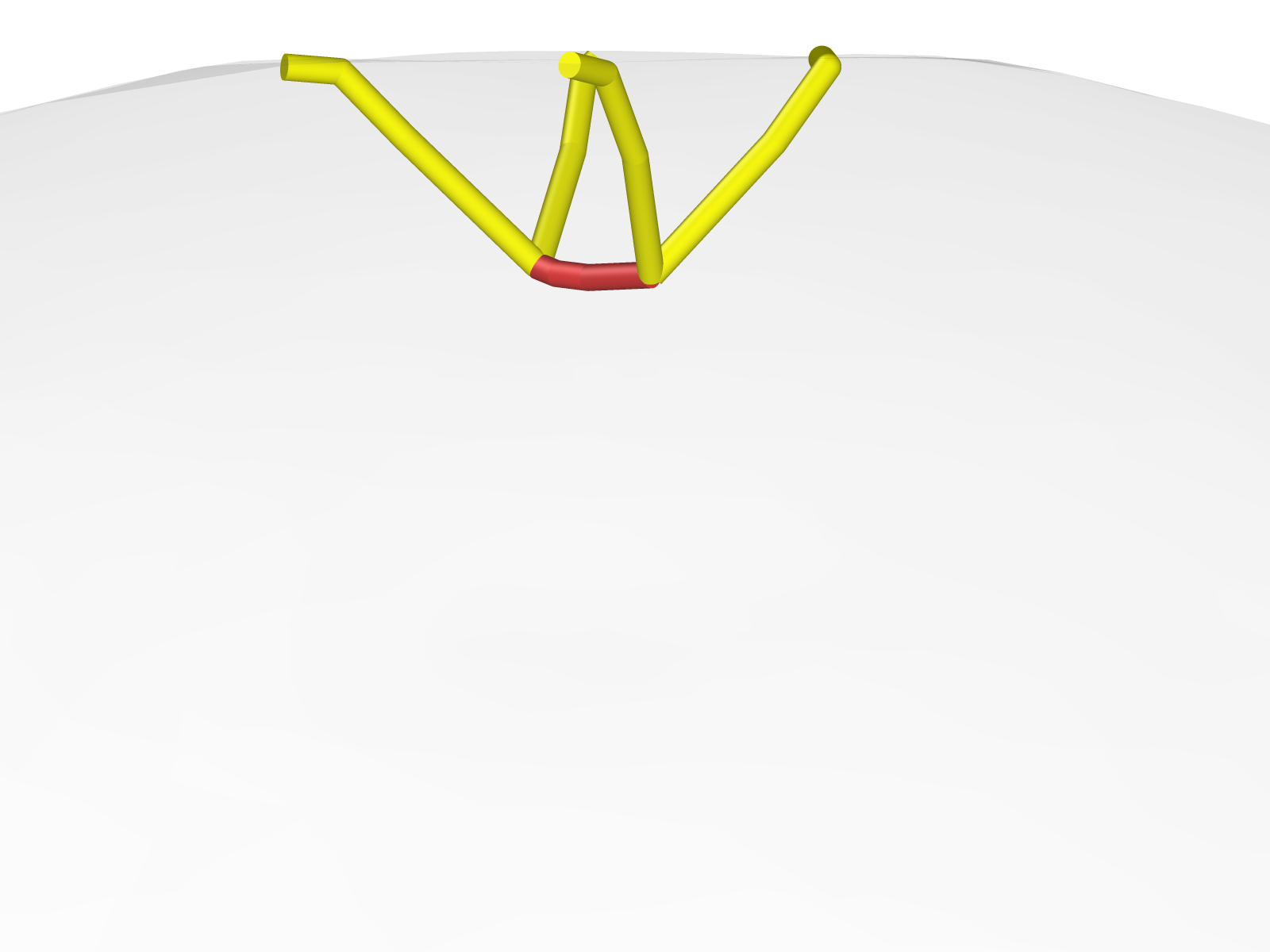}}
			\subcaption*{$\epsilon=2.6\%$}
		\end{subfigure}
		\begin{subfigure}[h]{0.24\textwidth}
			\centerline{\includegraphics[width=1\textwidth]{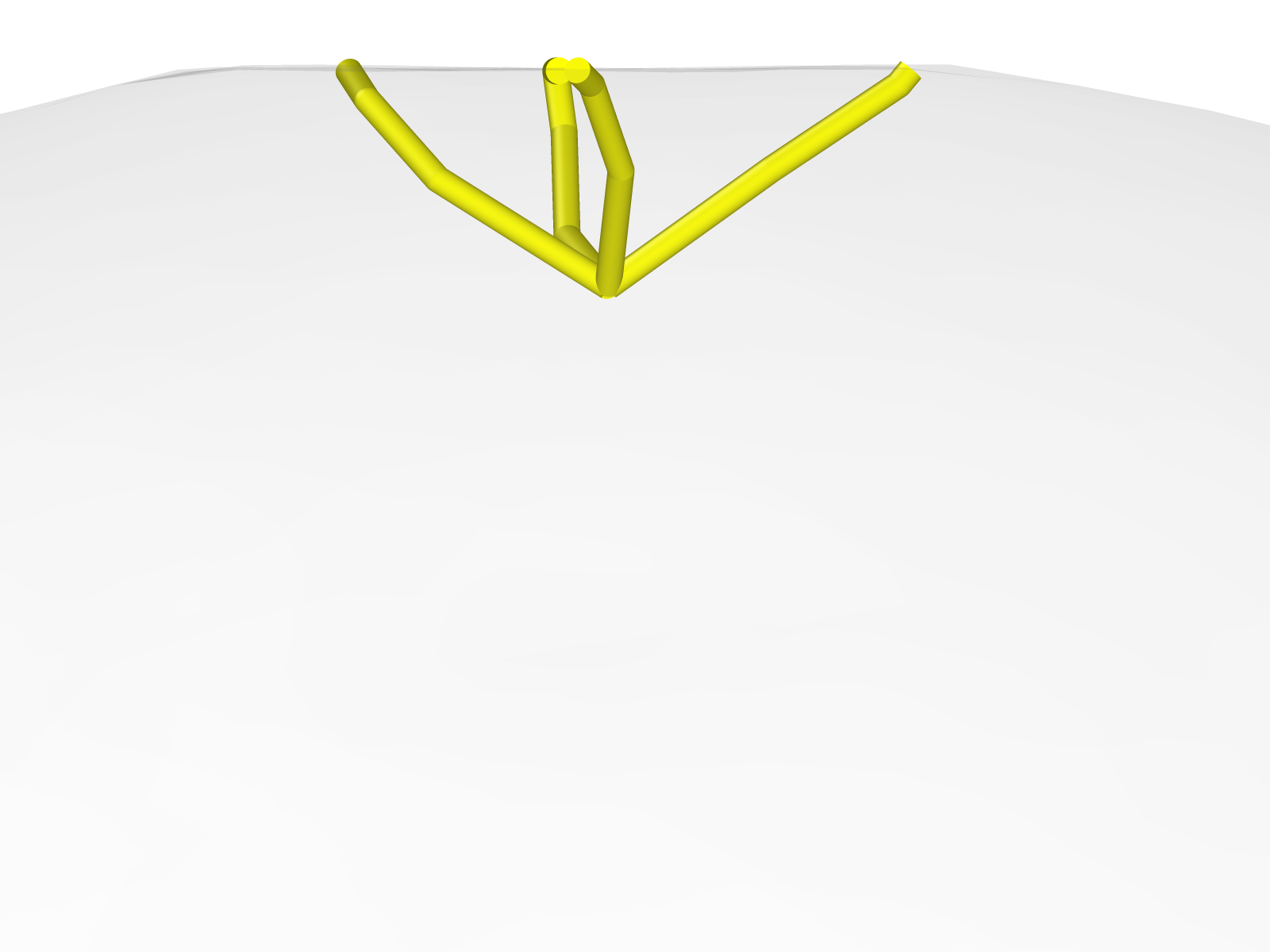}}
			\subcaption*{$\epsilon=3.4\%$}
		\end{subfigure}
		\begin{subfigure}[h]{0.24\textwidth}
			\centerline{\includegraphics[width=1\textwidth]{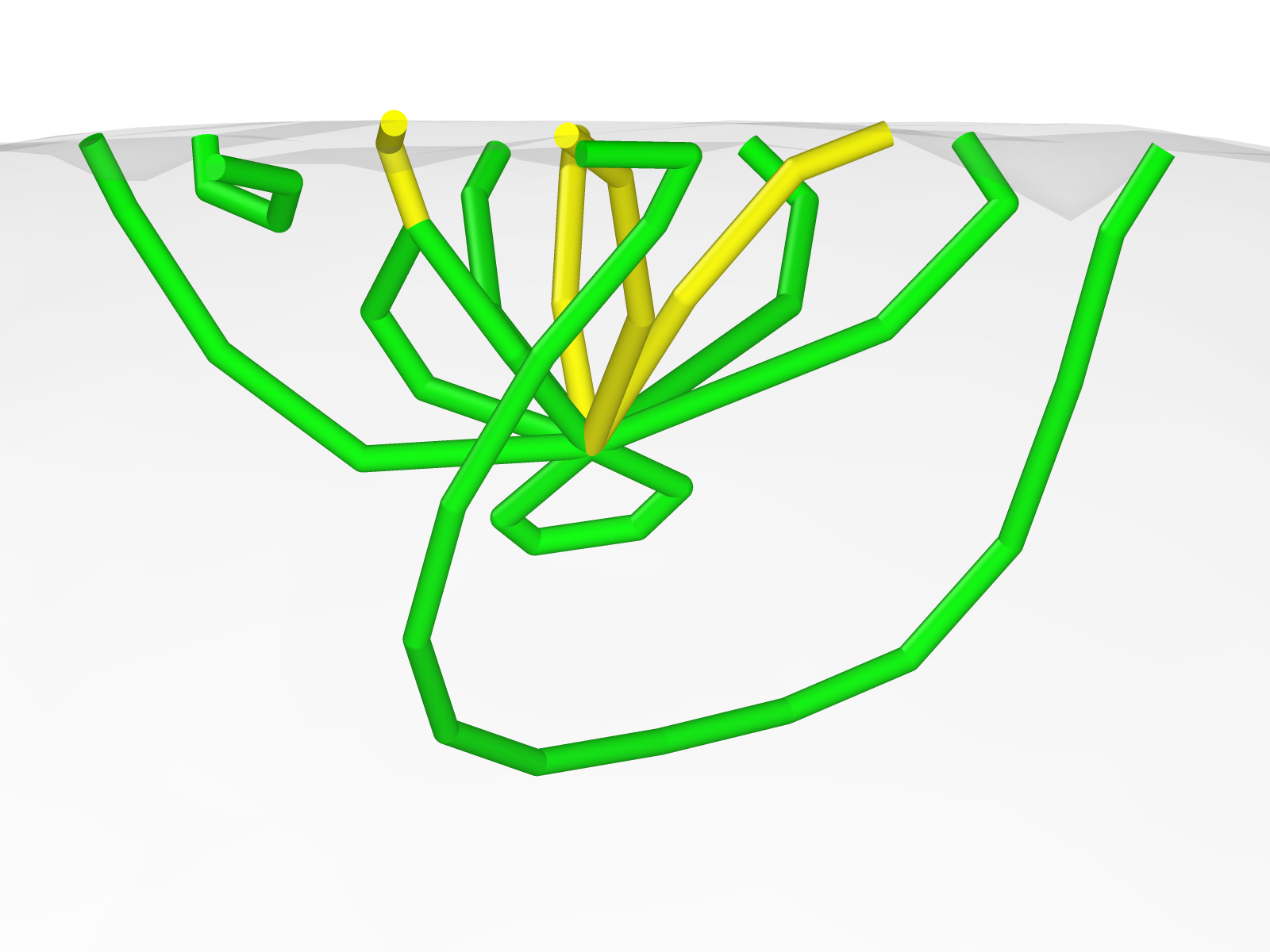}}
			\subcaption*{$\epsilon=6.1\%$}
		\end{subfigure}	
		\caption{Ag = 75\%}
	\end{subfigure}
	\begin{subfigure}[h]{1\textwidth}
		\centering
		\begin{subfigure}[h]{0.24\textwidth}
			\centerline{\includegraphics[width=1\textwidth]{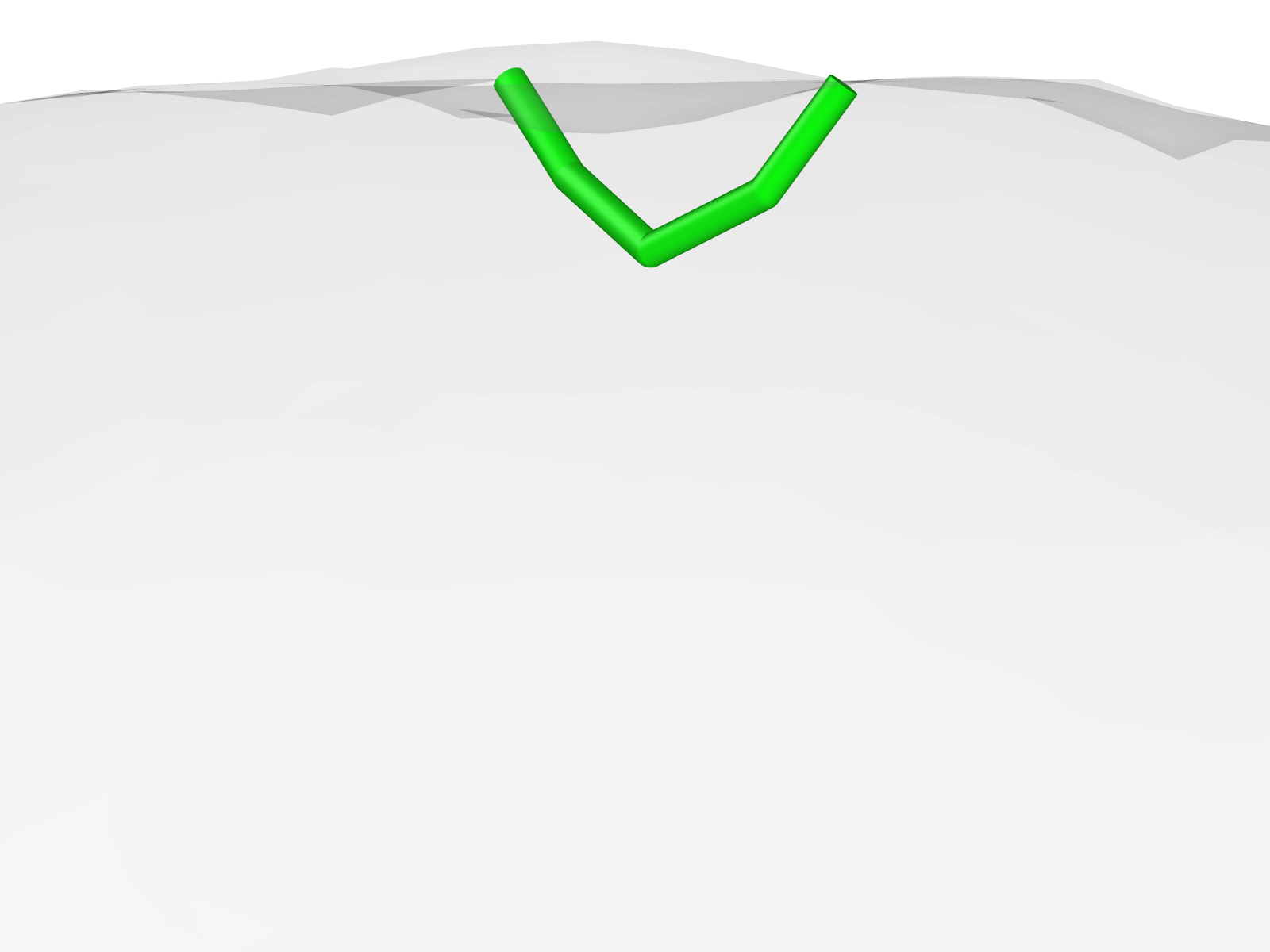}}
			\subcaption*{$\epsilon=1.6\%$}
		\end{subfigure}
		\begin{subfigure}[h]{0.24\textwidth}
			\centerline{\includegraphics[width=1\textwidth]{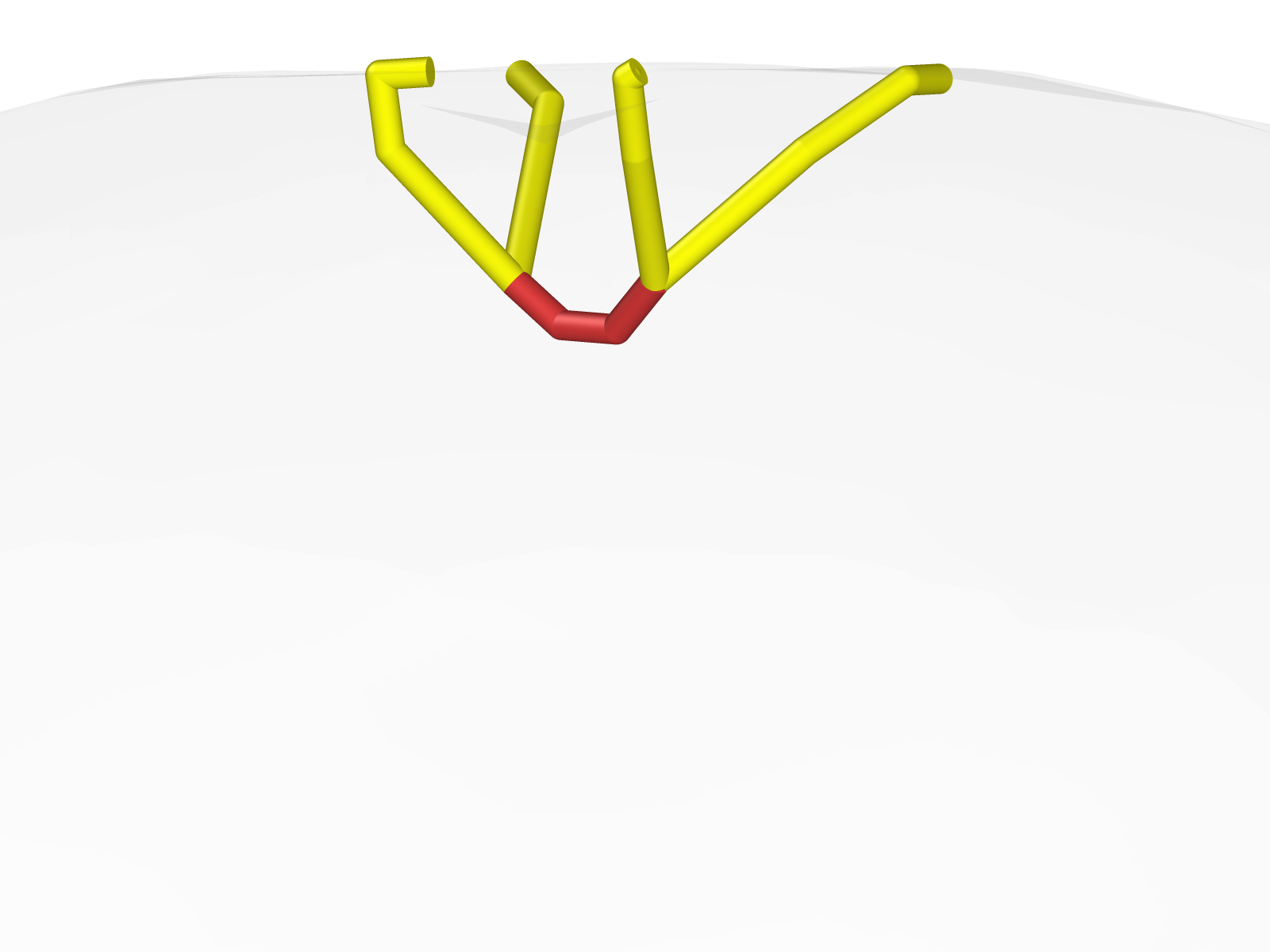}}
			\subcaption*{$\epsilon=2.7\%$}
		\end{subfigure}
		\begin{subfigure}[h]{0.24\textwidth}
			\centerline{\includegraphics[width=1\textwidth]{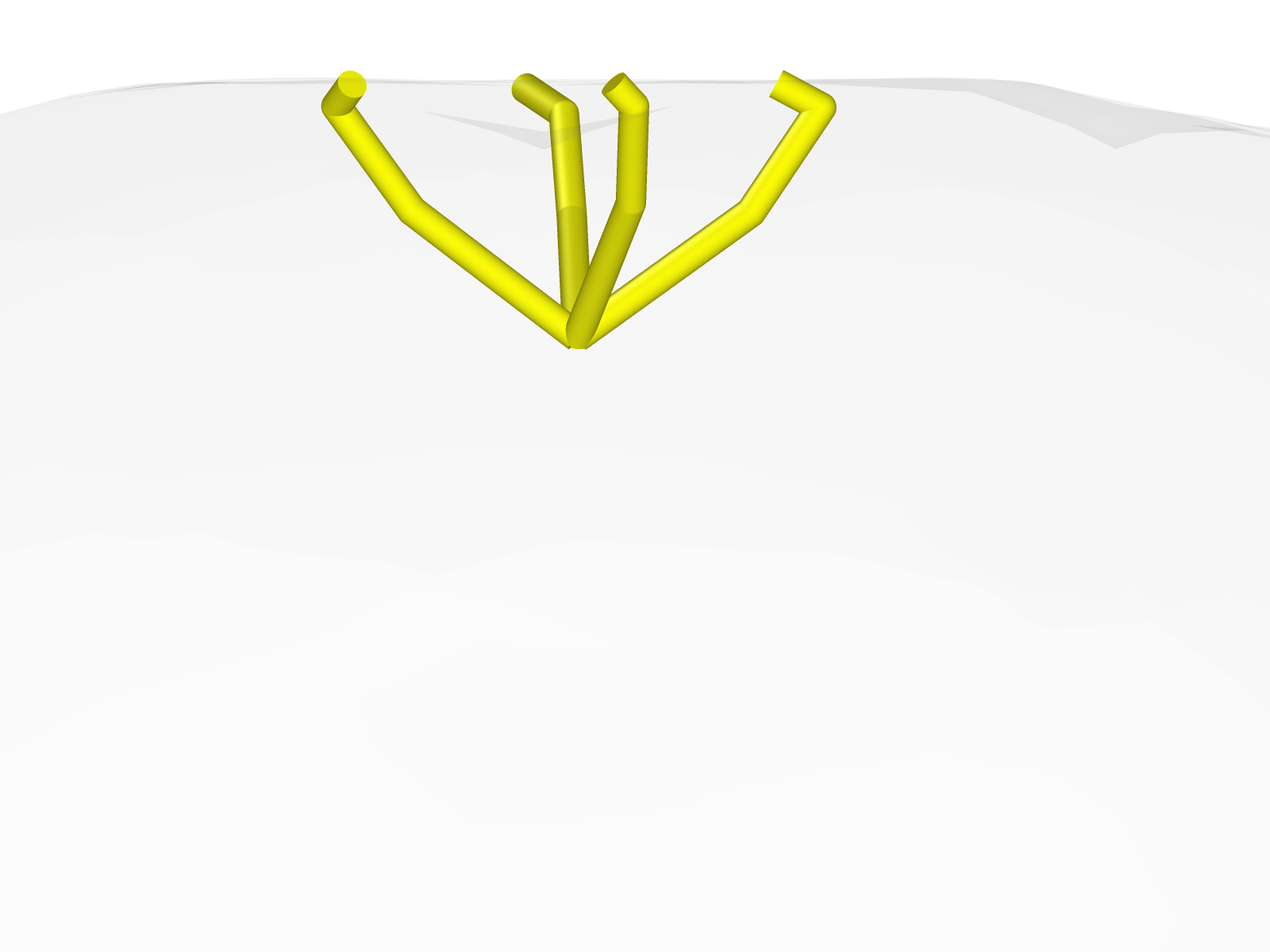}}
			\subcaption*{$\epsilon=3.5\%$}
		\end{subfigure}
		\begin{subfigure}[h]{0.24\textwidth}
			\centerline{\includegraphics[width=1\textwidth]{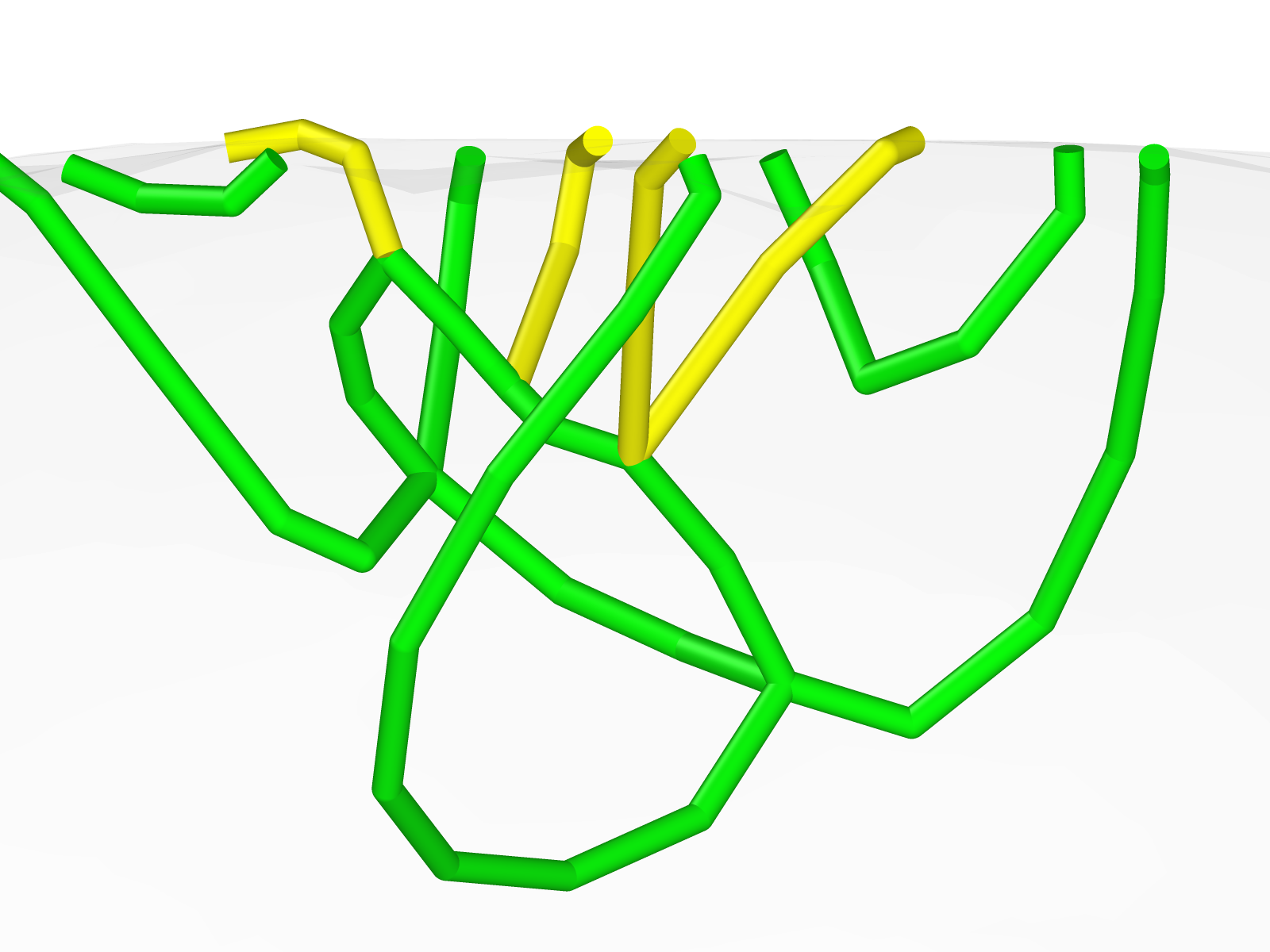}}
			\subcaption*{$\epsilon=6.1\%$}
		\end{subfigure}	
		\caption{Ag = 90\%}
	\end{subfigure}
	\caption{Dislocation structures during uniaxial compression of Coreshell nanosphere for different percentages of Ag. (Green line, Shockley partial dislocation with Burgers vectors $\frac{1}{6}<112>$; yellow line, hirth dislocation with Burgers vectors $\frac{1}{3}<100>$; Pink line, stair-rod dislocation with Burgers vectors $\frac{1}{6}<110>$; Red line indicates all other dislocation)}
	\label{fig13}
\end{figure}
For FGMs, dislocation initiation starts from the contact surface under compression at 1.6-1.7\% strain. Two Shockley partial dislocations are visible at 1.6\% strain for 10\%, 25\% of Ag but for 75\% \& 90\% of Ag dislocations are visible at 1.7\% strain as shown in Fig.\ref{fig12} as green line. So, we can conclude that dislocation initiation starts earlier for a lower percentage of Silver. However, at 1.5-1.6\% strain range in the upper surface, there is no Shockley Partial dislocation seen for 50\% of Ag. At 1.6\% strain, two Shockley Partial dislocations are shown on the lower surface. At 1.8\%strain, two Shockley Partial dislocations and one Hirth dislocation are visible on the upper surface. The inconsistent surface finish of FGM nanoparticles causes this phenomenon. With the initiation of dislocation, the stress drops a little. Subsequently, the stress increases rapidly and reaches a peak. The ultimate compressive strength was obtained at 2.6\% strain for 10\%, 25\% \& 90\% of Ag and at 2.7\% strain for 50\% of Ag and at 2.5\% strain for 75\% of Ag. At the ultimate compressive strength, four Hirth dislocations are visible which acts as a lock as shown in Fig. \ref{fig12} as yellow line. Hirth lock blocks the movement of the plane which leads to stress hardening. As the compression goes on, the stress abruptly falls to the lowest point at 3.1\% strain for 10\% \& 50\% of Ag, at 3.2\% strain for 25\% of Ag, at 3.3\% strain for 75\% of Ag, at 3.45\% strain for 90\% of Ag. In this stage, material failure occurs. Again, the stress starts to increase and reaches a peak point at 6\% strain for 10\%, 25\% \& 50\% of Ag, at 5.9\% strain for 75\%, at 6.1\% strain for 90\% of Ag. In this stage, Shockley partial dislocation propagates towards the center of the nanosphere as shown in Fig. \ref{fig12}.
For Core-shell, dislocation initiation starts from the contact surface under compression at 1.5-1.6\%. Two Shockley p dislocations are visible at 1.5\% strain for 10\%, 25\%, 50\%, 75\% of Ag but at 1.6\% strain for 90\% of Ag as shown in Fig. \ref{fig13}. So, we can say that dislocation initiation starts at the same strain rate for all percentages of Silver except 90\% of Ag. The ultimate compressive strength was obtained at 2.6\% strain for 10\%, 25\% \& 75\% of Ag and at 2.5\% strain for 50\% of Ag and at 2.7\% for 90\% of Ag. At ultimate compressive strength, four Hirth dislocations are visible which acts as a lock. As the compression goes on, the stress falls to the lowest point at 3.1\% strain for 10\% \& 25\% of Ag, at 3.2\% strain for 50\% of Ag, at 3.4\% strain for 75\% of Ag, at 3.5\% strain for 90\% of Ag. In this case, the strain rate is increased with the percentage increase of Silver. Again, the stress reaches a peak from the bottom at 6\% strain for 10\%, 25\% \& 50\% of Ag, at 6.1\% strain for 75\% \& 90\% of Ag.

\section{Conclusion}
In this work, MD simulations were performed to study the compressive deformation of Ag-Au functionally graded  and core-shell nanospheres oriented along [001]. We were especially interested in the effect of dislocations on the plasticity mechanisms and yield stress of varying Ag-Au \%wt. The nucleation of Shockley partial dislocations from the contact edge top and bottom surfaces appears to be the cause of first plastic deformation in many circumstances. Hirth dislocations form a pyramid-like structure and Shockley partial dislocation disappears. The pyramid-like structure may disappear immediately. With increasing compression, the Shockley partial dislocation becomes evident and propagates. With the visual of new partial dislocation, materials trends to behave like amorphism. We observed that the stress is mainly dependent on dislocation density. Initially, The stress level rises gradually, but with partial dislocation, the stress level reduces slightly. Then Hirth dislocation is visible and acts as a planar lock and helps to concentrate stress. As a result, we got the maximum strength in stress strain graph and observe the plasticity mechanisms. After that, stress drops abruptly and again starts to increase and now creates some wave-like stress pattern, indicating the amorphism of materials. We also found that the ultimate strength is higher in core-shell for different wt\% of Ag than FGMs. FGMs maintain a decreasing order with the increasing wt\% of Ag for the ultimate strength. But Core-shell does not follow the same trend. Core-shell gives an unusual trend, and ultimate strength increases upto a certain wt\% of Ag and then decreases. The phenomena may occur due to the core and shell effect on the nanosphere. To better understand the mechanical properties of nanospheres, we simulated the precisely [001] oriented compression of nanospheres with bare surfaces at low temperatures. However, such optimal settings are uncommon in real-life tests, which are typically conducted at room temperature with no precise orientations. The [001] orientation is directly responsible for the pyramidal-like formations in our simulation. A slight misorientation and the effects of temperature are expected to delay but not prevent their creation. In the end, a considerable number of additional calculations would be required to narrow the gap between experiments and theory. We will try to construct statistics for varied orientations and temperatures in a future investigation.

\section*{Acknowledgment}
The authors would like to acknowledge Multiscale Mechanical Modeling and Research Networks (MMMRN),ME Dept, BUET for their technical assistance to conduct the research.

\section*{Data availability}
The raw/processed data required to reproduce these findings cannot be shared at this time as the data also forms part of an ongoing study. 

\section*{CRediT authorship contribution statement}
\textbf{Prottay Malakar:} Methodology, Software, Data Curation, Visualization, Formal analysis, Writing - Original Draft.
\textbf{Md Al Rifat Anan:} Software, Data Curation, Visualization, Formal analysis, Writing - Original Draft.
\textbf{Mahmudul Islam:} Conceptualization,Formal analysis, Writing-Review \& Editing. 
\textbf{Md Shajedul Hoque Thakur:} Software, Writing-Review \& Editing.
\textbf{Satyajit Mojumder:} Project administration, Resources, Writing-Review \& Editing.

 \bibliographystyle{elsarticle-num} 
 \bibliography{Cite_nanosphere}





% \end{thebibliography}
\end{document}